\documentclass[letterpaper%
, twoside%
, 12pt%
, thesepararticles%
, english%
, creativecommons,hyperref, withAlgo2e %
]{_thETS_MyVersion}

\usepackage{amssymb}
\usepackage{siunitx}
\usepackage{tabularx}
\usepackage{amsmath}
\usepackage{graphicx}
\usepackage{subcaption} 
\usepackage{bm}
\usepackage{color, colortbl}
\usepackage{lineno}
\usepackage{setspace}
\usepackage{array, multirow}
\usepackage{diagbox}
\usepackage[ruled,vlined]{algorithm2e}
\usepackage{tabto}
\usepackage{accanthis}
\usepackage[utf8]{inputenc}
\usepackage[T1]{fontenc}
\definecolor{Gray}{gray}{0.9}
\definecolor{bulletblue}{RGB}{0,153,255}
\definecolor{bulletred}{RGB}{255,51,51}
\definecolor{bulletyellow}{RGB}{255,204,51}
\newcolumntype{Y}{>{\centering\arraybackslash}X}

\newcommand{\chapref}[1]{\hyperref[#1]{\textbf{Chapter~\ref{#1}}}} 

\makeatletter
\renewcommand\hyper@natlinkbreak[2]{#1}
\makeatother

\DeclareMathOperator*{\argmax}{argmax}

\newcites{refs}{LIST OF REFERENCES}

\title{Using Atom-Like Local Image Features to Study Human Genetics and Neuroanatomy in Large Sets of 3D Medical Image Volumes}

\author{Laurent \MakeUppercase{Chauvin}}
\authorcopyright{Laurent Chauvin}

\datesoutenance{July 14, 2022}

\datedepot{July 26, 2022}

\directeur{M. }{Matthew Toews}{Ecole de Technologie Superieure, Departement de Genie des Systemes}

\codirecteur{M.}{Jacques de Guise}{Ecole de Technologie Superieure, Departement de Genie des Systemes}

\jury{M.}{Ismail Ben Ayed}{Ecole de Technologie Superieure, Departement de Genie des Systemes}

\examinexterne{M.}{Jean-Baptiste Poline}{McGill University}

\president{M.}{Rafael Menelau Oliveira Cruz}{Ecole de Technologie Superieure, Departement de Genie Logiciel}

\begin{document}

\pagenumbering{Roman}

%%- Title page -%%
\maketitle

%%- Jury presentation -%%
\presentjury

%%- Foreword -%%
\begin{foreword}

%\lipsum[1] % Texte de remplissage pour donner un exemple de la mise en page

This thesis was the result of Laurent Chauvin's career path from the Surgical Planning Laboratory of the Harvard Medical School, where he assisted in a wide range of cutting edge research in medical robotics and image acquisition protocols related to magnetic resonance and ultrasound scanning projects of the human body.
In this thesis, he has ushered in an era in 2022 where we increasingly view objects in the 3D world and universe in terms of volumetric 3D images, a virtual reality or "metaverse", in which 3D structures such as the human body are naturally described by elementary particles defined by the most fundamental of mathematical properties, including the notions of atom-like particles within 3D cartesian space defined by three spatial variables $(x,y,z) \in R^3$ and a scale $\sigma$ defined by isotropic Einstein diffusion. A variety of arbitrary deep multi-channel neural network architectures will emerge to classify limited sets of labeled training data, however there will only ever be one scale-space $(x,y,z,\sigma)$ defined by the most elementary of differential operators $\nabla$ and $\nabla^2$ used by the likes of Newton, Leibnitz, Laplace, Gauss, Fourier, Fresnel, Huygens, Hamilton, Maxwell, Planck, Einstein, Schrodinger and Dirac from a single scalar field $I(x,y,z,\sigma) \in R^1$. Laurent Chauvin has expanded human knowledge in this theoretical domain by introducing discrete sign-parity invariance (SP-invariance) into the analysis of the human brain, equivalent to CP-invariance in the transition of elementary charged particles to antiparticles such as the proton to the antiproton, and in numerous discoveries regarding the nature of the human brain, of the potential human errors and the promise for equitable, unbiased healthcare of human subjects in large medical imaging datasets from fundamental image particles.\\~\\
- Matthew Toews

\end{foreword}
\chapter*{}
\vspace{15em}

{\textit{\hspace*{\fill} \large \guillemotleft You learn the most from things\\ \hspace*{\fill}you enjoy doing so much that\\ \hspace*{\fill}you don’t even notice time passing\guillemotright}\\}
\hspace*{\fill} - Albert Einstein

%%- Acknowledgements -%%
\begin{acknowledgements}
First, I want to express my most sincere gratitude to my supervisor Pr.Matthew Toews for his incredible pedagogy, his unwavering support and more importantly for sharing his infinite motivation, passion and wisdom with me. Thank you so much, it was such a great pleasure to work and learn from you. I couldn't have hoped for a better mentor.

I also want to express my gratitude to my lab mates and professors at ETS: Etienne, Mohsen, Herve, Christian, Kuldeep, Karthik, Jerome and Jose.

I also want to thank my friends for always being there, not only in good times (D\&D, Fiasco, Camping trips, etc...), but also, and especially in the bad ones to cheer me up. You have no idea how much that meant to me, so thank you Alex (my shooting star), Marmich, Nick, Laure, Romain, Clarisse, Gael, Joe, AnneMa, Eve, Baptiste, Shirley, Sam, Joel, Margaux, Valentin, Boris, Ted, Caroline, Lisa, Matthieu, Marine, David, Juliette, Florian, Sophie, Bastien, Celine, Pierre, Asli, Kristin, Sean, Alexis, Frank, Pierre-Yves, Thomas, Clement, Junichi, Sonia, Hatsuo, Mathieu, Fabien, Alan, Mik, Tonio, Tete, Coro, Vincent, Maxime. Please forgive me if I forgot some of you, but I am deeply thankful to each and every one of you.

A very very special thought to the Crou: Marion, Aude, Yahya, and Capucine, who have been my four pillars during this PhD. Your friendship means more to me than I can express and I'm so proud and grateful to have you in my life. I also want to thank Annick for her unconditional support, her patience, her faith in me, and whose laughs, courage and positivity inspired me.

I also want to deeply thank my parents and my brother, who always believed in me. None of this would have been possible without their trust and support during my whole life. Thank you for everything you have done for me all these years. You are a source of inspiration and pride to me.

A thought to Maya and Flora, my goddaughter and my niece. This work has been done in the hope of building a brighter future for you.

Finally, I want to dedicate my thesis to my grand-parents, that I always kept in my mind during this PhD, who gave me the strength and courage to never give up. I can only hope they are proud of what I accomplished thanks to them.

\end{acknowledgements}

%%- Summary -%%
\setcounter{chapter}{-1}
\begin{summary}{Utilisation de caractéristiques d'image inspirée par le modele atomique pour étudier la génétique et la neuroanatomie humaines dans de grands ensembles d'images médicales 3D}{IRM, Image Medicale Volumetrique, Point-clef Invariant d'Echelle, Large Jeux de Donnees, Invariance, Symmetrie Discrete, ADN, Neuroanatomie, Empreinte Cérebrale}

Les contributions de cette thèse découlent de la technologie développée pour analyser de grands ensembles d'images volumétriques en termes de caractéristiques locale, inspirée du modèle atomique, extraites dans l'espace d'image 3D, suivant la célèbre transformation de caractéristique invariante d'échelle (SIFT)~\citep{Lowe2004DistinctiveKeypoints} dans l'espace d'image 2D. De nouvelles propriétés de caractéristiques sont introduites, notamment un signe de caractéristique binaire $s \in \{-1,+1\}$, analogue à une charge électrique, et un ensemble discret d'états d'orientation de caractéristiques symétriques dans l'espace 3D. Ces nouvelles propriétés sont exploitées pour étendre l'invariance des caractéristiques en incluant la transformation d'inversion de signe et de parité (SP), analogue à la transformation de conjugaison de charge et de parité (CP) entre une particule et son antiparticule en mécanique quantique, ce qui permet de prendre en compte l'inversion du contraste d'intensité locale entre différente modalités d'imagerie et les réflexions d'axe dues à la symétrie de forme.
Un nouveau noyau exponentiel $\mathcal{K}(f_i,f_j) \in [0,1]$ est proposé pour quantifier la similarité d'une paire de caractéristiques $f_i$ et $f_j$ extraites dans différentes images à partir de leurs propriétés, notamment la localisation, l'échelle, l'orientation, le signe et l'apparence. Ce noyau peut être calculé à l'aide d'algorithmes rapides d'indexation des caractéristiques du plus proche voisin, réduisant la complexité de la recherche d'un ensemble de $N$ caractéristiques de $0(N)$ à $0(log~N)$, permettant à des algorithmes d'analyse d'images très efficaces de s'adapter à de large ensembles de données. Une nouvelle mesure intitulée le Jaccard flou $J(A,B)$ est proposée pour quantifier la similarité d'une paire d'ensembles de caractéristiques $A=\{f_i\}$ et $B=\{f_j\}$ sur la base de leur chevauchement ou intersection-sur-union $|A \cap B|/|A \cup B|$, où le noyau $\mathcal{K}(f_i,f_j)$ établit une équivalence non binaire ou douce entre une paire d'éléments de caractéristiques $(f_i,f_j)$. Le Jaccard flou peut être utilisée pour identifier des paires d'ensembles de caractéristiques extraites de même individus ou de même familles avec une grande précision, et un simple seuil de distance a permis de découvrir de manière surprenante des erreurs d'étiquetage d'individus et de familles auparavant inconnues dans d'importants ensembles de données publiques de neuroimagerie. La distance flou de Jaccard peut également être évaluée entre un ensemble de caractéristiques et un groupe d'ensembles de caractéristiques afin de réaliser une classification en fonction des étiquettes de données par groupe. Un nouvel algorithme est proposé pour recaler ou aligner spatialement une paire d'ensembles de caractéristiques $A$ et $B$, intitulé SIFT Coherent Point Drift (SIFT-CPD), en identifiant une transformation $T : B \rightarrow A$ qui maximise le Jaccard flou $J(A,T\circ B)$ entre un ensemble de caractéristiques fixe $A$ et un ensemble transformé $T\circ B$. L'algorithme SIFT-CPD permet d'obtenir un recalage plus rapide et plus précis que l'algorithme CPD original basé uniquement sur les informations de localisation des caractéristiques, dans une variété de contextes difficiles, notamment la tomographie thoracique, l'IRM cérébrale, les modalités multiples, la variabilité naturelle entre les sujets et les pathologies telles que les tumeurs et les maladies neurodégénératives.

\end{summary}

%%- Abstract -%%
\begin{abstract}{MRI, Volumetric Medical Images, Scale-invariant feature, Keypoints, Big data, Invariance, Discrete Symmetry, DNA, Neuroanatomy, Brain fingerprinting}

The contributions of this thesis stem from technology developed to analyse large sets of volumetric images in terms of atom-like features extracted in 3D image space, following the well-known Scale Invariant Feature Transform (SIFT)~\citep{Lowe2004DistinctiveKeypoints} in 2D image space. New feature properties are introduced including a binary feature sign $s \in \{-1,+1\}$, analogous to an electrical charge, and a discrete set of symmetric feature orientation states in 3D space. These new properties are leveraged to extend feature invariance to include the sign inversion and parity (SP) transform, analogous to the charge conjugation and parity (CP) transform between a particle and its antiparticle in quantum mechanics, thereby accounting for local intensity contrast inversion between imaging modalities and axis reflections due to shape symmetry. A novel exponential kernel $\mathcal{K}(f_i,f_j) \in [0,1]$ is proposed to quantify the similarity of a pair of features $f_i$ and $f_j$ extracted in different images from their properties including location, scale, orientation, sign and appearance. This kernel may be computed using fast approximate k-nearest neighbor feature indexing algorithms that reduce the computational complexity of searching a set of $N$ features from $O(N)$ to $O(log~N)$, allowing highly efficient image analysis algorithms to scale to arbitrarily large datasets. A novel measure entitled the soft Jaccard $J(A,B)$ is proposed to quantify the similarity of a pair of feature sets $A=\{f_i\}$ and $B=\{f_j\}$ based on their overlap or intersection-over-union $|A \cap B|/|A \cup B|$, where the kernel $\mathcal{K}(f_i,f_j)$ establishes non-binary or soft equivalence between a pair of feature elements $(f_i,f_j)$. The soft Jaccard may be used to identify pairs of feature sets extracted from the same individuals or families with high accuracy, and a simple distance threshold led to the surprising discovery of previously unknown individual and family labeling errors in major public neuroimage datasets. The soft Jaccard may also be evaluated between a feature set and a group of feature sets in order to achieve classification according to group-wise data labels. A new algorithm is proposed to register or spatially align a pair of feature sets $A$ and $B$, entitled SIFT Coherent Point Drift (SIFT-CPD), by identifying a transform $T: B \rightarrow A$ that maximizes the soft Jaccard $J(A,T\circ B)$ between a fixed feature set $A$ and a transformed set $T\circ B$. SIFT-CPD achieves faster and more accurate registration than the original CPD algorithm based on feature location information alone, in a variety of challenging contexts including chest CT, brain MRI, multiple modalities, natural inter-subject variability and pathologies such as tumors and neuro-degenerative disease.

\end{abstract}

%%- Table of contents -%%
\tableofcontents

%%- List of tables -%%
\listoftables

%%- List of figures -%%
\listoffigures

\listofalgorithms

%%- List of abbreviations -%%
\begin{listofabbr}[3cm]
\item [AUC] Area Under the Curve
\item [CNN] Convolutional Neural Network
\item [CPD] Coherent Point Drift
\item [DOH] Diffusion Orientation Histogram
\item [ICP] Iterative Closest Point
\item [ML] Machine Learning
\item [MRI] Magnetic Resonance Imaging
\item [PACS] Picture Archiving and Communication System
\item [SIFT] Scale-Invariant Feature Transform
\item [VBM] Voxel-based Morphometry
\item [HCP] Human Connectome Project
\item [ADNI] Alzheimer's Disease Neuroimaging Initiative
\item [OASIS] Open Access Series of Imaging Studies
\item [SM] Same Subject
\item [MZ] Monozygotic Twin
\item [DZ] Dizygotic Twin
\item [HS] Half Sibling
\item [FS] Full Sibling
\item [UR] Unrelated Individual
%\item [TMI] Transactions on Medical Imaging
%\item [CDMRI] Computational Diffusion in Magnetic Resonance Imaging
%\item [MICCAI] Medical Image Computing and Computer Assisted Intervention
%\item [IPMI] Information Processing in Medical Imaging
%\item [OHBM] Organization for Human Brain Mapping
\end{listofabbr}

%%- List of symbols -%%
%\begin{listofsymbols}[3cm]
%\item [a] Première lettre de l'alphabet
%\item [A] Première lettre de l'alphabet en majuscule
%\end{listofsymbols}

\cleardoublepage

\pagenumbering{arabic}

% Marginpar to the left of the document
\reversemarginpar

%%%%%%%%%%%%%%%%%%%%%%%%%%%%%%%%%%%%%%%%%%%%%%%%%%%
% THESIS EXAMPLE
%%%%%%%%%%%%%%%%%%%%%%%%%%%%%%%%%%%%%%%%%%%%%%%%%%%

% -------------------------------------------------
% Plan Introduction
% -------------------------------------------------

%   Context and Motivation
%       medical image analysis, big data, personalized medicine
%       currently high focus on CNN deep learning-based classification

%       CNNs, deep learning very popular 
%       Keypoints: special type of CNN (paper with JB, needs to be published soon)

%   Technical Challenges:
%       CNN: highly effective for many data, few label contexts
%              less effective for many labels O(N), few data.
%              require manual labels, which may be erroneous for specific people
%       Keypoints: effective distances measure

%   Contributions
%      O(N log N) algorithm,  keypoint extraction, representation.
% 1: Soft Jaccard similarity measure in keypoint space
% 2: individual identification in prominent neuroimaging datasets, neuroimage fingerprint
% 3: family member identification in HCP
% 4: Registration CPD

% -------------------------------------------------
\setcounter{chapter}{0}
\begin{introduction}
\chaptermark{Introduction}
\begin{figure}
    \centering
    \includegraphics[width=0.9\linewidth]{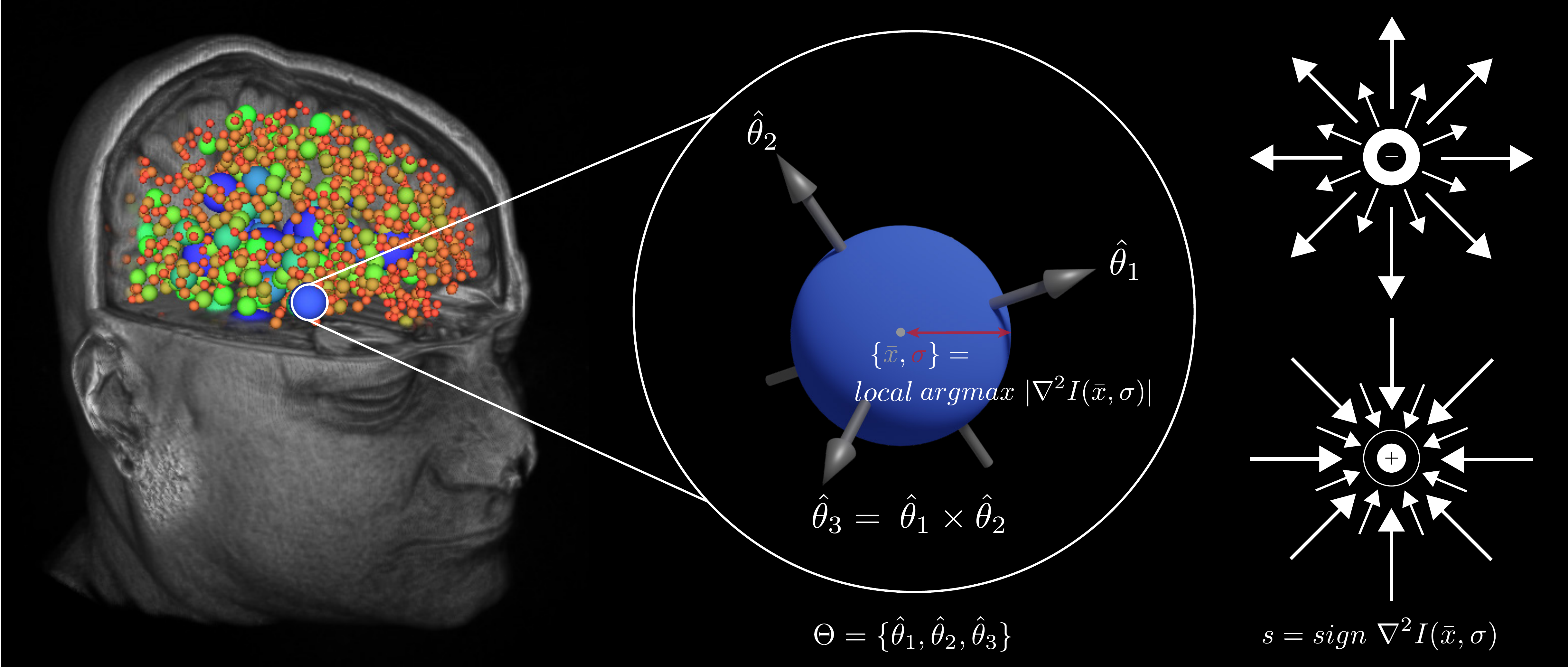}
    \captionsetup{width=0.9\textwidth}
    \caption{Illustrating atom-like image features extracted in a 3D magnetic resonance image (MRI) of the human head. Properties are shown for an individual feature, including location $\bar{x}$, scale $\sigma$, orientation axes $\Theta$ and binary sign $s = \{-,+\}$}
    \label{fig:sift_properties}
\end{figure}

The structure of matter in the 3D world is often described in terms of localized phenomena. In the 5th century BCE, Leucippus of Miletus and his disciple Democritus proposed the term 'atomos' to describe what they thought to be indivisible and fundamental particles of nature. Newton proposed that light might be composed of corpuscular particles~\citep{Newton1704OpticksLight}.  Einstein described the motion of particles as diffusing passively according to random Brownian motion~\citep{Einstein1905UberTeilchen}, equivalent to the heat diffusion equation ${\partial u}/{\partial t} \propto \nabla^2 u$. The modern atomic model of quantum chemistry in its most basic form is defined by Schrodinger's wave equation~\citep{Schrodinger1926UndulatoryMolecules}, equivalent to the heat equation and with complex-valued wave solutions permitting momentum, relating a potential field $\psi(x,t)$ over space and time coordinates $(x,t)$ to fundamental operators like the gradient $\nabla$ with respect to time and the Laplacian $\nabla^2$:
\begin{equation}
   i\hbar \frac{\partial \psi(x,t)}{\partial t} = -\frac{\hbar}{2m} \nabla^2 \psi(x,t),
\end{equation}
where $m$ is the mass of a particle and $\hbar$ is the reduced Planck constant.

The contributions of this thesis focuses on keypoint methods, where 3D image volumes such as MRI scans of the human brain are represented and analyzed as sets of highly informative atom-like image features, as shown in Figure~\ref{fig:sift_properties}. Features are identified and encoded via fundamental Gaussian, Laplacian and gradient operators~\citep{Toews2010FeaturebasedPatterns,Toews2013FeaturebasedImages,Rister2017VolumetricKeypoints}, following the highly successfully scale-invariant feature transform (SIFT)~\citep{Lowe2004DistinctiveKeypoints, Lindeberg1998FeatureSelection}, and used in highly efficient and robust imaging applications with no explicit training procedures, including registration~\citep{Luo2018UsingRegistration,Frisken2019PreliminaryUltrasound,Machado2018DeformableNeurosurgery}, segmentation~\citep{Wachinger2018KeypointSegmentation,Gill2014RobustApproach} and indexing~\citep{Toews2015FeatureBasedImages}. 
%The recent introduction of a modern GPU-based machine learning paradigm provides an alternative to fundamental operators, where vector operators may be derived from training data~\citep{Yi2016LIFTTransform}. 

\section{Problem statement and motivation}

%Problem: large medical image datasets, indexing specific subjects, fine-grained labels few examples per label (individuals, family members), memory storage of all data. Personalized medicine, curate data for individual subjects, accurate, subject-specific analysis without biased due to race or sex.

Large medical imaging datasets and machine learning algorithms provide new opportunities to increase scientific understanding of anatomy, function and connectivity from subject groups defined by factors such as genetics, demographics or disease status. However, levering information in large sets of volumetric image data is a challenging task, for example, maintaining and curating imaging data for individual human patients in a large, hospital PACS (Picture Archiving and Communication System) database. In the age of personalized medicine promising to tailor healthcare solutions to the unique characteristics of individuals~\citep{Hamburg2010PathMedicine}, can patients be assured of accurate diagnosis and treatment from their medical imaging data?

%Challenge: Machine learning is the default paradigm, however works best for large datasets with relatively few classes ,e.g. ImageNet 1000 classes 1000 examples per class. The main challenges are bias and lack of invariance in machine learning, need for training, large datasize. Bias is a particularly pernicious, unsolved phenomenon in machine learning, and recent work has shown that state-of-the-art models for face detection~\citep{buolamwini2018gender} and language modeling~\citep{bender2021dangers} learn built-in biases that negatively discriminate towards minority populations. There are currently few methods for automatically detecting errors in training set labels, which may lead to catastrophic errors for specific patients in the context of medical image data.

Currently, even for the most basic tasks of dataset curation and ensuring data integrity, there are few options for detecting potential errors at the level of individual subjects. For example, images with incorrect subject IDs may lead to bias in analyzing or training from large public datasets such as OASIS~\citep{Marcus2007OpenAdults}, ADNI~\citep{JackJr2008AlzheimerMethods} or HCP~\citep{VanEssen2013WUMinnOverview} and potentially lead to catastrophic misdiagnosis of individual patients. Storing and curating a set of $N$ images entails a $O(N^2)$ computational complexity for comparing all image pairs, and is generally intractible for naive algorithms and large datasets $N$. Computing pairwise whole-image (dis)similarity is a challenging task in and of itself, computational complexity challenges aside. Standard feature spaces, including atlases or parcellations, tend to simplify computational complexity while masking individual diversity. For example, the dominant computational paradigm and the {\em de facto} state-of-the-art solution to generic tasks such as classification is the deep neural network architecture~\citep{LeCun1989BackpropagationRecognition} coupled with gradient-based backpropagation training on graphics processing units (GPUs)~\citep{Krizhevsky2012ImageNetNetworks}. Of the numerous challenges to this paradigm, including training set bias~\citep{Torralba2011UnbiasedBias,Hermann2020OriginsNetworks,Buolamwini2018GenderClassification}, length training procedures~\citep{Bender2021DangersBig}, improperly calibrated outputs~\citep{Guo2017CalibrationNetworks}, lack of intrinsic invariance~\citep{Bardes2022VICRegLearning}, susceptibility to spoofing~\citep{Nguyen2015DeepImages} and attacks~\citep{Su2019OneNetworks}, the most relevant is training data sparsity, i.e. generalizing from few examples. In highly sparse contexts such as specific image retrieval, deep machine learning has not significantly improved upon classic methods for keypoint-based matching in 2D computer vision~\citep{Balntas2017HPatchesDescriptors,Bellavia2020ThereMatching}. Intuitively, machine learning seeks to generalized from training datasets, however in the process they tend to obscure or bias unique or idiosyncratic information required for fine-grained inferences at the level of individual images.

This thesis adopts an approach particularly well-suited to individual subject indexing. We adopt a memory-based (or instanced-based) model~\citep{Cover1967NearestClassification,Brin1995NeighborSpaces,Toussaint2002ProximityProgress} specifically to index all patient-specific information stored in memory, and thereby preserve patient-specific anatomical features such as cortical folding in brain MRI that may be unique to individuals or shared by close relatives. Each image is encoded as set of generic, salient localized characteristics or features, defined from nothing other than elementary mathematical operators or filters: the Gaussian $G = \{+1, +1\}$ the gradient $\nabla = \{-1, +1\}$ and the Laplacian $\nabla^2 = \{-1, +1\} * \{-1, +1\} = \{-1, +2, -1\}$. These elementary filters are fundamentally symmetric, rotation-invariant and unbiased (i.e. the Gaussian and Laplacian) or may be so via uniform sampling (i.e. the gradient). These may be used, without training, to identify generic image features and reliably recognised individual-specific structure when present in two or more images.

\section{Research objectives and contributions}
This thesis seeks to advance the theory of fundamental 3D image features by investigating the following research questions: What are the geometrical properties of local features extracted in 3D images, and how do these relate to fundamental properties of physical particles such as atoms, including symmetry, invariance and charge? What can local features tell us about the human brain and the genetic code that defines aspects such as family structure or biological sex?

The primary technological contributions of this thesis stem from characterizing the geometrical properties of 3D local features, in the context of the local invariant feature paradigm, where $O(N^2)$ data pairs may be evaluated in $O(N~log~N)$ computational complexity via highly robust and efficient local feature indexing, where inference scales gracefully to arbitrarily large datasets of generic image data with no explicit training procedures. These contributions may be listed as follows:
\begin{enumerate}
    \item We provide the first image feature description including a binary sign $s=\{-1,+1\}$ and discrete orientation state. This leads to feature descriptors that are invariant discrete symmetry, including charge conjugation and parity transforms (CP-symmetry), currently a topic of high interest in studying the transition of matter to anti-matter~\citep{Cronin1981CPOrigin,Pascoli2020MatterViolated,Georgescu2020CPViolation}. These are shown in Figure~\ref{fig:sift_properties}.
    \item We propose a kernel function to quantify the similarity of a pair features observed in different images, based on the variability of their properties including feature geometry and appearance. This leads to higher accuracy in tasks like probabilistic image registration and prediction.
    \item We propose a measure of (dis)similarity between a pair of feature sets based on their overlap or intersection-over union, quantified by a soft Jaccard-like distance. This provides a solution to a number of problems relating to massive datasets and personalized medicine, including curation and prediction.
\end{enumerate}

These technical contributions lead to a number of novel experimental discoveries:
\begin{enumerate}
    \item Individuals and family members including siblings may be rapidly identified via low Jaccard distance between MRI pairs, allowing personalized medicine to flag rare but potentially catastrophic errors in patient labels. Our approach was the first to identify labeling errors in large widely used medical imaging datasets, including images of different people labeled as the same person, images of the same person labeled as different people, images of family members labeled as unrelated due to inconclusive genotyping.
    \item The Jaccard distance between brain MRIs reflects the genetic separation between subjects, increasing between pairs of monozygotic twins, siblings and half-siblings, and unrelated individuals.
    \item Pairs of individuals of different race and sex may often have higher anatomical brain similarity than individuals of the same race and sex. This indicates that race and sex do not significantly factor in neuroanatomical similarity, just as they do not significantly impact overall genetic similarity in terms of shared single nucleotide polymorphism (SNP).
\end{enumerate}

\section{Publications}
The work presented here resulted in several publications in journal papers and conferences, detailed in the following sections.

\textbf{\underline{Journal Articles:}}
\begin{enumerate}
    \item \textbf{Laurent Chauvin}, Kuldeep Kumar, Christian Wachinger, Marc Vangel, Jacques de Guise, Christian Desrosiers, William Wells III, Matthew Toews, and Alzheimer’s Disease Neuroimaging Initiative. "Neuroimage signature from salient keypoints is highly specific to individuals and shared by close relatives." \textit{NeuroImage 204: 116208, 2020.} - \textbf{Accepted}\\
    (\textit{Figure selected as cover by the journal})
    \item \textbf{Laurent Chauvin}, Kuldeep Kumar, Christian Desrosiers, William Wells III, and Matthew Toews. "Efficient Pairwise Neuroimage Analysis using the Soft Jaccard Index and 3D Keypoint Sets." \textit{IEEE Transactions on Medical Imaging 2021} - \textbf{Accepted}
    \item \textbf{Laurent Chauvin}, William Wells III, and Matthew Toews. "A Model of Atom-like Image Structure using 3D SIFT and Discrete CP-Symmetry." \textit{IEEE Transactions on Pattern Analysis and Machine Intelligence 2022} - \textbf{Submitted}
\end{enumerate}

\textbf{\underline{Conference Articles:}}
\begin{enumerate}
    \item \textbf{Laurent Chauvin} and Matthew Toews. "Curating Subject ID Labels using Keypoint Signatures." \textit{MICCAI Workshop on Large-scale Annotation of Biomedical data and Expert Label Synthesis (LABELS), 2020.}
    \item \textbf{Laurent Chauvin}, Kuldeep Kumar, Christian Desrosiers, Jacques De Guise, William Wells, and Matthew Toews. "Analyzing brain morphology on the bag-of-features manifold." \textit{International Conference on Information Processing in Medical Imaging, pp. 45-56. Springer, Cham, 2019.} \textbf{(Oral Presentation)}
    \item \textbf{Laurent Chauvin}, Kuldeep Kumar, Christian Desrosiers, Jacques De Guise, and Matthew Toews. "Diffusion Orientation Histograms (DOH) for diffusion weighted image analysis." \textit{Computational Diffusion MRI, pp. 91-99. Springer, Cham, 2018.}
\end{enumerate}
\newpage
\textbf{\underline{Abstracts:}}
\begin{enumerate}
    \item \textbf{Laurent Chauvin}, Sukesh Adiga V, Jose Dolz, Herve Lombaert, Matthew Toews. "A Large-scale Neuroimage Analysis using Keypoint Signatures : UK Biobank". \textit{26th annual meeting of the Organization for Human Brain Mapping (OHBM), 2020}
\end{enumerate}

% % - Contributions:
% % -- First Author (7): Neuroimage, TMI, CDMRI (DOH), MICCAI, IPMI, OHBM, Registration paper
% % -- Co-Author (4): Kuldeep MICCAI, JB GPU, Kuldeep CDMRI?, Eleyine

\section{Thesis outline}
The work presented in this thesis is organized in 4 chapters, followed by a conclusion and recommendations for future work. In \chapref{chap1} local features and their applications will be described as well as key concepts necessary for a full understanding of the work presented here. \chapref{chap2} will lay the foundations of this work, presenting the theoretical model of extended 3D SIFT features, as well as a measure of the uncertainty of pairwise features arising from the same underlying distribution based on kernel density estimations over features appearance, location, scale and orientation. A practical application will be demonstrated by integrating this model in the Coherent Point Drift (CPD) algorithm, resulting in the capability of registering images in multiple difficult tasks (e.g. multi-modal images, tumors), more accurately and faster than the original implementation. In \chapref{chap3}, an application of our model for quantitatively measuring shared neuroanatomical structures across images, which led to multiple discoveries in widely used public neuroimaging datasets, will be presented. \chapref{chap4} is an extension of \chapref{chap3}, introducing a new geometrical kernel to reduce the impact of false keypoint matches, with similar appearance, but different geometry. A constraint induced by the geometrical kernel is the necessity for images to be aligned. However, this limitation can be lifted by registering images with the algorithm presented in \chapref{chap2}.
Finally, the main contributions of this thesis, the possible limitations and directions for future developments will be discussed.

\end{introduction}

%%- Uncomment the literature review for a thesis by articles -%%
\begin{literaturereview}
\label{chap1}
Image-based machine learning may be broadly defined in terms of artificial neural networks, where technical advances often entail lowering the difficulty of neural network training for large parameter spaces and enhancing the capacity to generalise from smaller training data sets. Artificial neural networks were designed to model and reproduce biological neural network processes, such as the Rosenblatt perceptron model, in which the output of a neuron is a weighted linear mixture of its inputs~\citep{Rosenblatt1958PerceptronBrain}. Soon after, the multi-layer perceptron (MLP)~\citep{Rosenblatt1961PrinciplesMechanisms} arose, which arranged perceptrons into numerous, feed-forward layers, assessed sequentially. Incorporation of mathematical invariance has been a frequent focus, enabling networks to learn using fewer data points or to generalise to previously unknown material. In the 1980s, LeCun designed the translation-invariant convolutional neural network (CNN) architecture~\citep{LeCun1989BackpropagationRecognition}, which, for a number of $N$ pixels, reduced the memory requirements of the weights from $O(N^2)$ for fully-connected MLP network layers to $O(N)$ for sets of weights computed locally and shared across the image in a way invariant to image translation, i.e. convolution filters. Despite its efficiency, CNN training using the backpropagation algorithm was too computationally complex for practical applications on CPU-based architecture computers until the development of highly parallel GPU-based convolution, allowing deep CNNs to be trained from 100000s of images, such as the ImageNet dataset~\citep{Krizhevsky2012ImageNetNetworks}, containing 1000 objects with 1000 training examples per category. However, the size of the dataset the network has been trained on has a significant impact on its performances~\citep{Luo2018HowPerformance}, especially for complex tasks, with many samples per class to be able properly generalize classes. Another possible issue is the biasing of the model toward its training set~\citep{Geirhos2019ImageNettrainedRobustness} while obscuring idiosyncrasies of samples to preserve only their shared characteristics.
Bias is a particularly pernicious, unsolved phenomenon in machine learning, largely as a result of random initialization and stochastic backpropagation, leading to unintended bias and asymmetry in trained filters, negatively affecting interpretation of both low-level of image textures~\citep{Torralba2011UnbiasedBias,Hermann2020OriginsNetworks} and high-level discrimination at the state-of-the-art deep models for face detection~\citep{Buolamwini2018GenderClassification,Serna2021InsideBiasBiometrics} and language~\citep{Bender2021DangersBig}
Although deep learning played a significant role in the development of medical imaging (e.g. automatic segmentation~\citep{Lai2015DeepSegmentation}, tumor detection~\citep{Baur2021AutoencodersStudy}, classification~\citep{Song2015LargeClassification}), personalized medicine requires the preservation of individual-specific information. By looking back in the early 2000s, we can find such technology in the field of computer vision and largely popularized by David Lowe~\citep{Lowe2004DistinctiveKeypoints}: the local handcrafted features (or keypoints).  

% Keypoints / SIFT
Keypoint algorithms transform image data $I(\bar{x})$ sampled on a regular $\bar{x}=\{x,y,z\} \in R^3$ pixel lattice into a sparse set of salient image observations or keypoints within a higher-dimensional geometrical space, i.e. $I(\bar{x},\Theta, \sigma)$ including local orientation $\Theta \in SO(3)$ and scale $\sigma \in R^+$, in addition to simple 3D location $\bar{x}$. Convolutional filters with special mathematical forms are used to discover Laplacian-of-Gaussian keypoints, ensuring invariance to image translation~\citep{Harris1988CombinedDetector}, scaling~\citep{Lindeberg1998FeatureSelection}, similarity~\citep{Lowe1999ObjectFeatures}, and affine deformations~\citep{Mikolajczyk2004ScaleDetectors}.
Until the development of programmable graphics processing units (GPUs), such as the Compute Unified Device Architecture (CUDA) development system, that could leverage its massive parallelization of the backpropagation learning algorithm~\citep{Krizhevsky2012ImageNetNetworks} to train full image-based CNNs, keypoint methods remained the de facto standard for image-based learning.
The SIFT algorithm's backbone, developped by David Lowe in 1999~\citep{Lowe1999ObjectFeatures}, is the Gaussian scale-space~\citep{Lindeberg1994ScalespaceScales} formed by recursive Gaussian convolution filtering, which may be effectively constructed using separable and even uniform 1D filters.
The SIFT algorithm has the following advantages: keypoints are identified in a way that is independent of image resolution, object pose, or intensity variations; there is no need for a training procedure or data; Gaussian and Gaussian derivative filters are rotationally symmetric and uniformly sampled; and the SIFT algorithm is unbiased by the specific training data used. Because of these factors, the SIFT keypoint algorithm and descriptor remain an effective and competitive solution in situations when few training examples are available, such as matching photographs of the same exact 3D objects~\citep{Bellavia2020ThereMatching} or scenes~\citep{Mishchuk2017WorkingLoss}.
The SIFT method bears some resemblance, on certain aspects, with the mammalian vision system, such as the retinal center-surround processing in the lateral geniculate nucleus and the directed gradient operators organised into hypercolumns as discovered by Hubel\&Wiesel in 1968~\citep{Hubel1968ReceptiveCortex}.
The scale-space resulting from Gaussian convolution operations $I(x,y,z,\sigma)=I(x,y,z)*N(\sigma)$ representing the process of isotropic diffusion of particles (such as the Brownian motion) or heat.
The Laplacian-of-Gaussian operator $\nabla^2 I(x,y,z,\sigma)$ is a scale-normalized operator that is equivalent to the Laplacian operator that forms the kinetic energy component of the Hamiltonian, such as in the Schrodinger equation.

% 3D SIFT
With the popularization of volumetric images, it became necessary to generalize the SIFT algorithm to 3D image data, used in a variety of applications including video processing~\citep{Scovanner20073dimensionalRecognition}, 3D object detection~\citep{Flitton2010ObjectVolumes}, and medical image analysis~\citep{Cheung2009SIFTTransform,Allaire2008FullAnalysis,Toews2013EfficientFeatures,Rister2017VolumetricKeypoints}. Our work here is based on the 3D SIFT-Rank approach~\citep{Toews2013EfficientFeatures}, which is unique in the way local keypoint orientation and descriptors are computed. Keypoint orientations may be parameterized via a suitable representation such as quaternions or 3x3 rotation matrices, generally determined from dominant gradients of the scale-space $\nabla I(x,y,z,\sigma)$ in a scale-normalized neighborhood around the keypoint. The 3D SIFT-Rank method identifies dominant peaks in a discrete spherical histogram of gradient, rather than solid angle histograms~\citep{Scovanner20073dimensionalRecognition,Allaire2008FullAnalysis}, partial 3D orientation information~\citep{Cheung2009SIFTTransform} or principal components of the local gradient~\citep{Rister2017VolumetricKeypoints}. This allows identification of multiple dominant orientations (3x3 rotation matrices) at each keypoint $(x,y,z,\sigma)$, leading to multiple descriptors per keypoint and providing robustness to noise, rather than single orientations per keypoint~\citep{Rister2017VolumetricKeypoints}. Descriptors generally encode the scale-space gradient $\nabla I(x,y,z,\sigma)$ in the neighborhood about the keypoint, following reorientation. The 3D SIFT-Rank method adopts a compact 64-element gradient orientation descriptor, where the local orientation space is sampled according to $x\times y\times z=2\times2\times2=8$ spatial bins and 8 orientation bins. Other approaches adopt descriptors that are orders of magnitude larger, e.g. 768-elements from $4^3$ spatial bins x 12 orientation bins, leading to similar image matching performance at a much greater memory footprint~\citep{Rister2017VolumetricKeypoints}. Finally, the SIFT-Rank descriptor is normalized by ranking~\citep{Toews2009SIFTRankCorrespondence}, offering invariance to monotonic variations in image gradient. 

% Applications
The 3D SIFT-Rank method has been used in a variety of keypoint applications analyzing 3D images of the human body. These include modeling the development of the infant human brain over time~\citep{Toews2012FeaturebasedMRI}, keypoint matching between different image modalities having multi-modal, non-linear intensity relationships~\citep{Toews2013FeaturebasedImages}, robust alignment of lung scans~\citep{Gill2014RobustApproach}, kernel density formulation for efficient memory-based indexing from large image datasets~\citep{Toews2015FeatureBasedImages}, efficient whole-body medical image segmentation via keypoint matching and label transfer~\citep{Wachinger2015KeypointSegmentation,Wachinger2018KeypointSegmentation}, alignment of 4D cardiac ultrasound sequences~\citep{Bersvendsen2016RobustSequences}, alignment of 3D ultrasound volumes for panoramic stitching~\citep{Ni2008VolumetricSIFT}, identifying family members from brain MRI~\citep{Toews2016HowFeatures}, and large-scale population studies using multiple neurological MRI modalities~\citep{Kumar2018MultimodalFramework}. The 3D SIFT-Rank method was also used for robust image alignment in the context of image-guided neurosurgery~\citep{Luo2018FeaturedrivenCompensation}, including non-rigid registration~\citep{Machado2018NonrigidMatching} based on regularization of deformation fields via thin-plane splines and finite element models~\citep{Frisken2019PreliminaryUltrasound,Frisken2020ComparisonResection}, and robust filtering of image-to-image correspondences using the variogram~\citep{Luo2018UsingRegistration}.

% Other descriptors
Alternative keypoint descriptors have been proposed based on vector data, e.g. diffusion MRI histograms of the human brain~\citep{Chauvin2018DiffusionAnalysis}.
In terms of keypoint descriptors, a number of 2D descriptors have been proposed, generally based on local image gradient orientation information, including gradient orientation~\citep{Lowe1999ObjectFeatures}, ORB~\citep{Rublee2011ORBSURF}, BRIEF~\citep{Calonder2011BRIEFFast}, typically these have not been extended to 3D. Rank-order normalization has been shown to improve upon standard 2D gradient descriptors~\citep{Toews2009SIFTRankCorrespondence}. Deep learning has been used to extract keypoints and descriptors, learned invariant feature transform (LIFT)~\citep{Yi2016LIFTTransform}, DISK~\citep{Tyszkiewicz2020DISKGradient}, LF-Net~\citep{Ono2018LFNetImages}, SuperPoint~\citep{Detone2018SuperPointDescription}, Hardnet~\citep{Mishchuk2017WorkingLoss}. Surprisingly, variants of the original SIFT histogram descriptor including SIFT-Rank~\citep{Toews2009SIFTRankCorrespondence}, DSP-SIFT~\citep{Dong2015DomainsizeDSPSIFT} or RootSIFT~\citep{Arandjelovic2012ThreeRetrieval} are still competitive in terms of keypoint matching performance~\citep{Balntas2017HPatchesDescriptors, Schonberger2017ComparativeFeatures}, particularly for non-planar objects and image retrieval~\citep{Bellavia2020ThereMatching}. SIFT keypoint extraction is based solely on fundamental, symmetric and uniform mathematical operators, e.g. the Gaussian, Laplacian, and uniformly sampled gradient operators. The resulting keypoints thus represent a class of highly informative patterns that are invariant to image scaling and rotation in addition to translation, and may be identified in any context with no explicit training procedure and nor bias towards specific training datasets used.

% Multimodal descriptors
However, most of the descriptors previously presented here are limited to single imaging modality.
In the case of multi-modal images, local features have to be adapted to account for the possible contrast inversion between images. The most popular approach consists in computing contrast-invariant descriptors from different image modalities, by using gradient and region reversal (also including the image intensity inversion). A first attempt has been made in~\citep{Kelman2007KeypointVariations}, where authors proposed a modification of the 2D SIFT descriptor, SIFT-GM, by mirroring gradients with an orientation larger than $\pi$, in order to reduce the range of the orientation histogram from $[0,2\pi)$ to $[0,\pi)$. This method is known has the gradient reversal. However, authors failed to consider that a gradient reversal might also induce a change of direction of the dominant gradient, hence resulting in a region reversal.
This issue has been fixed in the Symmetric-SIFT descriptor~\citep{Chen2009RealtimeDescriptor}, that accounts for gradient and region reversal. This work has been the base of multiple variant, such as
~\citep{Hossain2012EffectiveRegistration} and~\citep{Teng2015MultimodalDescriptors} that proposed to use the Symmetric-SIFT descriptor as a first step, before computing a global orientation difference between local features, and recomputing descriptors according to this new orientation. In~\citep{Hossain2010EnhancementRegistration} authors developed a variant by using the average of squared difference of gradient magnitude in the Symmetry-SIFT descriptor orientation histogram to overcome some limitation that may create similar descriptors for visually different regions.
\citep{LvG2019SelfsimilarityRegistration} propose to evaluate the angles between keypoints main orientation using pairs of triplets instead of pair of single features as proposed in the Symmetric-SIFT descriptor. However, to limit the complexity explosion of all triplets combinations, a subset of keypoints has to be selected first.
\citep{Bingjian2011ImageImages} proposed to extract SIFT features to identify salient regions in the image and compute the shape-context descriptor for each keypoint. After a RANSAC to eliminate outliers, a transform is evaluated using the euclidean distance between shape-context descriptors and a least mean square (LMS) approach.
In~\citep{Toews2013FeaturebasedImages}, authors propose a 3D implementation of the SIFT-Rank~\citep{Toews2009SIFTRankCorrespondence} with a gradient and region reversal, where point sets are registered through an iterative process based on the Hough transform for transform estimation. However, except for~\citep{Toews2013FeaturebasedImages,Rister2017VolumetricKeypoints}, all these methods focus on 2D descriptors.

% Medical Image Analysis
% FBM, VBM, Humunculous, modalities (fMRI, DTI, etc...)
% Feature-based Morphometry (Toews, 2010; Castellani, 2012; Chen, 2014)
% Voxel-based Morphometry (Ashburner, 2000; 
% Tensor-based Morphometry (Wang, 2013)
% Object-based Morphometry (Mangin, 2004)

% ICP-SIFT: (Sharp, 2002; Jiao, 2019)
% CPD-SIFT: (Xia, 2013)
% Non-rigid: (Franz, 2006; Urschler, 2006)
% Multi-modal: (Hossain, 2011) 
In terms of neuroimage analysis, local features have been widely used for a ranged of applications. In the context of inference and classification, feature-based morphometry has been used in identify group-related anatomical features relating to disease~\citep{Toews2010FeaturebasedPatterns}, to predict infant developmental age~\citep{Toews2012FeaturebasedMRI}, to classify neurodegenerative diseases~\citep{Chen2014DetectingApproach} or mental disorders~\citep{Castellani2012ClassificationMorphometry}.
In the context of image registration, local features have been combined with various point cloud registration algorithms such as the Iterative Closest Point (ICP)~\citep{Sharp2002ICPFeatures,Jiao2019Point3DSIFT} or the Coherent Point Drift (CPD)~\citep{Xia2013RobustDrift} methods. Local features have also been used to estimate non-rigid or deformable registration solutions~\citep{Franz2006AdaptiveRegistration,Urschler2006SIFTImages,Luo2018FeaturedrivenCompensation,Machado2018NonrigidMatching}, and in the context of multiple image modality registration~\citep{Hossain2011ImprovedRegistration}. The majority of the applications in this thesis focus on brain anatomy from structural MRI, however descriptor analysis has also been extended to diffusion MRI (dMRI)~\citep{Chauvin2018DiffusionAnalysis,Kumar2018MultimodalFramework,Cheung2009SIFTTransform} and functional MRI (fMRI)~\citep{Hardoon2007UnsupervisedCorrelation,Shahamat2015FeatureData}.

Scientific breakthroughs in understanding the human brain include the cortical homunculus~\citep{Penfield1937SomaticStimulation} mapping sensory inputs and motor outputs to the human body surface, the retinotopic neural structure of the visual cortex~\citep{Hubel1968ReceptiveCortex}, the invention of the magnetic resonance imaging (MRI) ~\citep{Lauterbur1973ImageResonance}, the functional MRI (fMRI) measuring the blood oxygenation level or BOLD signal in brain regions~\citep{Belliveau1991FunctionalImaging}, the diffusion MRI (dMRI) measuring water diffusion and axonal connectivity~\citep{Basser1994MRImaging}, the development of group analysis software including based on voxel-based morphometry (VBM) ~\citep{Ashburner2000VoxelbasedMethods} and functional and structural libraries (FSL)~\citep{Smith2004AdvancesFSL}, and finally large publicly available datasets including the Open Access Series of Imaging Studies (OASIS) (416 subjects for the cross-sectional study, 1098 for the longitudinal)~\citep{Marcus2007OpenAdults}, the Alzheimer's Disease Neuroimaging Initiative (ADNI) (1400 subjects, Alzheimer's disease) ~\citep{JackJr2008AlzheimerMethods}, the Human Connectome Project (1200 young adults, 405 families)~\citep{VanEssen2013WUMinnOverview} or the UK Biobank (100,000s of subjects) ~\citep{Sudlow2015UKAge}.

%However, aside from local features, other methods have been applied to neuroimage analysis.

\end{literaturereview}

\chapter{Registering Image Volumes using 3D SIFT \\ and Discrete SP-Symmetry}
\label{chap2}
\chaptermark{Registration using 3D SIFT and Discrete SP-Symmetry}

\articleAuthors{
{Laurent Chauvin\textsuperscript{a}}{William Wells III\textsuperscript{b,c}}{and Matthew Toews\textsuperscript{a}}
}{
{\setstretch{1.2}
\textsuperscript{a} Department of Systems Engineering, École de Technologie Supérieure, Québec, Canada\\
\textsuperscript{b} Brigham and Women's Hospital, Harvard Medical School, USA%, \\
%25 Shattuck Street, Boston, MA 02115, USA
\\
\textsuperscript{c} Computer Science and Artificial Intelligence Lab, Massachusetts Institute of Technology, USA%, \\
%77 Massachusetts Avenue, Cambridge, MA 02139, USA
\\~\\
Paper submitted in \textit{IEEE Transactions on Pattern Analysis and Machine Intelligence}, June 2022
}
}
\\
\textbf{Abstract}\\
This paper proposes to extend local image features in 3D to include invariance to discrete symmetry including inversion of spatial axes and image contrast. A binary feature sign $s \in \{-1,+1\}$ is defined as the sign of the Laplacian operator $\nabla^2$, and used to obtain a descriptor that is invariant to image sign inversion $s \rightarrow -s$ and 3D parity transforms $(x,y,z)\rightarrow(-x,-y,-z)$, i.e. SP-invariant or SP-symmetric. SP-symmetry applies to arbitrary scalar image fields $I: R^3 \rightarrow R^1$ mapping 3D coordinates $(x,y,z) \in R^3$  to scalar intensity $I(x,y,z) \in R^1$, generalizing the well-known charge conjugation and parity symmetry (CP-symmetry) applying to elementary charged particles. Feature orientation is modeled as a set of discrete states corresponding to potential axis reflections, independently of image contrast inversion. Feature descriptors augmented to include a binary sign are invariant to discrete coordinate reflections and intensity contrast inversions, in addition continuous 3D similarity transforms. Feature properties are factored in to probabilistic point-based registration as symmetric kernels, based on a model of binary feature correspondence. Experiments using the well-known coherent point drift (CPD) algorithm demonstrate that SIFT-CPD kernels achieve the most accurate and rapid registration of the human brain and CT chest, including multiple MRI modalities and abnormal local variations such as tumors or occlusions.

\newcolumntype{P}[1]{>{\centering\arraybackslash}p{#1}}
\makeatletter
\def\thickhline{%
  \noalign{\ifnum0=`}\fi\hrule \@height \thickarrayrulewidth \futurelet
   \reserved@a\@xthickhline}
\def\@xthickhline{\ifx\reserved@a\thickhline
               \vskip\doublerulesep
               \vskip-\thickarrayrulewidth
             \fi
      \ifnum0=`{\fi}}
\makeatother

\newlength{\thickarrayrulewidth}
\setlength{\thickarrayrulewidth}{3\arrayrulewidth}

\makeatletter
\algocf@newcmdside@kobe{LetIn@let}{%
    \KwSty{EM Optimization:} repeat until convergence:%
    \ifArgumentEmpty{#1}\relax{ #1}%
    \algocf@group{#2}%
    \par
}
\newcommand\EMOpti[1]{%
    \LetIn@let{#1}%
}

\algocf@newcmdside@kobe{EM@EStep}{%
    \KwSty{E-Step:} Compute matrix $P$:%
    \ifArgumentEmpty{#1}\relax{ #1}%
    \algocf@group{#2}%
    \par
}
\newcommand\EStep[1]{%
    \EM@EStep{#1}%
}

\algocf@newcmdside@kobe{EM@MStep}{%
    %\KwSty{M-Step:} Solve for \textbf{R},s,\textbf{t},$\sigma^2$:%
    \KwSty{M-Step:} Solve for optimal transformation $\mathcal{T}$%
    \ifArgumentEmpty{#1}\relax{ #1}%
    \algocf@group{#2}%
    \par
}
\newcommand\MStep[1]{%
    \EM@MStep{#1}%
}
\makeatother

% vector representation
\newcommand{\vect}[1]{\bm{#1}} % version 1

\section{Introduction}
3D structure is often described in terms of localized phenomena or features. In classical physics, fine-scale matter in the atomic model is characterized via the Schrodinger equation~\citep{Schrodinger1926UndulatoryMolecules}, where the symmetry of the Laplacian operator ensures that atomic properties remain invariant to changes in the viewpoint of the observer. Similarly, elementary 3D image features may be identified in generic image volumes via fundamental mathematical operators, e.g. the 3D Laplacian-of-Gaussian operator~\citep{Toews2013EfficientFeatures,Rister2017VolumetricKeypoints}, following from scale-space theory and the 2D scale-invariant feature transform (SIFT)~\citep{Lindeberg1998FeatureSelection,Lowe2004DistinctiveKeypoints}. These are used in highly robust and efficient memory-based applications, e.g. detection~\citep{Flitton2010ObjectVolumes}, registration~\citep{Machado2018NonrigidMatching,Luo2018FeaturedrivenCompensation}, segmentation~\citep{Wachinger2018KeypointSegmentation,Gill2014RobustApproach}, indexing~\citep{Chauvin2021EfficientSets,Toews2015FeatureBasedImages} in a manner invariant to transformations including translation, rotation and scaling, and are applicable in generic contexts with no training or calibration procedures.

Our work here seeks to advance the understanding of 3D keypoints observed in image volumes from the notion of symmetry, specifically discrete symmetry in feature orientation. A geometrical object may be said to exhibit a symmetry if its observed properties remain unaffected by (or are invariant to) a group of transforms, i.e., a symmetry group. In particle physics, discrete symmetries include axis reflections and also simultaneous charge conjugation (inversion), 3D parity transformation and time reversal (CPT-symmetry). The CPT transform models the transition of a charge-associated particle to its anti-particle, e.g. a proton to an anti-proton, and is currently a topic of high interest~\citep{Borchert202216partspertrillionRatio,Lehnert2022MirrorTwin} as symmetry violations have been observed in special cases including charge conjugation and parity (CP) transforms~\citep{Christenson1964EvidenceMeson,Sozzi2008DiscreteTheory}. Nevertheless in image analysis, discrete shape symmetries including reflections~\citep{Liu2010ComputationalGraphics,Cohen2019GaugeCNN} and image contrast inversion~\citep{Chen2009RealtimeDescriptor,Hossain2012EffectiveRegistration,Teng2015MultimodalDescriptors,Alexander2001AnalysisMRI,LvG2019SelfsimilarityRegistration} are typically represented as separate, unrelated phenomena, typically in 2D coordinate space where 3D parity does not apply. Indeed, image processing and graphics research focuses largely on representations and challenges stemming from 2D data, including photographs, surface meshes or projections, and achieving invariance to viewpoint changes. Despite the compelling analogy between local features and the atomic model, the notion of a local feature sign, analogous to a particle charge, has not been proposed nor linked to the 3D parity transform in the context of image processing.
\begin{figure*}[ht]
    \centering
    \includegraphics[width=1\textwidth]{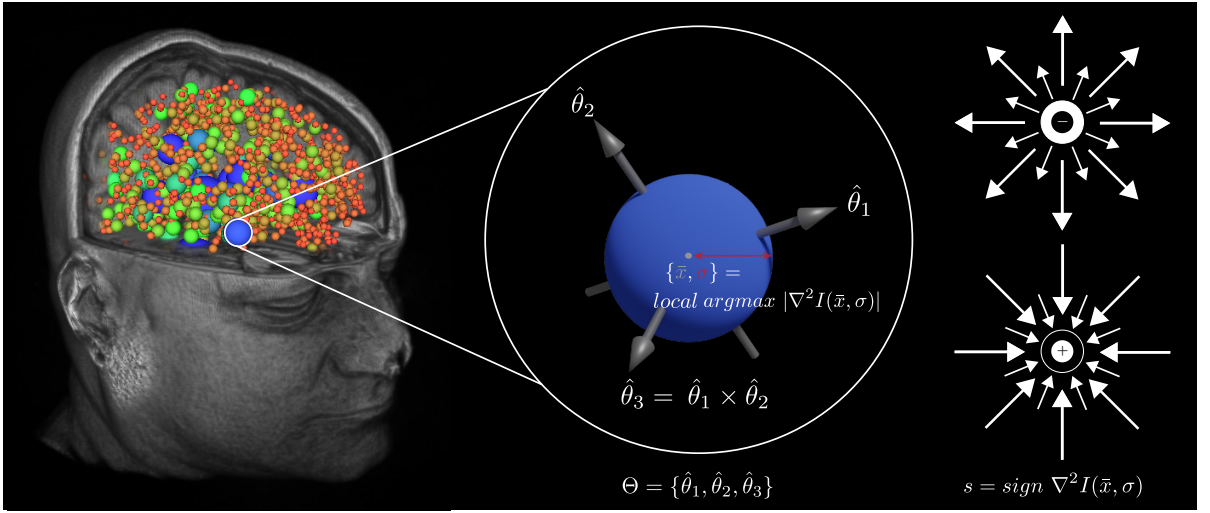}
    \caption{Illustrating the geometrical properties of 3D features (colored spheres) extracted from an MRI image of the human brain. Properties include location $\vect{x}$, orientation $\Theta$, scale $\sigma$, we introduce a binary feature sign $s = \{-,+\}$ based on the sign of the Laplacian-of-Gaussian}
    \label{fig:properties}
\end{figure*}

Our primary contribution is the first model of discrete symmetry including intensity inversion in the context of local feature processing as shown in Figure~\ref{fig:properties}. We note that a CP-like transform occurs in the context of image processing, where a local feature observed in different imaging modalities may exhibit simultaneous contrast inversion and gradient reversal. We refer to this as a sign inversion and parity transform (SP-transform), which applies generally to arbitrary scalar image fields $I: R^3 \rightarrow R^1$ mapping 3D coordinates $(x,y,z) \in R^3$ to a scalar image intensity $I(x,y,z) \in R^1$, and we seek SP-invariance. Local features are endowed with a binary sign $s \in \{-1,+1\}$ and a set of discrete orientation states, allowing feature descriptors and image registration to be computed in a manner invariant to reflections and intensity contrast inversions, thus achieving SP-invariance. Volumetric data allows physical matter and local image features to be observed in situ within isotropic 3D space and free from projective distortion, thus facilitating the analogy between local image features and atoms in terms of charge. Local feature orientation is defined by a 3D rotation matrix or coordinate reference frame, where two primary axes are estimated from dominant image gradient directions and the third vector is defined by the cross product, ensuring feature reference frames are restricted to $SO(3)$ despite potential parity transform due to contrast inversion. Four discrete orientation states are defined, corresponding to inversions of the two primary axis vectors, due to a combination of symmetric image gradient patterns and intensity contrast inversion. Gradient-based descriptors may be inverted via sign change, independently of orientation state.

Our secondary contribution is to incorporate invariant feature properties into a probabilistic point-based registration framework. The variability of feature scale, orientation and location between images is modeled via an exponential kernel function, approximating a zero-mean noise model, invariant to feature reflection and intensity contrast inversion. Experiments adapt our model to the well-known Coherent Point Drift (CPD)~\citep{Myronenko2010PointDrift} algorithm, and demonstrate that our SIFT-CPD model of enhanced feature geometry leads to faster and more accurate inter-subject registration, in a diverse variety of image data including multiple T1 and T2 weighted MRI modalities of the human brain, tumors and CT images of the human torso. We also demonstrate that our model may be used to achieve SP-invariance independently of the method used to estimate feature orientation, e.g. via principal component analysis or maximum gradient directions.

The remainder of this paper describes related work, method and experiments. Throughout the paper, we aim to provide an accessible yet complete presentation of related concepts in 3D geometry, image processing and particle physics. 

\section{Related work}

Our work focuses on characterizing 3D features extracted from volumetric images, and extending invariance to account for discrete symmetry due to both axis reflection and intensity contrast inversion.

\subsection{Local Image Features}

In computer vision and image processing, informative image structures are typically identified via mathematically-defined filters or convolution operators. First order gradient $\nabla I$ operators~\citep{Harris1988CombinedDetector,Shi1994GoodTrack} may be used to identify corner-like structures, in a manner invariant to translation and rotation. Second-order derivative operators such as the Laplacian $\nabla^2 I$ may be used to identify blob-like structures, and to extend feature invariance to scale or size based on Gaussian scale-space theory~\citep{Lindeberg1998FeatureSelection}, as in the seminal scale-invariant feature transform (SIFT)~\citep{Lowe2004DistinctiveKeypoints}. The gradient and Laplacian operators may be regarded as minimal filter representations required to localize singular patterns in a scalar image $I: \mathcal{R}^D\rightarrow\mathcal{R}^1$ defined over a $D$-dimensional coordinate space. Extended representations may be used to achieve affine invariance~\citep{Mikolajczyk2005PerformanceDescriptors}, or make use of alternative representations, eg. non-linear scale spaces~\citep{Alcantarilla2012KAZEFeatures}, information theory~\citep{Kadir2001SaliencyDescription,Toews2010MutualinformationImages}, integral images for efficiency~\citep{Bay2008SpeededupSURF}. Over-complete filters banks, ie. convolutional neural networks (CNNs), may be trained for the purpose of identifying point correspondences~\citep{Jiang2021COTRImages,Ono2018LFNetImages,Noh2017LargeScaleFeatures}, possibly from training labels generated from fundamental operators~\citep{Yi2016LIFTTransform,Detone2018SuperPointDescription}.

Local features incorporate descriptors of the image content in a region or receptive field. The Gaussian scale-space~\citep{Lindeberg1998FeatureSelection} may be viewed as a CNN with a single Gaussian channel per layer~\citep{Carluer2021GPUDescriptor}, where the scale or receptive field size is defined explicitly by the Gaussian standard deviation $\sigma$, and is sampled in 3 logarithmic increments between $[\sigma,2\sigma]$ prior to subsampling. The receptive fields of deep network activations are a function of the filter sizes and depth~\citep{Luo2016UnderstandingNetworks}, and may be relatively large and dependent on the image contents due to non-linear operations including max pooling and ReLu~\citep{Noh2017LargeScaleFeatures}. Traditional descriptors have made use of gradients $\nabla I$, including gradient orientation histograms SIFT~\citep{Lowe2004DistinctiveKeypoints}, binary comparisons~\citep{Calonder2011BRIEFFast} and variants such as rank ordering~\citep{Toews2009SIFTRankCorrespondence}, RootSIFT~\citep{Arandjelovic2012ThreeRetrieval}. Descriptors may be learned via triplet loss and gradient analysis~\citep{Ono2018LFNetImages,Tyszkiewicz2020DISKGradient}. While GPU-based deep learning is the defacto state-of-the-art for image classification~\citep{Krizhevsky2012ImageNetNetworks}, variants of the traditional gradient representations remain competitive with learned variants~\citep{Balntas2017HPatchesDescriptors, Schonberger2017ComparativeFeatures}, specifically for matching images of non-planar objects~\citep{Moreels2007EvaluationObjects} and image retrieval~\citep{Bellavia2020ThereMatching}.

The SIFT algorithm has been generalized to 3D volumetric images. A body of work investigates SIFT-like detectors in 3D video coordinates $(x,y,t)$ defined by 2D space $x,y$ and 1D time $t$~\citep{Laptev2005SpaceTimePoints,Scovanner20073dimensionalRecognition}. In the context of 3D space $x,y,z$, applications include object detection~\citep{Flitton2010ObjectVolumes}, medical image analysis~\citep{Cheung2009SIFTTransform,Allaire2008FullAnalysis,Toews2013EfficientFeatures,Rister2017VolumetricKeypoints}, segmentation~\citep{Gill2014RobustApproach,Wachinger2018KeypointSegmentation}, alignment~\citep{Bersvendsen2016RobustSequences}, image stitching~\citep{Ni2008VolumetricSIFT}, large-scale indexing~\citep{Toews2015FeatureBasedImages}, population studies~\citep{Toews2016HowFeatures,Kumar2018MultiModalMRI}. Most recently, the Jaccard distance between feature sets was introduced to automatically flag errors in large public training MRI datasets~\citep{Chauvin2019AnalyzingManifold,Chauvin2020NeuroimageRelatives}, and to identify family members from brain MRI~\citep{Chauvin2021EfficientSets}, capabilities facilitated by highly efficient feature indexing. We adopt the 3D SIFT-Rank method~\citep{Toews2013EfficientFeatures} using a GPU-optimized implementation~\citep{Carluer2021GPUDescriptor} first used in the context of brain MRI analysis~\citep{Pepin2020LargeScaleMasking}. 

\subsection{Image Registration}
Registration is a fundamental image processing task, and seeks to identify an optimal transform $T: \Omega_1 \rightarrow \Omega_2$ between the coordinate systems $\Omega_1 \in R^3$, $\Omega_2 \in R^3$ of the same object or scene, often in three spatial dimensions, based on a pair of observed images $(I_1$,$I_2)$. Registration generally makes use of intensity and geometry information~\citep{Saiti2020ApplicationMethods}, our work focuses on properties of local feature geometry, specifically location, orientation, scale and sign.

Point-based registration approaches are most generally applicable to local features, where the loss function minimizes the distance between pairs of points or points and a model. Examples of point cloud registration algorithms~\citep{Pomerleau2015ReviewRobotics} include iterative algorithms such as the Iterative Closest Point (ICP) algorithm~\citep{Besl1992MethodShapes} where points contribute uniformly to a solution, or the Coherent Point Drift (CPD) algorithm~\citep{Myronenko2010PointDrift} where points contribute according to a probabilistic weighting. 3D SIFT keypoint correspondences have been used to achieve point-based registration the context of image-guided neurosurgery~\citep{Luo2018FeaturedrivenCompensation}, including non-rigid image registration via thin plate splines~\citep{Machado2018NonrigidMatching}, finite element methods~\citep{Frisken2019PreliminaryUltrasound} and the CPD algorithm~\citep{Jiao2019Point3DSIFT}, however these have made use of keypoint locations and not orientation and scale properties as we propose.

Feature-based registration methods generally consider points with properties beyond simple location. Points on a surface model may be endowed with properties such as mass~\citep{Golyanik2016GravitationalRegistration} or charge~\citep{Jauer2019EfficientClouds}, however these properties tend to be assigned algorithmically, eg. uniformly assigned across points, and not derived from the image content itself. Local properties including orientation, scale and affine deformation may be used to enhance point-based registration~\citep{Riggi2006FundamentalParameters,Ma2017RemoteMatching}. Smooth deformations may be computed in a manner consistent with local rigid or affine reference frames~\citep{Arsigny2009FastRegistration}. The variability of geometrical misalignment is related to feature scale $\sigma$~\citep{Chauvin2019AnalyzingManifold,Toews2007StatisticalVariability}, consistent with uncertainty due to Gaussian blur. The localization accuracy of SIFT correspondences has been compared to that of manual human labeling in 2D and 3D ~\citep{Toews2007StatisticalVariability,Machado2018NonrigidMatching}, showing similar accuracy, where experts preferred automatic SIFT correspondences in 80\% cases~\citep{Machado2018NonrigidMatching} in the case of brain imaging. Our 3D SIFT-CPD algorithm proposed in the following section accounts for feature properties of orientation and scale, in addition to point locations.

Multi-modal image registration is a major challenge, where intensities may vary locally in a non-linear manner between modalities. Training may be used to approximate an intensity mapping for a specific domains, e.g. MRI and CT~\citep{Hu2021EndtoendLearning}, however this mapping may not be functional or stationary throughout the image. In the general case with no specific training domain or data, intensity-based registration must adopt statistical similarity measures such as mutual information~\citep{Viola1997AlignmentInformation}. In the case of 2D image keypoints, invariance to local contrast inversion may be achieved by transforming the image intensity into a contrast-invariant format, including phase congruency~\citep{Xia2013RobustDrift} or Laplacian~\citep{Wachinger2012EntropyRegistration} images. As contrast inversion leads to image gradient reversal, attempts have been made to reverse aspects linked to the gradient, including descriptor orientation 
~\citep{Kelman2007KeypointVariations,Chen2009RealtimeDescriptor,Hossain2012EffectiveRegistration,Bingjian2011ImageImages}. Combinations of multiple keypoints~\citep{Teng2015MultimodalDescriptors} including self-similarity~\citep{LvG2019SelfsimilarityRegistration} may be adopted. In preliminary work, we proposed to account for 3D gradient reversal~\citep{Toews2013FeaturebasedImages}, here we extend this to include reflections and SP-symmetry via the notion of a binary sign $s \in \{-1,+1\}$.

\subsection{Discrete Symmetry}

In a physical system, a symmetry refers to a property that remains unchanged under set of transforms, i.e. an invariant property. The symmetry group is the Lie group of transforms under which a geometrical object such as a particle is invariant~\citep{Liu2010ComputationalGraphics}. A symmetry may be described as continuous or discrete. Continuous symmetry pertains to continuous parameters of pose and scale, eg. translation and orientation relative to the reference frame of an image acquisition device, and are analogous to a consequence of Noether's theorem for Lagrangian mechanical systems~\citep{Noether1971InvariantProblems,Tanaka2021NoetherNetworks}. Continuous symmetry has been investigated in the computer vision literature in terms of differential invariance~\citep{Florack1992ScaleImages,Olver1994DifferentialApproach,Lindeberg1998FeatureSelection}. Classical neural networks such as CNNs are generally invariant to translations but not to reflections or scale changes~\citep{LeCun1989BackpropagationRecognition,Krizhevsky2012ImageNetNetworks}. Robustness to certain transforms may be improved via regularization~\citep{Bardes2022VICRegLearning}. Recently models of invariance or equivariance in trainable neural networks have emerged, including SO(3)~\citep{Esteves2018LearningCNNs,Spezialetti2019LearningSupervision} rotations, 3D rigid transforms~\citep{Moyer2021EquivariantImaging} non-linear transforms~\citep{Spezialetti2019LearningSupervision}, probabilistic symmetries~\citep{Bloem-Reddy2020ProbabilisticNetworks} and gauge theory~\citep{Cohen2019GaugeCNN}. Much of this work has been limited to 2D image modalities, i.e. surfaces or projective images, and has not considered the 3D notion of SP-symmetry including a binary sign $s \in \{-1,+1\}$.

A discrete symmetry generally involves discrete displacements in location and/or orientation, e.g. translations, rotations, reflections, glide reflections or repetitive crystal structure. Charge conjugation and parity symmetry (CP-symmetry)~\citep{Sozzi2008DiscreteTheory}, and generally space-time reversal in CPT-symmetry including time relating to the transition of a particle to its anti-particle~\citep{Borchert202216partspertrillionRatio,Lehnert2022MirrorTwin} are important aspects of discrete symmetry relating to 3D reflection of elementary physical particles. CP-symmetry is closely related to reflection symmetry, which exists in a function exhibiting one or more planes of symmetry, i.e. a  reflection about a single axis $x=0$ such that $f(x)=f(-x)$, or in a 3D system, the parity coordinate transform $f(x,y,z)=f(-x,-y,-z)$, both characterized in 3D by a rotation matrix $R$ with determinant $det(R)=-1$. The effect of the parity transform may be seen by first representing $f(x)$ in terms of $f(x) = f_s(x) + f_a(x)$, the sum purely symmetric and anti-symmetric image components $f_s(x)$ and $f_a(x)$ defined by $f_s(-x)=f_s(x)$ and $f_a(-x)=-f_a(x)$. Symmetric or even functions $f_s(x)$ include the Gaussian and Laplacian-of-Gaussian operators, wave functions associated with bosonic particles such as the photon, the cosine function, etc. Anti-symmetric or odd functions $f_a(x)$ include the gradient operator $\nabla I$, wave functions associated with fermionic particles such as electrons, and the sine function, etc. As the gradient is an anti-symmetric operator, an image contrast inversion $-\nabla I(x)=\nabla I(-x)$ is equivalent to an axis reflection. This phenomenon is analogous to charge conjugation and parity (CP) symmetry, which was thought to be an inviolable in nature until the discovery of special cases beginning with the composite meson particle~\citep{Christenson1964EvidenceMeson}. In our work here, charge is analogous to image sign or contrast, which may be inverted, and we propose SP-invariance to register images of differing modality.

Image processing methods specifically relating to our work have analyzed discrete shape symmetry (ie. reflections)~\citep{Liu2010ComputationalGraphics} and image contrast inversion~\citep{Chen2009RealtimeDescriptor,Hossain2012EffectiveRegistration,Teng2015MultimodalDescriptors,Alexander2001AnalysisMRI,LvG2019SelfsimilarityRegistration} as separate, unrelated phenomena. Our work is the first to consider these within the unified framework of discrete SP-symmetry, including a binary sign and discrete states of local reference frame orientation. Our method generalizes to various contexts where orientation is estimated from dominant local image gradients $\nabla I$. For example, Toews et al. identify maxima in a 3D spherical gradient histogram~\citep{Toews2013EfficientFeatures} such that $\vect{\hat{\theta}}_1 =  \underset{\vect{\hat{\theta}}}{\mathrm{argmax}}~|\nabla I\cdot \hat{\theta}|$,  $\vect{\hat{\theta}}_2 =  \underset{\vect{\hat{\theta}}}{\mathrm{argmax}}~|\vect{\hat{\theta}} \times (\nabla I \times \vect{\hat{\theta}}_1)|$, with a third axis defined by the cross product $\vect{\hat{\theta}}_3 = \vect{\hat{\theta}}_1 \times \vect{\hat{\theta}}_2$. Rister et al. compute the eigenvectors of the local gradient structure tensor, ie. the $3\times3$ gradient correlation matrix~\citep{Rister2017VolumetricKeypoints}. We demonstrate that both such representations may be modeled in terms of discrete SP-symmetry.

\section{Method}

Our method provides a means of characterizing 3D image features in a manner invariant to discrete symmetry transforms, including symmetric image patterns and sign inversion and parity (SP-symmetry) transforms due to image contrast inversion. We begin by describing the properties of a single local image feature, which we augment to include a binary sign and a set of discrete axis reflections. This allows us to compute a descriptor that is invariant to reflections and changes in sign due inversion of intensity contrast. We then define a kernel function $\mathcal{K}(f_n,f_m)$ to quantify the similarity of a pair of features $(f_n,f_m)$ potentially arising from the same underlying structure in two different images. Finally, this kernel is  incorporated into probabilistic image registration in order to estimate a transform $T: f_m \rightarrow f_n$ aligning the coordinate systems of two images.

\subsection{Single-Feature Properties}
\label{sec:particle}
A feature is a distinctive spherical region localized in 3D image space, as illustrated in Figure~\ref{fig:properties}. An individual feature $f=\{g,\vect{a},s\}$ is characterized within the image by its geometry $g$, a descriptor of the image appearance $\vect{a}$ surrounding the feature and a binary sign $s$. The feature geometry $g=\{\Theta,\sigma,\vect{x}\} \in SO(3)\times R^+ \times R^3$ is a scaled coordinate reference frame in 3D defined by 7 parameters. These include a scale $\sigma \in \mathbb{R}^+$, a coordinate location $\vect{x} \in \mathbb{R}^3$ and a reference frame $\Theta \in SO(3)$. The reference frame may be defined as a set $\Theta = \{\vect{\hat{\theta}}_1,\vect{\hat{\theta}}_2,\vect{\hat{\theta}}_3\}$ of orthonormal unit vectors $\vect{\hat{\theta}}_i, i \in [1,2,3]$ where $\vect{\hat{\theta}}_i \cdot \vect{\hat{\theta}}_j = 0, i \ne j$. Note that alternative representations for rotation could be used, eg. $SU(2)$ Pauli matrices or unit quaternions. Feature appearance $\vect{a}\in \mathbb{R}^M$ is an M-dimensional descriptor of the image, resampled according to geometry $g$. The sign $s \in \{-1,1\}$ is used to achieve invariance to image contrast inversion. 

Let $I: \mathbb{R}^3 \rightarrow \mathbb{R}^1$ be a scalar image sampled over 3D coordinates where $I(\vect{x})$ represents a voxel measurement at location $\vect{x}$. Let $I(\vect{x},\sigma)=I(\vect{x})*G(\vect{x},\sigma)$ be a scale-space defined by convolution of the image with a Gaussian filter $G(\vect{x},\sigma)$ of parameter $\sigma$~\citep{Lindeberg1998FeatureSelection}. Features are detected as points in scale-space $\{\vect{x_m},\sigma_m,s_m\}$ at which the magnitude of the scale-normalized Laplacian operator is maximized, similarly to the original SIFT method~\citep{Lowe2004DistinctiveKeypoints}, here $m$ is a feature index. Furthermore, they are endowed with a binary sign as follows.
\begin{align}
    \{\vect{x_m},\sigma_m\} &=  \underset{\vect{x},\sigma}{\mathrm{ local~argmax}}~|\nabla^2 I(\vect{x},\sigma)|, \notag \\
    s_m &= {\mathrm{sign}} \left( \nabla^2 I(\vect{x}_m,\sigma_m) \right).
\end{align}
Scale-space loci $\{\vect{x_m},\sigma_m,s_m\}$ represent generic spherical regions in which the divergence of the local gradient field $\nabla I(\vect{x},\sigma)$ is maximized. Feature sign $s_m$ is novel to our characterization, and is defined as the sign of the Laplacian-of-Gaussian.

Feature orientation $\Theta$ is derived from the image gradient $\nabla I$ surrounding the feature origin $\vect{x}$. Orientation axis vectors $\vect{\hat{\theta}}_1,\vect{\hat{\theta}}_2,\vect{\hat{\theta}}_3$ may be determined according dominant gradient orientations within a window $w(\vect{x},\sigma) \in [0,1]$ centered upon $\vect{x}$ with spatial extent proportional to scale $\sigma$. As $\Theta \in SO(3)$, the determinant $det(\Theta)=1$ is positive, ie. no reflections. Axes are thus defined by two primary axis vectors $\vect{\hat{\theta}}_1$ and $\vect{\hat{\theta}}_2$, with the third constrained according to a handedness convention, ie. the right-hand rule via the cross product $\vect{\hat{\theta}}_3 = \vect{\hat{\theta}}_1 \times \vect{\hat{\theta}}_2$. Furthermore, we assume that axis vectors may be ordered uniquely according to the gradient magnitude along their directions, i.e. $[| \nabla I\cdot\vect{\hat{\theta}}_1|] > [| \nabla I\cdot\vect{\hat{\theta}}_2|]$, where $[ | \nabla I\cdot\vect{\hat{\theta}}_i|] = \int w(\vect{x},\sigma)| \nabla I(\vect{u})\cdot\vect{\hat{\theta}}_i| du$ is the expected value of the gradient magnitude along axis $\vect{\hat{\theta}}_i$.

Our primary contribution is a descriptor that is invariant to discrete SP-transforms, i.e. is SP-symmetric, based on binary sign $s$. We first note that the sign of the orientation axes $\vect{\hat{\theta}}_1,\vect{\hat{\theta}}_2$ may generally be ambiguous due to a variety of factors including bilateral symmetry of the image pattern or its local gradient distribution (e.g. an ellipsoid or rectangular box), an image contrast inversion (e.g. different imaging modalities), or the algorithm used to estimate orientation (e.g. principal component analysis has an inherent eigenvector sign ambiguity). Sign ambiguity may be represented by maintaining a discrete set of possible orientation reference frames $\Theta$. We consider four potential reference frame states $\{\pm \vect{\hat{\theta}_1},\pm \vect{\hat{\theta}_2}\}$ (with $\vect{\hat{\theta}_3}=\vect{\pm\hat{\theta}_1} \times \pm\vect{\hat{\theta}_2}$) resulting from axis inversion, i.e. due to either intensity contrast inversion or pattern symmetry, which are equivalent to the identity in addition to rotations of $\pi$ about each of three axes. By defining the third axis vector $\vect{\hat{\theta}_3}$ using the cross product, we ensure feature orientation follows the right-hand rule. Figure~\ref{fig:workflow} illustrates the four possible orientation states, including an initial reference frame and rotations of $\pi$ about each of the 3 axes.

Feature sign $s$ is used to generate a novel contrast inversion-invariant descriptor $\vect{a}_m$ for each orientation state, based on the image content surrounding location $\vect{x_m}$ at scale $\sigma_m$ and orientation $\Theta_m$. Note that the geometry $g$ defines a 7-parameter similarity transform from the local feature reference frame to a canonical reference frame, including rotation, translation and scaling. This this is used to resample the image according to a characteristic reference frame $\hat{I}(\vect{u}) = I(\sigma_m\Theta_m\vect{u}+\vect{x_m})$ that is invariant to global similarity transforms of the image. A variety of descriptors may be computed, eg. spherical harmonics~\citep{DiMaio2009SphericalharmonicMaps}, we adopt a computationally efficient variant of the gradient orientation histogram~\citep{Lowe2004DistinctiveKeypoints,Toews2013EfficientFeatures,Rister2017VolumetricKeypoints}. As in~\citep{Toews2013EfficientFeatures}, 3D space surrounding the feature origin $\vect{x}_m$ is quantized uniformly into eight discrete octants $\vect{r}=[\pm 1,\pm 1,\pm 1]$, and the image gradient at each point $\vect{u}$ into eight discrete symmetric directions $\vect{\phi}=[\pm 1,\pm 1,\pm 1]$. These are combined into an $8\times8=64$-element appearance descriptor, by accumulating the gradient magnitude into a histogram over spatial location and orientation
\begin{align}
a(\vect{r},\vect{\phi}) = \sum_{\vect{u} \in \vect{r}} \| \nabla \hat{I}(\vect{u})\cdot \vect{\phi}  \|[\vect{\phi}=s\vect{\phi}_{\vect{u}}],
\label{eq:descriptor}
\end{align}
where $\vect{\phi}_{\vect{u}}= \underset{\vect{\phi}}{\mathrm{argmax}}
\{ \nabla \hat{I}(\vect{u}) \cdot s\vect{\phi} \}$ is the direction of maximum gradient at image location $\vect{u}$, which is inverted in the case of $s=-1$ by the Iverson bracket $[\vect{\phi}=s\vect{\phi}_{\vect{u}}]$  evaluating to 1 upon equality and 0 otherwise. Thus descriptor orientation elements $\vect{\phi}$ are invariant to reversal due to sign change $a(\vect{r},\vect{\phi}) = a(\vect{r},s\vect{\phi})$, and parity transforms of orthants $\vect{r} \rightarrow -\vect{r}$ are accounted for by feature orientation states $\{\pm \vect{\hat{\theta}_1},\pm \vect{\hat{\theta}_2}\}$, as illustrated in Figure~\ref{fig:workflow}.

\begin{figure*}
    \centering
    \includegraphics[width=0.9\textwidth]{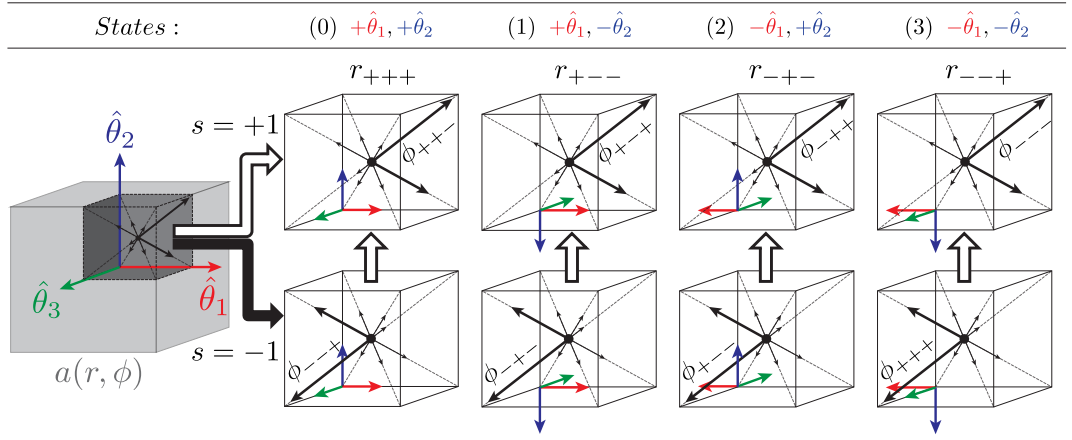}
    \caption{Illustrating an appearance descriptor $a(r,\phi)$ and four discrete orientation states given principal axes $\hat{\theta}_1$, $\hat{\theta}_2$ (left to right) and feature sign $s$ (upper and lower). The descriptor content is shown for an example octant ($r_{+++}$) and gradient orientation ($\phi_{++-}$). Axis orientation is defined independently of sign (left to right), and negative sign $s=-1$ inverts the descriptor gradient (lower row) achieving invariance to contrast inversion}
    \label{fig:workflow}
\end{figure*}

\subsection{Two-Feature Observations}

Image registration requires identifying pairs of features $(f_n,f_m)$ arising from the same underlying structure in different images. To do this, we propose a kernel function $K(f_n,f_m)\in [0,1]$ to quantify the similarity of two features $f_n$ and $f_m$ potentially sampled from the same distribution. We construct $\mathcal{K}(f_n,f_m)$ as a product of squared exponential kernels, ensuring that it is positive, symmetric $\mathcal{K}(f_m,f_n)=\mathcal{K}(f_n,f_m)$, and proportional to a Gaussian density. It may be expressed as the product of three factors
\begin{align}
    \mathcal{K}(g_{n},g_{m}) = &\mathcal{K}_{\sigma}(\sigma_{n},\sigma_{m})\mathcal{K}_{\Theta}(\Theta_{n},\Theta_{m})\mathcal{K}_{x}(\vect{x_{n}},\vect{x_{m}}), \label{eq:geo_kernel}
\end{align}
quantifying the variability in scale, orientation and location, respectively. As in the notions of linear and angular momentum from classical physics, deviations of location and orientation are modeled as orthogonal and independent.

The first factor in Equation~\eqref{eq:geo_kernel},  $\mathcal{K}_{\sigma}(\sigma_{n},\sigma_{m})$, is a kernel penalizing the log scale difference:
\begin{equation}
   \mathcal{K}_{\sigma}(\sigma_{n},\sigma_{m}) = \exp \left( - \| \log \sigma_n - \log \sigma_m \|^2 \right),
   \label{eq:scale_kernel}
\end{equation}
and is proportional to a log normal density about mean $\log 1=0$ with unit variance $1$. The log maps multiplicative variations in feature scale on the range $\sigma \in [0,\infty]$ to additive variations on the range $\log \sigma \in [-\infty,\infty]$, which may be modelled as symmetric and Gaussian.

The second factor $\mathcal{K}_{\Theta}(\Theta_{n},\Theta_{m})$ quantifies the variability of feature orientations based on the angular displacement of the axes:
\begin{equation}
    \mathcal{K}_{\Theta}(\Theta_{n},\Theta_{m}) = \exp \left(-3 + \sum_{i=1,2,3} \vect{\hat{\theta}_{in}}\cdot \vect{\hat{\theta}_{im}} \right).
    \label{eq:orient_kernel}
\end{equation}
In Equation~\eqref{eq:orient_kernel}, the scalar product
$\vect{\hat{\theta}_{in}}\cdot \vect{\hat{\theta}_{im}}=\cos(\angle( \vect{\hat{\theta}_{in}}, \vect{\hat{\theta}_{im}}))$ quantifies the angular separation between axis unit vectors $\vect{\hat{\theta}_{in}}$ and  $\vect{\hat{\theta}_{im}}$ on the range $[-1,1]$. This is equivalent to modeling the angular difference between features independently in each axis $i$ as proportional to a von Mises density over the unit circle.

The third factor $\mathcal{K}_{x}(\vect{x_{n}},\vect{x_{m}})$ is a kernel that quantifies the similarity of feature locations based on their squared displacement:
\begin{equation}
    \mathcal{K}_{x}(\vect{x_{n}},\vect{x_{m}}) = \exp \left( -\frac{||\vect{x_n} - \vect{x_m}||^2}{k~\sigma_{n}\sigma_{m}+\sigma_{T}^2} \right),
    \label{eq:loc_kernel}
\end{equation}
where the denominator normalizes the spatial displacement $||\vect{x_n} - \vect{x_m}||^2$ via a linear function of the product of feature scales $\sigma_{n}\sigma_{m}$.
The first term $k~\sigma_{n}\sigma_{m}$ embodies intrinsic variance due to the scale of the observed feature, where $\sigma_{n}\sigma_{m}$ represents the square of the geometric mean of feature scales and $k$ is a positive proportionality constant. This allows quantifying the deviation in location in a manner invariant to keypoint scale. The second term $\sigma_{T}^2$ represents the minimum achievable variance in mapping $T$ given the particular imaging context, which dominates in the case of small-scaled features, i.e. $k~\sigma_{n}\sigma_{m} < \sigma_{T}^2$. Note that in previous work~\citep{Chauvin2021EfficientSets} $k=1$ and $\sigma_{T}=0$, here we use $k$ and $\sigma_{T}$ to estimate a more accurate linear relationship between scale and observed variability. These are empirically set to provide a generally useful weighting, e.g. here we use $k=12$, $\sigma_T^2=200$.

\subsection{Multi-Feature Registration}

Registration seeks a transform $\mathcal{T}$ mapping the coordinate system of a fixed image $I_f$ to that of a moving image $I_m$, here from sets of features extracted in the fixed $\mathcal{F}=\{f_{n}\}$ and moving $\mathcal{M}=\{f_{m}\}$ images. The transform maps the properties of features from one coordinate space to the next $\mathcal{T}:\vect{x_m} \rightarrow \vect{x_n$}, and here is taken to be a global similarity transform, followed by independent zero-mean deviations for individual features. Note that the similarity transform is a 7-parameter representation  $\mathcal{T}=\{d\Theta,d\sigma,d\vect{x}\} \in SO(3)\times R^+ \times R^3$ equivalent that of an individual feature.

Let $g_m'$ represent the geometry of feature $f_m$ as transformed by $\mathcal{T}$, ie. $g_m'=T \circ g_m$, note that $T$ may act on location, scale and orientation parameters in a consistent manner. We propose using our kernel function $\mathcal{K}(g_n,g_m')$ to account for feature properties including scale and orientation in a standard point-based registration framework. Our kernel lends itself naturally to a probabilistic algorithm such as Coherent Point Drift (CPD) in Algorithm~\ref{alg:cpd}, which seeks a maximum likelihood solution based on the Expectation-Maximization (EM) algorithm~\citep{Dempster1977MaximumAlgorithm}. The expectation step (E) intend estimates the log-likelihood based on the current parameters, while the maximization step (M) maximizes the log-likelihood over the parameters using the result of the E step. CPD registration is based on a probability map $p_{mn}$ between each pair of points $(\vect{x_{\mathit{n}}},\vect{x'_{\mathit{m}}})$, and takes the form of a softmax function. We propose to bias this probability map based on a kernel function $\mathcal{K}(g_n,g_m')$ as follows:
\begin{equation}
    p_{mn}=\frac{\exp\left(-\frac{||\vect{x_{\mathit{n}}}-\vect{x'_{\mathit{m}}}||^2}{2\lambda^2}\right)\mathcal{K}(g_{n},g'_{m})}{\sum_{k=1}^{M}\exp\left(-\frac{||\vect{x_{\mathit{n}}}-\vect{x'_{\mathit{k}}}||^2}{2\lambda^2}\right)\mathcal{K}(g_{n},g_{k}')+\eta},
    \label{eq:pij}
\end{equation}
where $\eta = (2\pi\lambda^2)^{D/2}\frac{w}{1-w}\frac{M}{N}$ is a constant ensuring the denominator $p_{mn}$ is non-zero, where $w$ is parameter accounting for the relative probability of a uniform background density not associated with any specific feature. Equation~\eqref{eq:pij} includes an exponential expression with data variance parameter $\lambda^2$ and our kernel function $\mathcal{K}(g_n,g_m')$, note that with $\mathcal{K}(g_n,g_m')=1$ it is equivalent to the original CPD algorithm~\citep{Myronenko2010PointDrift}. Thus in our proposed SIFT-CPD, two exponential factors operate on point coordinates $\vect{x}$ including Equation~\ref{eq:loc_kernel}.

Iterative algorithms such as EM must be initialized within a neighborhood of the solution in order to converge correctly. We initialize registration via a global 3D Hough transform between feature sets $\mathcal{F}$ and $\mathcal{M}$ in 3D~\citep{Toews2013EfficientFeatures} analogously to the classic 2D SIFT algorithm~\citep{Lowe2004DistinctiveKeypoints}, where all features vote independently as to the predicted transform, after which a most likely transform is identified\footnote{The Hough transform method was originally proposed to track particles in bubble chamber photographs~\citep{Hough1959MachinePictures}}. Here, each feature $f_{n} \in \mathcal{F}$ votes for a transform $\mathcal{T}_{nm}: g_n \rightarrow g_m$ mapping the geometry of feature $f_{n}$ to that of feature $f_{m}$, where $f_{m}=\underset{f}{\mathrm{argmin}}~\| \vect{a}_n-\vect{a} \|$ is the nearest neighbor (NN) of $f_{n}$ in terms of the Euclidean distance $\| \vect{a}_n-\vect{a}_m \|$ between appearance descriptors. Matches are identified between all features and discrete orientation states $\{\pm\hat{\theta}_1,\pm\hat{\theta}_2\}$ including parity transforms, and in a manner invariant to intensity contrast inversions due to the binary sign $s \in \{-1,+1\}$ in Equation~\eqref{eq:descriptor}. A dominant Hough transform $\mathcal{T}^* \in \{\mathcal{T}_{nm}\}$ is then identified as the transform consistent with the largest number of matching features or inlier correspondences. A matching feature pair $(f_{n},f_{m})$ is an inlier of the transform $\mathcal{T}^*$ if their transform $\mathcal{T}_{nm}: g_n \rightarrow g_m$ differs by less than a threshold $\|\mathcal{T}_{nm}-\mathcal{T}^*\| < Thres$. $\mathcal{T}^*$ is thus defined as:
\begin{align}
    \mathcal{T}^* =
    \underset{T}{\mathrm{argmax}}~
    |\{T: \| \mathcal{T}_{nm}-\mathcal{T} \| < Thres\} |,
    \label{eq:thres}
\end{align}
where the thresholding operation in Equation~\eqref{eq:thres} may be defined as the logical conjunction of thresholds independently applied in rotation, scaling and squared displacement
\begin{align}
  \| \mathcal{T}_{nm}-\mathcal{T} \| < Thres ~~
  \rightarrow ~& \cos(\vect{\hat{\theta}}_i-\vect{\hat{\theta}}_{inm}) < ~\epsilon_{cos\theta} \\
& \wedge~|\log \sigma-\log\sigma_{nm}| < ~\epsilon_{\log\sigma}  \notag \\
&\wedge~\|\vect{x}-\vect{x_{nm}}\|^2/\sigma\sigma_{nm} < \epsilon_{x/\sigma}, \notag
\end{align}
using thresholds $(\epsilon_{cos\theta},\epsilon_{\log\sigma},\epsilon_{x/\sigma})$. These may be set generously to apply in a wide variety contexts, here we use
$(\epsilon_{cos\theta},\epsilon_{\log\sigma},\epsilon_{x/\sigma}) = (0.7,\log 1.5, 0.25)$. The Hough transform may be implemented efficiently using approximate NN search~\citep{Muja2014ScalableData} $\mathcal{T}_{nm}$ similar to the mean-shift algorithm~\citep{Comaniciu2002MeanAnalysis}. Our modified CPD algorithm is provided in Algorithm~\ref{alg:cpd}, and the $solve$ functions for similarity transforms is found in Algorithm~\ref{alg:solve_rigid} of Appendix~\ref{app:solve}. The algorithm iterates until the variance $\lambda^2$ converges.

\begin{algorithm}
    \SetKwInOut{InHere}{Inputs}
    \SetKwInOut{OutHere}{Outputs}
    \SetKwInOut{Init}{Initialization}
    \SetKwInOut{Opti}{EM Optimization}
    \DontPrintSemicolon
    
    \InHere{ $\mathcal{F}=\{f_{n}\}$~\qquad\text{Fixed Features} \\  $\mathcal{M}=\{f_{m}\}$\qquad\text{Moving Features}
    }
    \OutHere{ $\mathcal{T}:\mathcal{M} \rightarrow \mathcal{F}$~\quad\text{Transform}
    }
    
    \Init{$0\leq w \leq 1$\;
    $\lambda^2 = \frac{1}{DNM}\sum_{n=1}^{N}\sum_{m=1}^{M}||\vect{x_n}-\vect{x_m}||^2$\;
    $\mathcal{T} \leftarrow \mathcal{T}^*$}
    
    \EMOpti{
    \EStep{
        $p_{mn}=\frac{\exp^{-\frac{1}{2\lambda^2}||\vect{x_{\mathit{n}}}-\vect{x'_{\mathit{m}}}||^2}\mathcal{K}(g_{n},g'_{m})}{\sum_{k=1}^{M}\exp^{-\frac{1}{2\lambda^2}||\vect{x_{\mathit{n}}}-\vect{x'_{\mathit{k}}}||^2}\mathcal{K}(g_{n},g'_{m})+\eta}$}
    \MStep{
        $\{\mathcal{T},\lambda^2\} = solve(\mathcal{F},\mathcal{M},P)$
    }
    }
    The aligned point set is $\mathcal{T}(\mathcal{M})$ 
    \\ The correspondence probability is $P$.
    
    \caption{The probabilistic SIFT-CPD registration algorithm adapted from~\citep{Myronenko2010PointDrift} to include kernel $\mathcal{K}(g_n,g'_m)$}
    \label{alg:cpd}
\end{algorithm}
\vspace{\baselineskip}
\noindent

\section{Experiments}

We hypothesize that our kernel function $\mathcal{K}(g_n,g_m)$ incorporating feature scale and orientation will improve the accuracy of registration beyond point information alone, and that feature orientation state and binary sign will provide invariance to intensity contrast inversion and reflections, independently of the method used to identify feature orientation $\Theta$. Experiments validate these hypotheses in the context of 3D medical imaging data, which provide the opportunity to investigate features arising from realistic, diverse anatomical structure observed across subjects and 3D imaging modalities.

Three sets of experiments are performed. The first is based on synthetic similarity transforms, and establishes baseline accuracy of methods in the case of known ground truth. The second consists of inter-subject image registration trials, first between brain MRI and then between chest CT volumes of different people, with confounds including brain tumors and contrast variations due to multiple T1 and T2-weighted MRI modalities, demonstrating superior algorithm performance in a diverse set of contexts. Variability is estimated by repeating trials following synthetic transforms of images and multiple subjects. The third investigates orientation state changes in registration trials for different feature orientation estimation methods, i.e., principal components~\citep{Rister2017VolumetricKeypoints} or maximum orientation directions~\citep{Toews2013EfficientFeatures}.

Note that the true mapping $T$ between the anatomies of different subjects is generally non-linear and may not exist throughout the image, due to aspects of anatomy specific to individuals or in the case of occlusion or missing structure. As mentioned, features are assumed to follow a transform $T$ that is globally linear followed by random feature-specific deviations as specified by our kernel $\mathcal{K}(g_n,g_m)$. For inter-subject registration, ground truth is established from inlier correspondences identified via the Hough transform, which are visually validated for correctness and rejected or manually adjusted if necessary. This follows the work of~\citep{Toews2007StatisticalVariability,Machado2018NonrigidMatching}, where automatic and manually labeled correspondences exhibit the same error range. The average Point Registration Error (PRE) measure is used to quantify registration performance, based on the sum of 3D point differences between the registration solution and ground truth.

In all experiments, feature sets are extracted using a GPU implementation of the SIFT-Rank algorithm for computational efficiency~\citep{Carluer2021GPUDescriptor}, and registered using five point-cloud registration methods: ICP with 20 and 100 iterations (respectively ICP20 and ICP100), the original CPD algorithm using feature centers $\vect{x}$, SIFT-CPD using full feature geometry and our kernel $\mathcal{K}(g_n,g_m)$, and SIFT-CPD* using only inlier features from the Hough transform. All algorithms are initialized to the Hough transform solution prior to iterative registration in order to ensure a fair comparison.

\subsection{Synthetic Image Registration}

A preliminary experiment was first performed establish the baseline performances of algorithms against known ground truth, here synthetic similarity transformations of a single image. A T1w brain image from the Human Connectome Project (HCP)~\citep{VanEssen2012HumanPerspective} dataset was selected (see Table~\ref{tab:experiments} for more details), and 100 different synthetic transforms were applied, where each was generated by a random rotation about each axis on an angle range of $\rho \in \pm [10^{\circ},30^{\circ}]$ followed by a translation $\delta \in \pm[0~mm,10~mm]$. The registration error in rotation and translation was evaluated relative to the known transform applied, and the results are presented in Table~\ref{tab:samesubject_registration}. As expected, the lowest error is achieved for SIFT-CPD* (inliers alone), and this serves as a minimum error baseline. SIFT-CPD and CPD perform similarly, with slightly less error for SIFT-CPD. ICP results are generally poor, failing to converge in 8 and 21 cases for 20 and 100 iterations.

\begin{table*}
\large
\centering
\caption{Comparison of five registration algorithms based on 100 synthetic transforms of a single T1w MRI brain volume. Error is listed terms of individual rotation and translation axes, overall point registration error (PRE), and the sum of squared intensity differences (SSD) following registration}
\resizebox{1\columnwidth}{!}{
\bgroup
\def\arraystretch{1.8}
\begin{tabular}{|cc|ccc|ccc|cc|cc}
    \hline
    \multicolumn{2}{|P{3.5cm}|}{\multirow{2}{*}{\backslashbox[3.9cm]{Algorithm}{Error}}}&\multicolumn{3}{|c|}{\textit{Rotation Error (deg)}}&\multicolumn{3}{|c|}{\textit{Translation Error (mm)}}&\multicolumn{2}{|c|}{\textit{PRE (mm)}}&\multicolumn{2}{|c|}{\textit{SSD ($e^{11}$)}}\\ 
    \cline{3-12}
    \multicolumn{2}{|P{3.5cm}|}{}&\multicolumn{1}{|c|}{X-Axis}&\multicolumn{1}{|c|}{Y-Axis}&\multicolumn{1}{c|}{Z-Axis}&\multicolumn{1}{|c|}{X-Axis}&\multicolumn{1}{|c|}{Y-Axis}&\multicolumn{1}{|c|}{Z-Axis}&\multicolumn{2}{|c|}{Mean $\pm$ Std.Dev.}&\multicolumn{2}{|c|}{Mean $\pm$ Std.Dev.}\\
    \thickhline
    \multicolumn{2}{|P{3.5cm}|}{ICP20} & \multicolumn{1}{|c|}{$25.7 \pm 22.0$} & \multicolumn{1}{|c|}{$16.8 \pm 14.8$} & \multicolumn{1}{|c|}{$24.8 \pm 17.2$} & \multicolumn{1}{|c|}{$14.2 \pm 19.6$} & \multicolumn{1}{|c|}{$12.8 \pm 29.1$} & \multicolumn{1}{|c|}{$17.4 \pm 29.0$} & \multicolumn{2}{|c|}{$16.37 \pm 8.41$} &
    \multicolumn{2}{|c|}{$7.62 \pm 1.03$}\\
    \hline
    \multicolumn{2}{|P{3.5cm}|}{ICP100} & \multicolumn{1}{|c|}{$8.1 \pm 18.9$} & \multicolumn{1}{|c|}{$6.3 \pm 18.8$} & \multicolumn{1}{|c|}{$8.5 \pm 18.5$} & \multicolumn{1}{|c|}{$9.0 \pm 24.0$} & \multicolumn{1}{|c|}{$9.2 \pm 32.8$} & \multicolumn{1}{|c|}{$8.5 \pm 23.2$} & \multicolumn{2}{|c|}{$8.16 \pm 17.15$} &
    \multicolumn{2}{|c|}{$4.89 \pm 2.31$}\\
    \hline
    \multicolumn{2}{|P{3.5cm}|}{CPD} & \multicolumn{1}{|c|}{$0.06 \pm 0.03$} & \multicolumn{1}{|c|}{$0.06 \pm 0.03$} & \multicolumn{1}{|c|}{$0.06 \pm 0.03$} & \multicolumn{1}{|c|}{$0.8 \pm 0.6$} & \multicolumn{1}{|c|}{$1.3 \pm 1.1$} & \multicolumn{1}{|c|}{$0.7 \pm 0.5$} & \multicolumn{2}{|c|}{$1.81 \pm 0.88$} &
    \multicolumn{2}{|c|}{$1.38 \pm 0.39$}\\
    \hline
    \multicolumn{2}{|P{3.5cm}|}{\textbf{SIFT-CPD}} & \multicolumn{1}{|c|}{$\bf{0.02 \pm 0.01}$} & \multicolumn{1}{|c|}{$\bf{0.01 \pm 0.01}$} & \multicolumn{1}{|c|}{$\bf{0.02 \pm 0.01}$} & \multicolumn{1}{|c|}{$\bf{0.9 \pm 0.5}$} & \multicolumn{1}{|c|}{$\bf{0.8 \pm 0.6}$} & \multicolumn{1}{|c|}{$\bf{0.5 \pm 0.4}$} & \multicolumn{2}{|c|}{$\bf{1.05 \pm 0.32}$} &
    \multicolumn{2}{|c|}{$\bf{1.43 \pm 0.33}$}\\
    \hline
    \multicolumn{2}{|P{3.5cm}|}{\textbf{SIFT-CPD*}} & \multicolumn{1}{|c|}{$\bf{0.02 \pm 0.01}$} & \multicolumn{1}{|c|}{$\bf{0.01 \pm 0.01}$} & \multicolumn{1}{|c|}{$\bf{0.02 \pm 0.01}$} & \multicolumn{1}{|c|}{$\bf{0.9 \pm 0.4}$} & \multicolumn{1}{|c|}{$\bf{0.9 \pm 0.4}$} & \multicolumn{1}{|c|}{$\bf{0.4 \pm 0.3}$} & \multicolumn{2}{|c|}{$\bf{0.68 \pm 0.15}$} &
    \multicolumn{2}{|c|}{$\bf{0.96 \pm 0.39}$}\\
    \hline
\end{tabular}
\egroup
}
\label{tab:samesubject_registration}
\end{table*}

\subsection{Inter-subject Registration}

Four sets of experiments quantified the accuracy of inter-subject registration, i.e. registration of images of different individuals. The goal was to evaluate the accuracy and speed of registration in a diverse set of contexts, including different aspects of anatomy (brain, chest), abnormalities (tumor) and modalities (T1w and T2w MRI, CT). Information regarding datasets is provided in Table~\ref{tab:experiments}. Note that minimum registration error is generally non-zero due to differences in anatomy, and the goal is to identify a robust transform minimizing the PRE for sets of inlier correspondences.

\begin{table*}
\centering
\caption{Inter-subject registration experimental data including the numbers of features per image and Hough transform inlier matches}
\resizebox{1\columnwidth}{!}{
\bgroup
\def\arraystretch{1.8}
\begin{tabular}{|c|c|c|c|c|c|c|}
    \hline
    \textbf{Dataset} & \textbf{Anatomy} & \textbf{Modality} & \textbf{Resolution} & \textbf{Voxel Size (mm)}& \textbf{Features} & \textbf{Inliers}\\
    \thickhline
    \citep{VanEssen2012HumanPerspective} & Brain & T1w MRI &$260\times260\times311$&$0.7\times0.7\times0.7$&$6000$&$350$\\
    \hline
    \citep{Regan2011GeneticDesign} & Chest & CT &$280\times280\times235$&$1.2\times1.2\times1.2$&$3000$&$70$\\
    \hline
    \citep{Menze2015MultimodalBRATS} & Brain Tumor & T1w MRI &$240\times240\times155$&$1.0\times1.0\times1.0$&$1200$&$70$\\
    \hline
    \citep{VanEssen2012HumanPerspective} & Brain & T1w-T2w MRI &$260\times260\times311$&$0.7\times0.7\times0.7$&$5000$&$115$\\
    \hline
\end{tabular}
\egroup
}
\label{tab:experiments}
\end{table*}

\subsubsection{T1w Brain Images}

\begin{figure}[!h]
    \centering
    \includegraphics[width=1\textwidth]{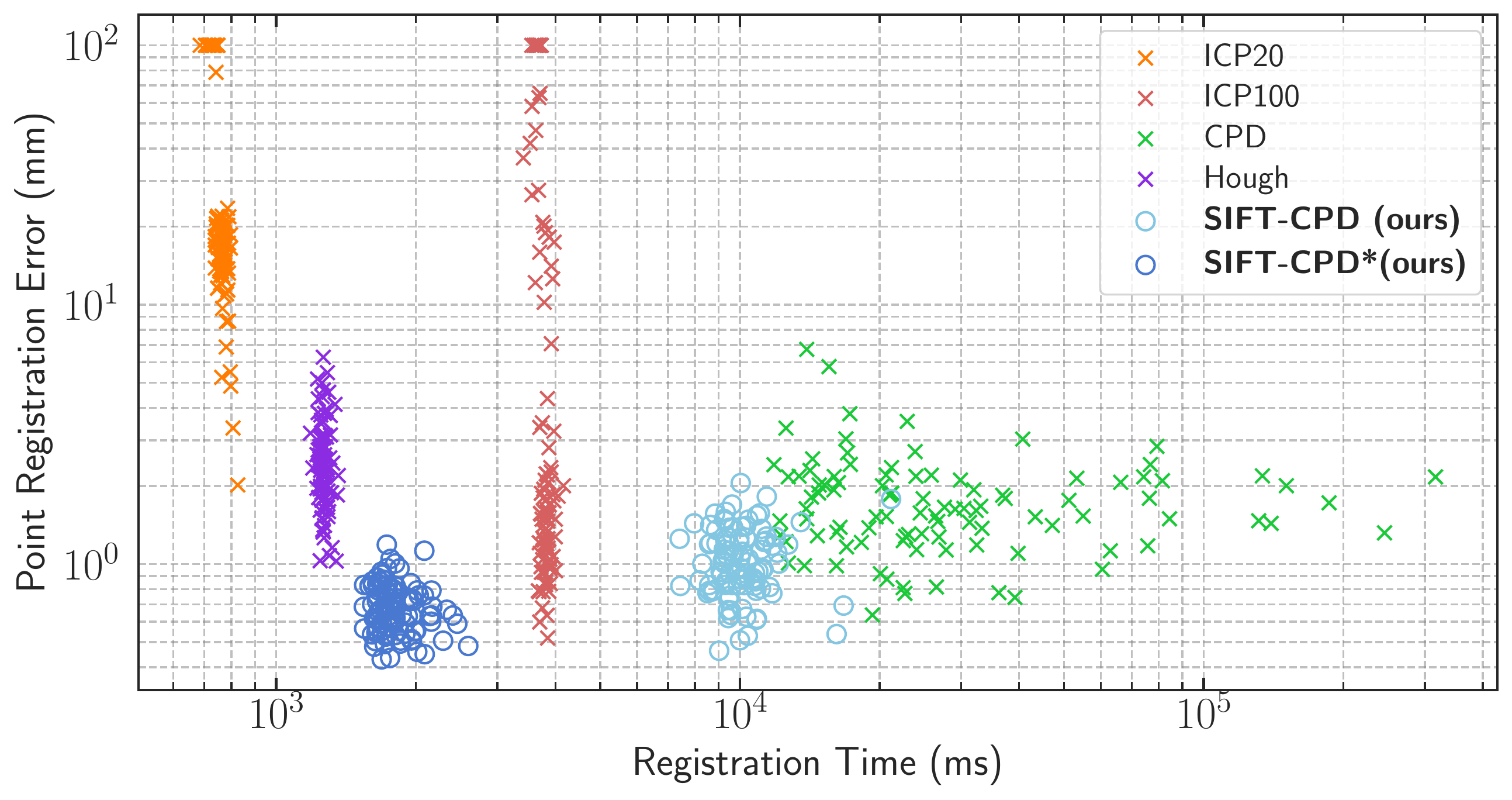}
    \caption{Inter-subject registration performance for healthy brain T1w MRI}
    \label{fig:synth_data_brain}
\end{figure}

The following experiment quantified inter-subject registration accuracy between brain images of a pair of different subjects. As in the previous experiment, one image was fixed, while the other was subjected to a random transform from the previously mentioned ranges $\rho$ and $\delta$. A pair of monozygotic twin subjects was selected, which generally share more inlier correspondences in comparison to unrelated subjects due to higher neuroanatomical similarity~\citep{Chauvin2020NeuroimageRelatives}. The results for 100 random transforms are shown in Figure~\ref{fig:synth_data_brain}, these are generally consistent across all registration experiments. ICP variants which weight all points uniformly do not correctly converge (red and orange). CPD and SIFT-CPD (green and light blue) both converge consistently to acceptable results in term of PRE, both lower than that the initial robust Hough transform (purple). SIFT-CPD is consistently lower in terms of error and computation time in comparison to CPD, demonstrating that geometrical properties improve estimation, specifically in the case where the majority of features may be outliers with no valid correspondence between images. Overall, the SIFT-CPD* (dark blue) leads to the lowest error and computation time, as it refines the original set of Hough transform inliers.

\subsubsection{CT Chest Images}
\begin{figure}
    \centering
    \includegraphics[width=1\textwidth]{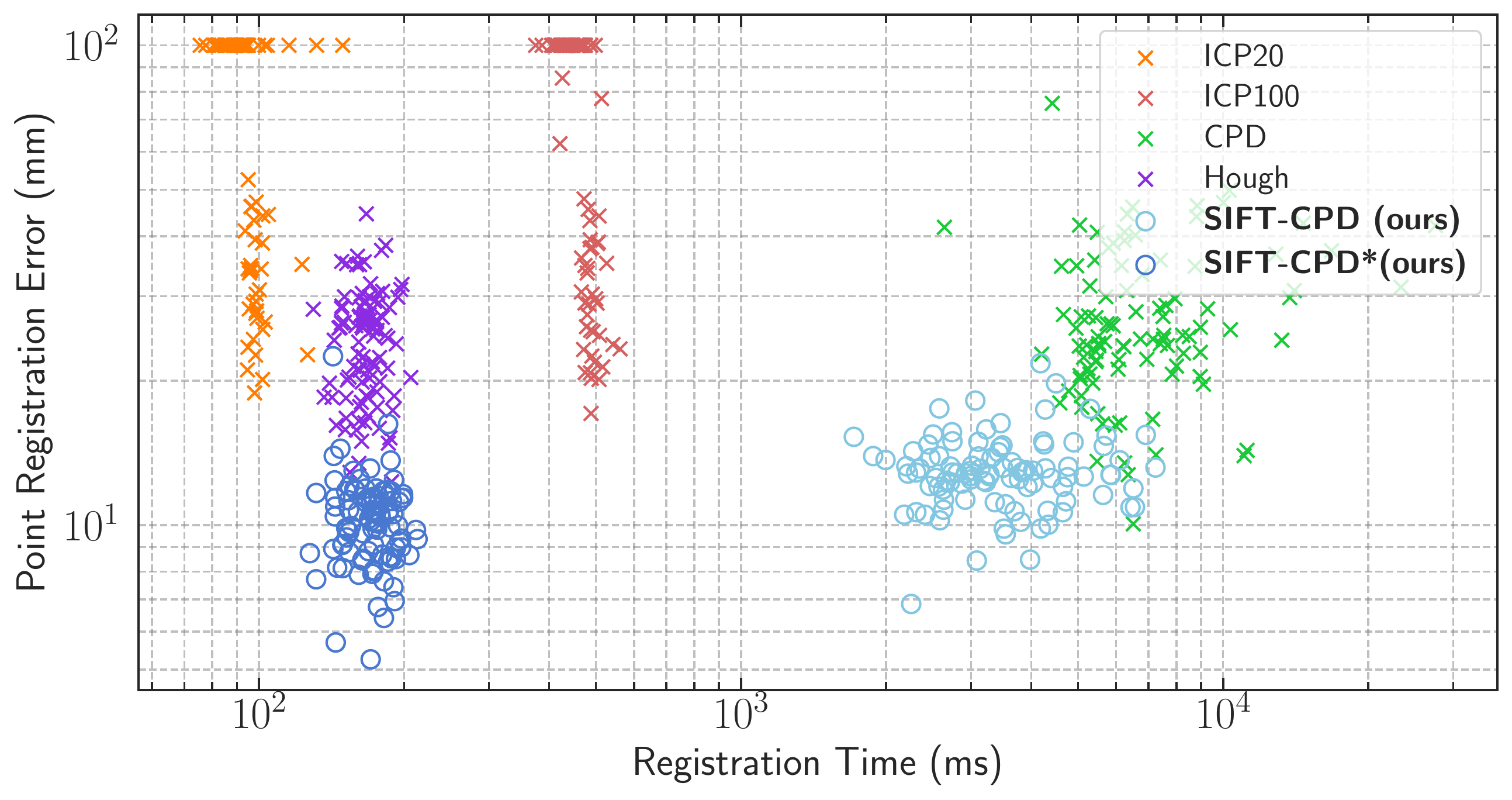}
    \caption{Inter-subject registration performance for chest CT images}
    \label{fig:synth_data_lungs}
\end{figure}

In order to show that our method is not tied to specific contexts or imaging modalities, we performed inter-subject registration of CT chest images from the Chronic Obstructive Pulmonary Disorder Genetics (COPDGene) dataset~\citep{Regan2011GeneticDesign}. Two different subject images in expiration breathing state were randomly selected (see Table~\ref{tab:experiments} for more details).
As in the previous experiment, we generated 100 random transform with the same $\rho$ and $\delta$ parameters, applied the transform to one image, and registered it using different registration approaches. The PRE for each registration is shown in Figure~\ref{fig:synth_data_lungs}. Although the computational complexity of SIFT-CPD is comparable to that of CPD in Figure~\ref{fig:synth_data_lungs}, SIFT-CPD* is significantly faster.

\subsubsection{Abnormal Variations and Occlusion}
\begin{figure}
    \centering
    \includegraphics[width=1\textwidth]{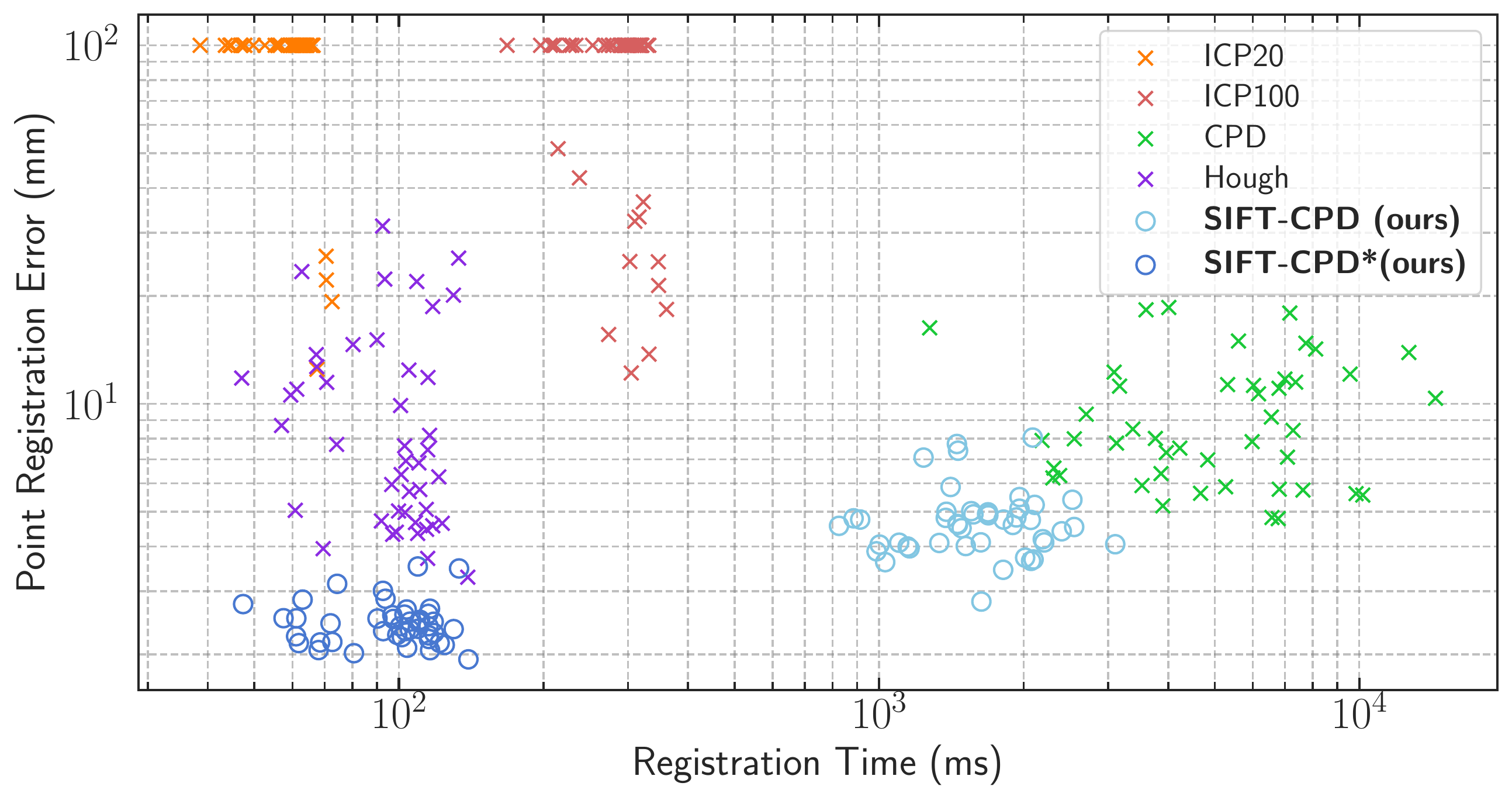}
    \caption{Inter-subject registration performance for brains in T1w MRI, with occlusions due to tumors~\citep{Menze2015MultimodalBRATS}}
    \label{fig:occlusion_results}
\end{figure}

Practical image analysis requires coping robustly with occlusions and a lack of one-to-one correspondence, due to both normal inter-subject variations and abnormal pathological structure such as tumors. Our method is based on local feature properties, and is thus particularly robust to occlusion and missing structure. We used 10 T1w MRI images of different subjects with tumors from the BraTS dataset~\citep{Menze2015MultimodalBRATS} in order to evaluate registration in the presence of occlusion (see Table~\ref{tab:experiments} for more details). Nine subject images are randomly selected, and transformed 5 times each, and registered to the remaining subject image. Results are presented in Figure~\ref{fig:occlusion_results}.

\subsubsection{Multi-Modal Images}
\begin{figure}[!h]
    \centering
    \includegraphics[width=1\textwidth]{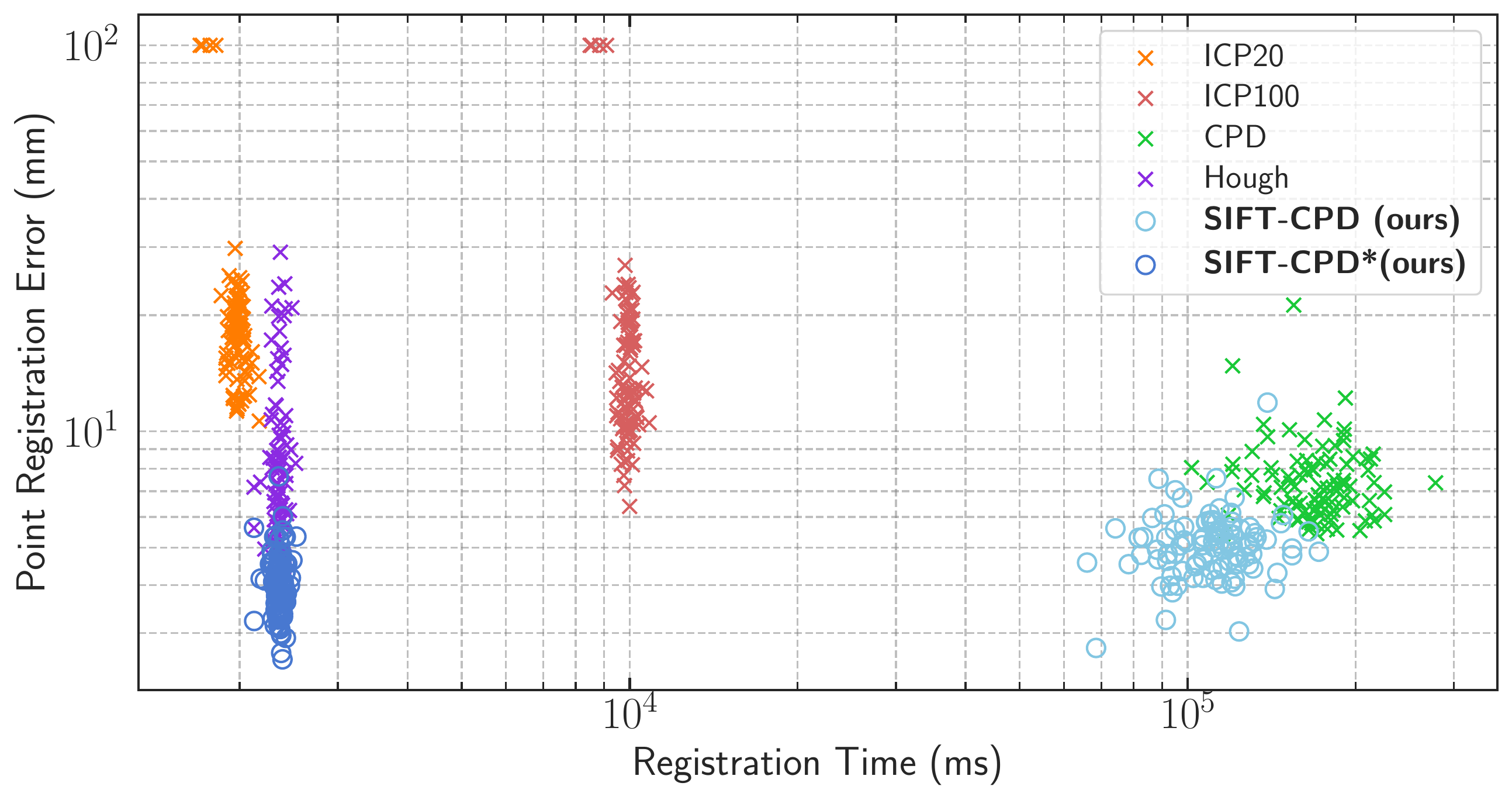}
    \caption{Inter-subject registration performance for healthy brain images between T1w and T2w MRI modalities~\citep{VanEssen2012HumanPerspective}, exhibiting intensity contrast inversion}
    \label{fig:synth_data_multimodal_brain}
\end{figure}

\begin{figure}[!h]
    \centering
    \includegraphics[width=1\textwidth]{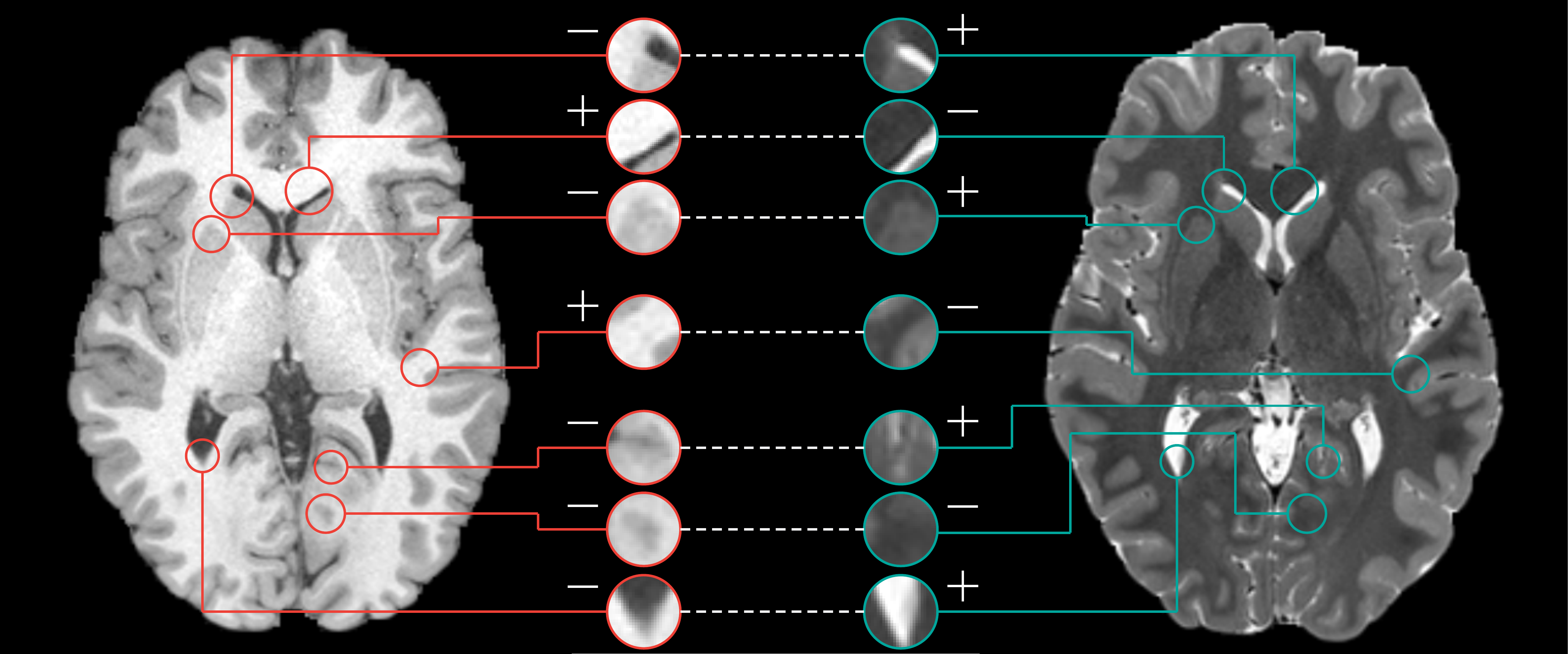}
    \caption{Examples of features matched between T1w and T2w images, showing feature signs, note the sign changes do not occur everywhere}
    \label{fig:multimodal_matches}
\end{figure}

To evaluate the registration performance of our approach on multi-modal images, a random pair of brain images was selected from HCP dataset, one with a T1w image and the other one with a T2w (see Table~\ref{tab:experiments} for more details). T1w and T2w images are acquired with different MRI parameters and highlight different tissue properties, where T2w imaging involves longer repetition time (TR) and time-to-echo (TE) parameters. The T2w image was randomly transformed and registered as in previous experiments, with 100 trials. Results are presented in Figure~\ref{fig:synth_data_multimodal_brain}. Note that without SP-invariant descriptors as proposed in Equation~\eqref{eq:descriptor}, multi-modal registration fails completely due to a lack of correspondence. Inlier feature correspondences between different modalities are shown in Figure~\ref{fig:multimodal_matches}, these are investigated in greater detail in the following experiment section.

\subsection{Feature Sign and Orientation State}

Here we investigated the binary sign $s$ and orientation state $(\theta_1,\theta_2)$ of local features in registration experiments, using two different methods for estimating feature orientation $\Theta$: eigenvectors of the local gradient structure tensor matrix~\citep{Rister2017VolumetricKeypoints} or maximum orthogonal gradient directions~\citep{Toews2013EfficientFeatures}. Both methods have been demonstrated to achieve effective local feature correspondences between images of the same modality, but not different modalities where discrete orientation state changes may occur.

\begin{figure}
    \centering
    \includegraphics[width=1\textwidth]{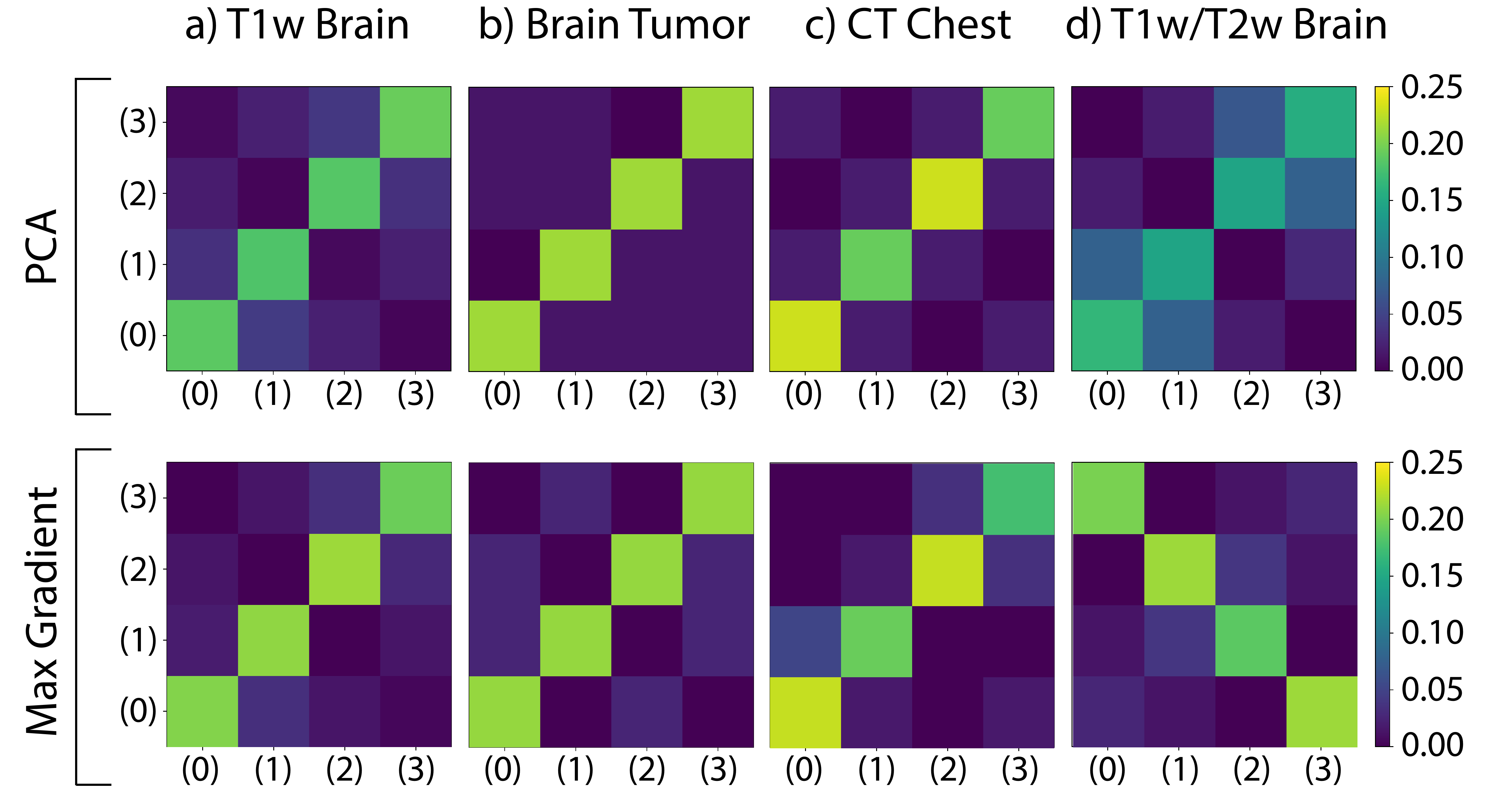}
    \caption{Orientation state transition histograms for principal gradient orientation estimation (see~\citep{Rister2017VolumetricKeypoints}) and max gradient orientation estimation (see~\citep{Toews2013EfficientFeatures}) for: a) T1w brain, b) T1w brain tumor~\citep{Menze2015MultimodalBRATS}, c) CT chest~\citep{Regan2011GeneticDesign}, d) Multi-modal T1w/T2w brain. States
    described in Figure~\ref{fig:workflow}, are combinations of discrete axis reflections $\{\pm\hat{\theta}_1,\pm\hat{\theta}_2\}$ as 0:$\{++\}$,
    1:$\{+-\}$,
    2:$\{-+\}$,
    3:$\{--\}$. Bins along the diagonal indicate no state change, off-diagonals indicate single-axis reflections and inverse diagonal indicates parity transform
    }
    \label{fig:states}
\end{figure}

\begin{figure}
    \centering
    \includegraphics[width=1\textwidth]{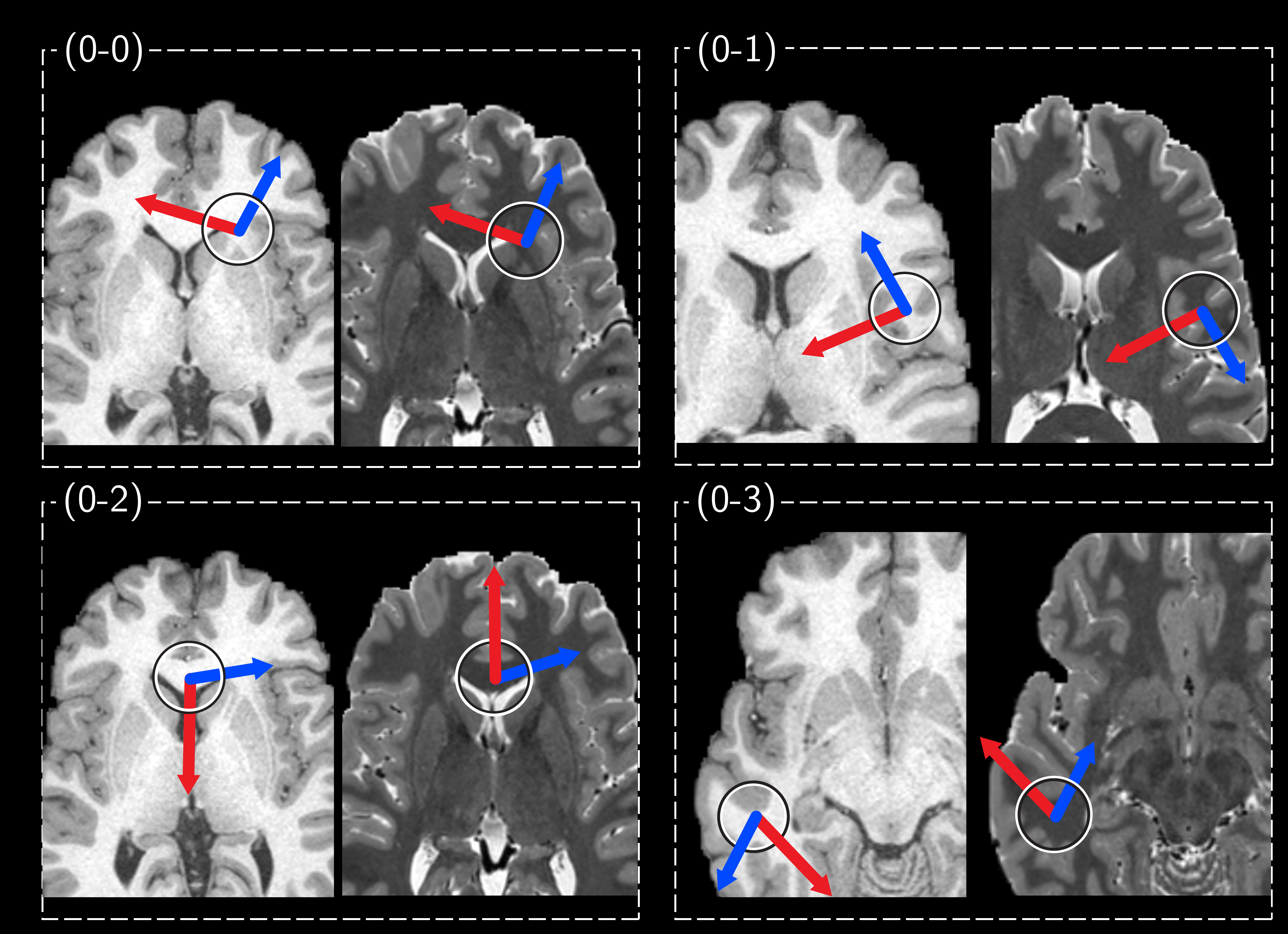}
    \caption{Examples of feature orientation state transitions observed between T1w and T2w modalities. Circles indicate feature location $\vect{x}$ and scale $\sigma$ in the plane, sign $s \in \{-1,+1\}$ is indicated by white-black or black-white color. Red and blue arrows indicate the primary $\hat{\theta}_1$ and secondary $\hat{\theta}_2$ orientation axes projected onto the (x,y) image plane}
    \label{fig:symmetries}
\end{figure}

Figures~\ref{fig:states} show distributions of orientation state transitions of inlier correspondences between images. For same modality registration (a, b, c) there are relatively few state changes (main diagonal), however in the case of multi-modal registration (d) noticeable differences exist. In the case of orientation determined via principal component analysis (PCA) of gradients~\citep{Rister2017VolumetricKeypoints}, approximately 40\% of all correspondences exhibit single-axis reflections (Figure~\ref{fig:states} upper d) rather than SP transforms, this is due to eigenvector sign ambiguity, which is accounted for in our algorithm. In the case of orientation determined by maximum gradient directions, correspondences generally exhibit full SP-transforms, i.e. sign inversion and parity (Figure~\ref{fig:states} lower d). Figure~\ref{fig:symmetries} shows examples of corresponding features and state transitions in the case of sign inversion, including identity (0-0), single axis reflections of major (0-1) and minor axes (0-2), and parity transforms (0-3).

\section{Discussion}
This paper proposed to extend local feature methods to account for SP-symmetry and to achieved 3D multi-modal image registration. We introduced the notion of discrete SP-symmetry in the of case features extracted from a scalar intensity image $I(\vect{x}) \in R^1$ over $\vect{x} \in R^3$ space. We proposed to model the geometrical properties of 3D image features in a manner invariant to SP-transforms observed in multi-modal image registration. We proposed to include a feature sign $s \in \{-,+\} = {\text sign}~\nabla^2~I(\vect{x},\sigma)$ analogous to an electric charge, and maintain a discrete set of four orientation states including reflections of primary axes and parity transforms. Sign is used to invert gradient descriptor elements in the case of intensity contrast inversion, i.e. charge conjugation. Feature geometry and appearance descriptors are thus invariant to SP transforms, in addition to continuous rotations, scalings and translations as in the 3D SIFT method~\citep{Toews2013EfficientFeatures,Rister2017VolumetricKeypoints}.

We integrated feature properties into a well-known probabilistic point-based image registration framework, the CPD algorithm~\citep{Myronenko2010PointDrift}, via a kernel function $\mathcal{K}(f_n,f_m)$, leading to a highly robust and stable SIFT-CPD registration algorithm. Image registration experiments are performed using a range of volumetric image modalities of the human brain and chest, showing that additional properties consistently improve the accuracy of registration in comparison to point locations alone. Registration error and computation time is consistently lower for our proposed SIFT-CPD vs. standard point-based CPD frameworks, indicating that feature scale and orientation improve the estimation of image-to-image transforms. An optimized version of SIFT-CPD functioning solely on sparse inlier feature correspondences exhibits the lowest error, refining the solution achieved via a robust but coarse initial Hough transform. Experiments also demonstrated how SP-symmetry may be applied despite the method used to estimate feature orientation estimation, here principal gradient orientations~\citep{Rister2017VolumetricKeypoints} and maximum gradient orientations~\citep{Toews2013EfficientFeatures}.  Invariant feature appearance descriptors are used to identify inliers and to initialize registration, there were not explicitly incorporated into the SIFT-CPD kernel function based solely on geometrical properties here, however this possibility is left for future work.

We note that the theory of SP-symmetry we present applies generally to individual channels of convolutional neural networks, and could be used within the general neural network learning framework to achieve SP-invariance, in addition to continuous invariance~\citep{Bardes2022VICRegLearning,Esteves2018LearningCNNs,Spezialetti2019LearningSupervision,Cohen2019GaugeCNN}. Here, 3D invariant features have been identified via unbiased, symmetric operators with no explicit training procedure, which could be used as is in general imaging contexts, or to train or validate domain-specific detectors via neural networks~\citep{Yi2016LIFTTransform,Detone2018SuperPointDescription}. All code required to reproduce our results will be available at \href{https://github.com/3dsift-rank/SIFT-CPD}{https://github.com/3dsift-rank/SIFT-CPD}.

%%- First demo chapter -%%
\chapter{Neuroimage signature from salient keypoints is highly specific to individuals and shared by close relatives}
\label{chap3}
\chaptermark{Neuroimage signature from salient keypoints}

\articleAuthors{
{Laurent Chauvin\textsuperscript{a}}{Kuldeep Kumar\textsuperscript{b}}{Christian Wachinger\textsuperscript{c,f}}{Marc Vangel\textsuperscript{d}}{Jacques de Guise\textsuperscript{a}}{Christian Desrosiers\textsuperscript{a}}{William Wells III\textsuperscript{e,f}}{and Matthew Toews\textsuperscript{a}{ for the Alzheimer’s Disease Neuroimaging Initiative}}
}{
{\setstretch{1.2}
\textsuperscript{a} Department of Systems Engineering, École de Technologie Supérieure, Canada%,\\
%1100 Notre-Dame Ouest, Montréal, Québec, Canada H3C 1K3
\\
\textsuperscript{b} CHU Sainte-Justine Research Center, University of Montreal, Canada%,\\
%3175 Chemin de la Côte-Sainte-Catherine, Montreal, Québec, Canada H3T 1C5
\\
\textsuperscript{c} Laboratory for Artificial Intelligence in Medical Imaging, University of Munich, Germany%,\\
%Geschwister-Scholl-Platz 1, 80539 Munich, Germany
\\
\textsuperscript{d} Massachusetts General Hospital, Harvard Medical School, USA%,\\
%55 Fruit Street, Boston, MA 02114, USA
\\
\textsuperscript{e} Brigham and Women's Hospital, Harvard Medical School, USA%, \\
%25 Shattuck Street, Boston, MA 02115, USA
\\
\textsuperscript{f} Computer Science and Artificial Intelligence Lab, Massachusetts Institute of Technology, USA%, \\
%77 Massachusetts Avenue, Cambridge, MA 02139, USA
\\~\\
Paper published in \textit{NeuroImage}, September 2019\\ Our figure has been selected to make the cover of the journal issue
}
}
\\
\textbf{Abstract}\\
Neuroimaging studies typically adopt a common feature space for all data, which may obscure aspects of neuroanatomy only observable in subsets of a population, e.g. cortical folding patterns unique to individuals or shared by close relatives. Here, we propose to model individual variability using a distinctive keypoint signature: a set of unique, localized patterns, detected automatically in each image by a generic saliency operator. The similarity of an image pair is then quantified by the proportion of keypoints they share using a novel Jaccard-like measure of set overlap.  Experiments demonstrate the keypoint method to be highly efficient and accurate, using a set of 7536 T1-weighted MRIs pooled from four public neuroimaging repositories, including twins, non-twin siblings, and 3334 unique subjects. All same-subject image pairs are identified by a similarity threshold despite confounds including aging and neurodegenerative disease progression. Outliers reveal previously unknown data labeling inconsistencies, demonstrating the usefulness of the keypoint signature as a computational tool for curating large neuroimage datasets.

% \linenumbers

\section{Introduction}
\label{sec1}

The human brain is a highly complex organ in terms of both structure and function, that is widely studied in vivo through magnetic resonance imaging (MRI)~\citep{Lerch2017StudyingMRI}. To what degree is neuroanatomy, as observed in MRI, unique to individuals? Can individuals be reliably distinguished from close relatives, i.e. siblings or monozygotic twins sharing 50-100\% of their polymorphic genes, despite natural aging, neurodegenerative disease, or noise due to the data measurement process? To what degree are unique aspects of neuroanatomy shared by close relatives? These questions are motivated by increasingly personalized modern medical practices and the need to accurately curate growing sets of clinical and research neuroimaging data. We address them in this paper using a unique computer vision method.

A number of studies have investigated the variability of individuals rather than populations~\citep{Valizadeh2018IdentificationFeatures,Finn2015FunctionalConnectivity,Miranda-Dominguez2014ConnectotypingConnectome}, where a common theme has been to encode data in terms of a unique neuroimage ``fingerprint'' or brainprint. Although the specific encodings used are data-dependent, the accuracy with which individuals can be identified based on their brainprint may indicate the degree to which inferences may be drawn from unique, subject-specific observations~\citep{Finn2015FunctionalConnectivity}. Brainprint investigations have been performed from a variety of MRI modalities, including structural~\citep{Wachinger2015BrainPrintMorphology,Takao2015BrainInformationa}, diffusion~\citep{Valizadeh2018IdentificationFeatures, Kumar2017FiberprintAnalysis, Yeh2016QuantifyingFingerprints} and functional~\citep{Colclough2017HeritabilityActivity,Finn2015FunctionalConnectivity,Miranda-Dominguez2014ConnectotypingConnectome,Chen2018IndividualNetwork,Amico2018QuestConnectomes} MRI, and in non-MRI data such as EEG~\citep{Armstrong2015BrainprintBiometrics}. Our work here is the first to investigate individual identification from multiple, large-scale public MRI datasets used by the neuroimaging community, including the Human Connectome Project (HCP)~\citep{VanEssen2013WUMinnOverview}, the Alzheimer's Disease Neuroimaging Initiative (ADNI)~\citep{JackJr2008AlzheimerMethods} and the Open Access Series of Imaging Studies (OASIS)~\citep{Marcus2007OpenAdults}, and first to report perfect accuracy in individual identification experiments.

The challenge in establishing a distinctive brainprint can perhaps be best illustrated by the convoluted neocortex, arguably the most distinguishing aspect of human neuroanatomy. Cortical folding patterns are highly unique to individuals and generally exhibit higher correlation between twins than unrelated individuals~\citep{VanEssen2016ParcellationsCortex, Thompson2001GeneticStructure}, suggesting a link between subtle neuroanatomic structure and shared genetics. A pair-wise image correlation analysis could potentially distinguish individuals and relatives, however such an approach is generally impractical for large datasets as the number of pair-wise operations including image registration is quadratic $N(N-1)/2$ in the number of images $N$. Most studies of individual variability have interpreted all data in terms of a standard feature set, e.g. spatially registered voxel-wise measurements~\citep{Takao2015BrainInformationa}, neuroanatomic segmentations~\citep{Wachinger2015BrainPrintMorphology}, cortical parcellations~\citep{Colclough2017HeritabilityActivity,Finn2015FunctionalConnectivity,Miranda-Dominguez2014ConnectotypingConnectome, Fischl2012FreeSurferFreeSurfer} and related volume or thickness measurements~\citep{Sabuncu2016MorphometricityTrait}. While standard measurements are invaluable in interpreting data and reducing computational complexity, a number of authors have noted that a one-size-fits-all representation may obscure or average out aspects of anatomy unique to individuals~\citep{Finn2015FunctionalConnectivity,Gordon2017IndividualspecificCorrelations,Chen2018IndividualNetwork} or close relatives, for example in the case where a one-to-one mapping between images may be ill-defined or nonexistent due to individual variability.

\begin{figure}[ht]
	\centering
	\includegraphics[width=1\textwidth]{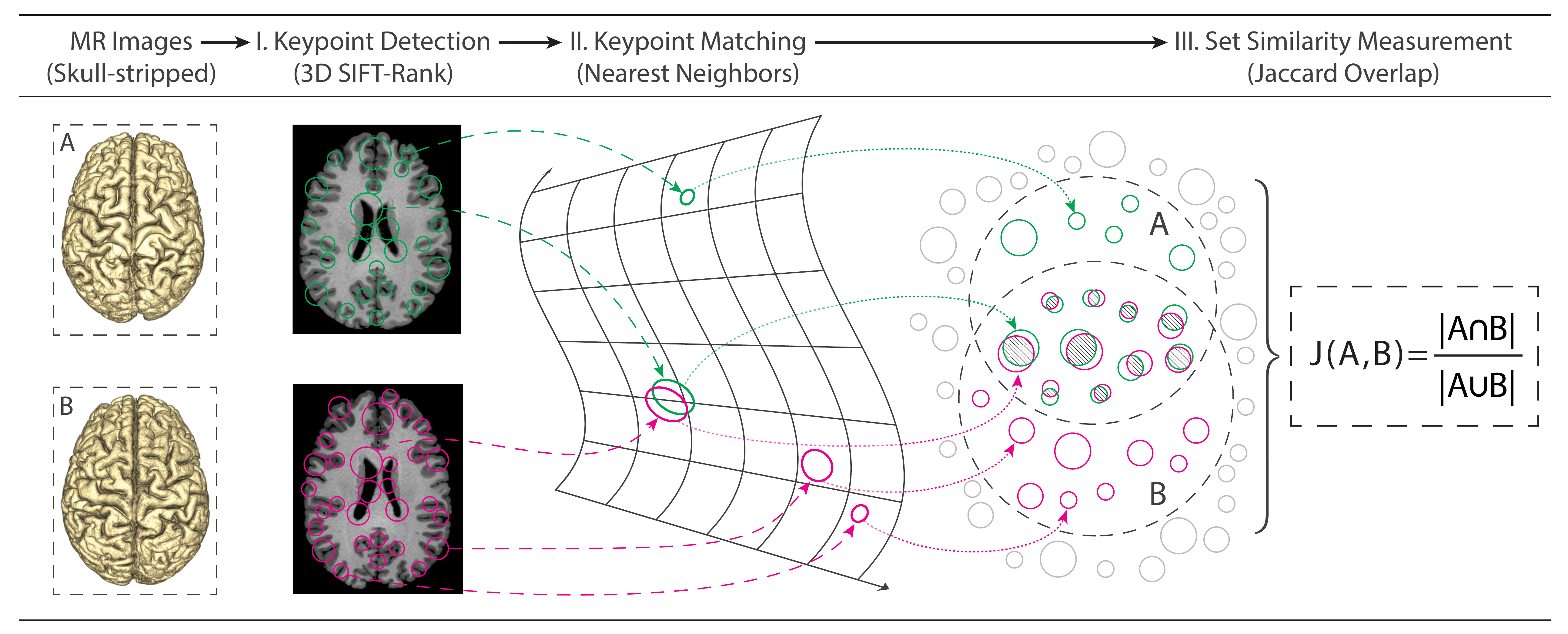}
	\caption{The workflow for computing the Jaccard overlap $J(A,B)$ similarity score between two images A and B. \textbf{Step I.} SIFT-Rank keypoints are extracted from skull-stripped MRI data. \textbf{Step II.} Similar keypoints are identified between images using a K-nearest neighbor search. \textbf{Step III.} The Jaccard overlap is computed as ratio of the intersection vs. the union of keypoint sets}
	\label{fig:method_workflow}
\end{figure}

An alternative is to encode the image as a unique set of informative localized features or keypoints, that can be detected efficiently from generic image content and identified robustly when present in different images. For example, the highly successful scale-invariant feature transform (SIFT)~\citep{Lowe2004DistinctiveKeypoints} from the field of computer vision uses highly efficient K-nearest neighbor (KNN) keypoint indexing to identify correspondences between generic intensity patterns in large image sets. Inspired by this work, we developed the 3D SIFT-Rank keypoint method~\citep{Toews2013EfficientFeatures} for analyzing volumetric image data as illustrated in Figure~\ref{fig:method_workflow}. Toews et al.~\citep{Toews2016HowFeatures} showed that the proportion of detected keypoints common to a pair of brain MRIs, as quantified by the Jaccard measure of set overlap~\citep{Levandowsky1971DistanceSets}, could be used to identify MRI pairs of siblings with high reliability. The Jaccard measure defines a similarity matrix $J(A,B)$ between all keypoints and image pairs $(A,B)$ of a dataset, that can be used in learning-based MRI analysis. Toews et. al.~\citep{Toews2010FeaturebasedPatterns} identified class-informative keypoint clusters in the similarity matrix $J(A,B)$ for MRI-based disease classification. Kumar et al.~\citep{Kumar2018MultimodalFramework} combined similarity matrices derived from multiple MRI modalities in a low-rank manifold embedding in order to study correlations between MRIs of siblings, and found that keypoints outperformed a number of baseline representations including MRI intensities and FreeSurfer derived measures (Volume + Area + Cortical Thickness). While learning procedures may be useful in analyzing a fixed dataset or group analysis, they are difficult to adapt to previously unseen data or classes, i.e. to account for images of previously unobserved individuals. This paper is the first to investigate the task of individual identification using the keypoint representation.

These results lead us to hypothesize that SIFT-Rank keypoint sets serve as a highly specific encoding of unique neuroanatomic structure. Here, we propose a novel generalisation of the Jaccard score $J(A,B)$ to account for probabilistic rather than hard set equivalence between keypoints. This leads to a highly efficient instance-based inference model, allowing new MRI data to be compared to a large dataset on-the-fly in order to identify all previous scans of the same individual despite myriad potential confounds including noise, atrophy due to neurodegeneration.
This paper reports the first comprehensive investigation of keypoint signatures for individual identification from MRI data. Our results based on several large-scale, public neuroimaging datasets demonstrated that all same-subject image pairs can be identified by a simple threshold on Jaccard overlap. A visual analysis of outlier cases revealed all were due to data labeling inconsistencies previously unknown to the neuroimaging community, demonstrating the practical importance of the approach in curating increasingly large data collections. 

\section{Material \& Methods}
\label{sec2}

\subsection{Data}
Experiments are based on a large, multi-site dataset pooled from 4 public neuroimaging repositories HCP Q4, ADNI 1, OASIS 1 and OASIS 3. The FreeSurfer v6.0 pre-processing pipeline was used to remove non-brain content such as skull from MRIs while preserving cortical and subcortical structures. Out of 8152 images, 616 failed the pre-processing pipeline (FreeSurfer error code, no output image generated) resulting in a final dataset of N=7536 MRIs of 3334 unique subjects. Pre-processing failures typically occurred in Talairach alignment or in skull stripping steps, and visual inspection revealed that all cases were due to noticeable image artifacts or noise. Table~\ref{tab:database_info} lists demographic and statistical information for this dataset.

\begin{table}[ht]
    \centering
	\caption{Dataset demographic and statistical information}
	\label{tab:database_info}
	    \resizebox{\columnwidth}{!}{
		\begin{tabular}{|c|ccc|ccc|}
		    \hline
            \textbf{Dataset} & \textbf{Subjects} & \textbf{Gender} & \textbf{Age} & \textbf{Images} & \textbf{Voxel Size} & \textbf{keypoints}\\
            & & \textbf{(M / F)} & \textbf{(Min/Avg/Max)} & & \textbf{(mm)} & \textbf{(Avg/Image)}\\
            \hline
            \textbf{HCP Q4} & 1011 & 469 / 542 & 22 / 29 / 36 & 1053 & 0.7 & 3633\\
            \textbf{ADNI 1} & 844 & 488 / 356 & 55 / 75 / 91 & 3291 & 1.0 & 1879\\
            \textbf{OASIS 1} & 416 & 160 / 256 & 18 / 53 / 96 & 416 & 1.0 & 2143\\
            \textbf{OASIS 3} & 1063 & 470 / 593 & 42 / 70 / 97 & 2776 & 1.0 & 1856\\
            \hline
            \textbf{Total} & \textbf{3334} & \textbf{1587 / 1747} & \textbf{18 / 56 / 97} & \textbf{7536} & \textbf{-} & \textbf{2130}\\
            \hline
		\end{tabular}
		}
\end{table}	

Our analysis is based on a pair-wise comparison of $N(N-1)/2=28,391,880$ possible image pairs. Each pair is assigned a relationship label based on database metadata, for five possible sibling relationships: same subject (SM), monozygotic twins (MZ), dizygotic twins (DZ), full-sibling (FS) or unrelated subjects (UR). Subjects from different data sets are na\"{i}vely assumed to be unrelated, due to a lack of information across databases. It is important to note that these datasets were acquired under different protocols and over different periods of time, resulting in potential bias due to within-dataset similarity. For example, the time interval between scans is under a year for HCP, in comparison to 11 years for OASIS 3, 3 years for ADNI and 2 years for OASIS 1, and pairs of HCP scans may thus exhibit higher similarity than others. Our method is nevertheless robust the ranges of inter- and intra-dataset variability of these data, as we mention later in the discussion.\\ 

\subsection{Processing}
Assessing the pairwise similarity of images in a large data set generally requires comparing measurements at spatially homologous locations throughout the images. Since image data are noisy and the precise spatial mapping between images may be unknown or nonexistent, an effective comparison requires a combination of image registration and/or feature extraction methods. Na\"{i}ve similarity assessment for all image pairs is generally intractable for large datasets as the number of pairwise operations including image registration is quadratic $N(N-1)/2$ in the number of images $N$, incurring a computational complexity of $O(N^2)$. Assessment based on a standard feature set such as spatially aligned cortical parcellations can reduce computational complexity, but may be insufficiently specific to capture subtle neuroanatomic patterns only observable in small subsets of a population, e.g. family members.

To address these challenges, we developed a method based on keypoint indexing~\citep{Lowe2004DistinctiveKeypoints,Toews2013EfficientFeatures}, where the image is represented as a set of generic features detected throughout the image via a saliency operator. Keypoints arise from generic neuroanatomical structure, and can be detected repeatably in a manner invariant to locally linear intensity shifts and global similarity transforms (i.e. 3D rigid transform + isotropic scaling) of image geometry~\citep{Toews2013EfficientFeatures}. In practice, keypoint extraction is highly robust to variations in MRI intensity and geometry, e.g. in the case of images acquired from different devices or sites, exhibiting artifacts such as low frequency MRI inhomogeneity effects or changes in patient position. It is also robust to partial occlusions, where locally missing or deformed image content will not significantly impact keypoints identified in other unaffected image regions. Once detected, image content associated with keypoints is encoded into informative descriptors that can be used to identify similar keypoints via highly efficient indexing methods. Specifically, divide-and-conquer algorithms can achieve $O(N~log~N)$ computational complexity using search trees to identify sets of similar keypoints, thus sidestepping the need for pairwise image comparisons and scaling gracefully to large datasets. Our method consists of image keypoint detection (Figure~\ref{fig:method_workflow}, Step I), keypoint matching (Step II) and finally computation of the Jaccard overlap similarity score (Step III). These three steps are described in greater detail below.\\

\noindent
\textbf{Keypoint Detection} seeks to transform each image into a set of salient image keypoints, where keypoints are encoded as informative descriptors for efficient image content indexing. A keypoint is as a spherical image region defined by a centroid $\bar{x} = [x,y,z]$ and a scale (or size) $\sigma$ within the image, and associated with a descriptor $\bar{f}$ encoding local image appearance. A deterministic two-step detection procedure is used, including 1) salient keypoint localization and 2) keypoint encoding with the so-called SIFT-Rank approach~\citep{Toews2009SIFTRankCorrespondence,Toews2013EfficientFeatures}. Keypoints are first localized by searching the image for regions that maximize an image saliency operator, signifying informative, local image patterns. A variety of such operators exist, here we use the 3D Laplacian-of-Gaussian (LoG) operator~\citep{Marr1980TheoryDetection} that responds to generic blob-like image patterns, reminiscent of center-surround simple cell retinal processing units in the mammalian visual system~\citep{Hubel1962ReceptiveCortex}. The LoG operator can be approximated efficiently by the difference-of-Gaussian (DoG) operator popularized by the SIFT algorithm in 2004~\citep{Lowe2004DistinctiveKeypoints} (see Equation~\eqref{eq:dog}). For each image $I$, a set of keypoints $\{~(\bar{x}_i,\sigma_i)~\}$ is identified as:
\begin{equation}
    \{~(\bar{x}_i,\sigma_i)~\} \, = \, \underset{\bar{x},\sigma}{\operatorname{argmax}} \ |~I*G(\bar{x},\sigma) \, - \, I*G(\bar{x},\kappa\sigma)~|
    \label{eq:dog}
\end{equation}
where in Equation~\eqref{eq:dog} $I$ is the 3D image, $G(\bar{x},\sigma)$ is the Gaussian function with isotropic variance $\sigma^2$, and $\kappa$ is a constant representing the multiplicative difference in scale. Note that the keypoint scale $\sigma$ is defined by the size of the Gaussian filters for which the DoG saliency operator in Equation~\eqref{eq:dog} is maximized.

Once keypoint regions are localized, the image content within each region is rescaled and rotated to a characteristic coordinate system, then encoded as a descriptor $\bar{f}$ representing the local image content as a fixed-length vector. The rescaling factor is defined by the keypoint scale $\sigma$ and the rotation by the 3D orientation of local image gradients, thus the descriptor is invariant to global scaled rigid transforms (i.e. similarity transforms) of image geometry, e.g. in the cases of variable patient scanning position or unregistered images. Local image gradient analysis is also used to reject keypoints arising from image patterns that cannot be reliably localized in 3D, e.g. smooth surfaces.

Distinctive image patterns generally arise from boundaries between regions of differing intensity contrast, e.g. white and grey matter. Descriptors here encode image content as histograms of local image gradient information, estimated via finite difference operators and quantized into discrete bins over local 3D location and orientation, an approach know as the histogram-of-oriented gradient (HoG) descriptor~\citep{Lowe2004DistinctiveKeypoints,Dalal2005HistogramsDetection}. Robustness to minor shifts or deformation of image geometry is achieved by the use of relatively coarse spatial bins, where small variations in gradient location or direction do not significantly impact descriptor values. The HoG descriptor is currently among the most effective and widely used descriptors for image keypoint matching, and is analogous of so-called orientation hypercolumns of complex cells that encode image structure in the mammalian visual cortex~\citep{Hubel1962ReceptiveCortex}. In our method, local image patches are cropped and rescaled to $11^{3}$ voxels, after which image gradient histograms are computed to encode the image content. For encoding, local 3D space and orientation are quantized uniformly into 2$\times$2$\times$2=8 spatial regions and 8 orientation histogram bins, thereby producing an 8$\times$8=64 element HoG image descriptor for each keypoint. Finally, this descriptor is rank-ordered to provide invariance to monotonic variations in image gradients~\citep{Toews2009SIFTRankCorrespondence}, e.g. due to variations in image contrast.

\noindent
\textbf{Keypoint Matching} seeks to identify pairs of keypoints arising from similar anatomical structure in different images, based on the similarity (or dissimilarity) of their descriptors. Descriptors from the same structure in different images are rarely identical, but differ by varying amounts due to imaging variations, noise, etc. The dissimilarity of a pair of descriptors $(\bar{f_i},\bar{f_j})$ is quantified by the Euclidean distance or $L2$-norm between their elements $d(\bar{f_i},\bar{f_j}) = \|\bar{f_i}-\bar{f_j}\|_{2}$. Assuming an additive Gaussian noise model and independent and identically distributed (IID) descriptor elements, the smaller the distance $d(\bar{f_i},\bar{f_j})$, the higher the likelihood that the descriptors arise from the same underlying anatomical structure.

A K-nearest neighbor (K-NN) search is used to identify sets of similar descriptors as follows. For each descriptor $\bar{f_i}$, a set of the $K^{th}$ closest or most similar descriptors $NN_k(\bar{f_{i}})$ is identified as
\begin{equation}
    NN_{k}(\bar{f_{i}}) \, = \, \{\bar{f_j} \, : \, d(\bar{f_i}, \bar{f_j}) \leq d_{i,K}\}
    \label{eq:knn}
\end{equation}
where $d_{i,K} = d(\bar{f_i},\bar{f_K})$ is the distance between $\bar{f_i}$ and the $K^{th}$ closest descriptor $\bar{f_K}$. Enumerating $NN_{k}(\bar{f_{i}})$ for each $\bar{f_i}$ in a set of $N$ descriptors via na\"{i}ve pairwise comparisons incurs a computational cost of $O(N^2)$. However rapid approximate K-NN algorithms using efficient tree-based search structures can perform this enumeration in $O(N\,\log\,N)$ computational complexity. We use a set 8 randomized search trees as proposed in~\citep{Muja2014ScalableData}, where the descriptor search space is divided according to the descriptor elements exhibiting the highest variance. Parameter $K$ can be set generously to at least the number of expected matches in the dataset, here we used $K=30$. Note that a relatively large percentage of matches may be spurious or incorrect due to noise, however these tend to be distributed randomly between unrelated subjects and have negligible impact. With these parameters, matching features of one image to all other 7535 images (represented by approximately 16,000,000 features) requires approximately 0.35 seconds on an Intel Xeon Silver 4110@2.10Ghz, demonstrating the high computational efficiency of the method. Note that $O(N^2)$ operations are required to explicitly enumerate matches between all pairs in a set of $N$ images for a complete analysis (i.e. 28,381,485 unrelated subjects in Figure~\ref{fig:jaccard_distributions}), however small numbers of closely related subjects may identified in $O(N\,\log\,N)$ complexity.

\noindent
\textbf{Set Similarity Measurement} seeks to quantify the likelihood that two independent keypoint sets arise from the same underlying object. As our data consist of sets of discrete keypoints, potentially arising from equivalent neuroanatomic regions in different subject scans, we use the Jaccard overlap measure from set theory. The Jaccard overlap score $J(A,B)$ between a pair of feature sets $A$ and $B$ is defined as
\begin{equation}
    J(A,B) \, = \, \frac{|A \cap B|}{|A \cup B|} \, = \, \frac{|A \cap B|}{|A| + |B| - |A \cap B|}
    \label{eq:Jaccard}
\end{equation}
\noindent
where, in Equation~\eqref{eq:Jaccard}, $|A|$ and $|B|$ represent the cardinalities (sizes) of sets $A$ and $B$, and $|A \cap B|$ represents the cardinality of their intersection, i.e. the set of elements shared by $A$ and $B$. The Jaccard measure has several desirable properties, for example it ranges from [0,1] for disjoint to identical sets respectively, and can be transformed into a distance metric~\citep{Levandowsky1971DistanceSets} $1-J(A,B)$ in the abstract space of variable-sized keypoints sets.

The intersection in Equation~\eqref{eq:Jaccard} is defined as the subset of elements shared by both sets:
\begin{equation}
    A \cap B \, = \, \{\bar{f_i} \, : \, \bar{f_i} \in A \land \bar{f_i} \in B \}
    \label{eq:intersection}
\end{equation}
Evaluating the intersection requires identifying pairs of equivalent set elements, i.e. $\bar{f_i} \in A$ and $\bar{f_j} \in B$ such that $\bar{f_i} = \bar{f_j}$. These are defined by the union of all unique NN keypoint matches identified between $A$ and $B$. As set equivalence is binary, each unique keypoint match thus $(\bar{f_i},\bar{f_j})$ contributes 1 element to the intersection $A \cap B$, and the cardinality $|A \cap B|$ may be computed as the sum of matching elements~\citep{Kumar2018MultimodalFramework,Toews2016HowFeatures}. Binary equivalence of keypoint descriptors is difficult to justify however, as statistical variations in image content generally introduce uncertainty into keypoint descriptors. We thus consider $(\bar{f_i},\bar{f_j})$ as contributing a soft value ranging from $[0,1]$ to $|A \cap B|$, which is proportional to the likelihood of $(\bar{f_i},\bar{f_j})$ arising from a Gaussian density of variance $\alpha_i^2$ and mean $\bar{f_i}$. The cardinality of the set intersection is then evaluated as: 
\begin{equation}
    |A \cap B| = \sum_{\bar{f_{i}} \in A} \max_{\bar{f_{j}} \in NN_{k}(\bar{f_{i}})\\ \cap B}  \exp\{-d^{2}(\bar{f_{i}},\bar{f_{j})}/2\alpha_i^{2}\}
    \label{eq:cardinality}
\end{equation}
In Equation~\eqref{eq:cardinality}, variance parameter $\alpha_i^2$ is set automatically for each keypoint descriptor $\bar{f_{i}}$ as the squared distance to the closest NN keypoint within the entire training set. This allows the method to adjust to variable sampling density in the keypoint descriptor space about sample $\bar{f_{i}}$ in a manner similar to adaptive or variable kernel density estimation~\citep{Terrell1992VariableEstimation}, thereby downweighting the contribution of unlikely matches to the cardinality $|A \cap B|$. A related benefit is the reduced sensitivity to the parameter $K$ in the evaluation of $|A \cap B|$. Note that the cardinality for standard binary set equivalence is computed via Equation~\eqref{eq:cardinality} by taking $d(\bar{f_{i}},\bar{f_{j}}) = 0$ for all $\bar{f_{j}} \in NN_{k}(\bar{f_{i}})$. Given that the nearest neighbor relationship is not strictly symmetric, the cardinality as computed via Equation~\eqref{eq:cardinality} depends generally on the feature set over which sum is computed. In practice, we notice little difference in Jaccard values, and symmetry may be imposed.

\section{Results}
\label{sec3}
We performed experiments to quantify the variability of T1-weighted MRI data acquired from individuals and close relatives using keypoint signatures. A large set (N=7536) of T1w MRIs of 3334 unique subjects was pooled from four public neuroimaging datasets (HCP Q4, ADNI 1, OASIS 1 and OASIS 3), where each image pair bears a unique pair-wise relationship label: same subject (SM), monozygotic twin (MZ), dizygotic twin (DZ), non-twin full sibling (FS) or unrelated (UR) subjects. Relationship information for pairs of MR images were provided by individual datasets, while relationships between image pairs from different datasets were na\"{i}vely assumed to be unrelated. We evaluated the Jaccard overlap derived from the image content for all $N(N-1)/2=28,391,880$ possible image pairs, and studied the distributions of Jaccard measurements conditioned on relationship labels. Since the Jaccard overlap was derived from the proportion of features shared between images, we expected it to decrease with the degree of genetic and environmental separation in the pairwise relationship, i.e. decreasing in the order of SM, MZ, DZ, FS, UR pairs.

Figure~\ref{fig:method_workflow} illustrates the workflow for evaluating the Jaccard similarity score $J(A,B)$ between an image pair $(A,B)$. First, a one-time pre-processing step was applied to each image, where non-brain structure was removed using the Freesurfer software~\citep{Fischl2012FreeSurferFreeSurfer}, after which keypoint features were detected using the authors' publicly available software implementation. Detection required approximately 15 seconds and identified approximately 2000 keypoints per MRI. After pre-processing, nearest-neighbor (NN) keypoint matches were enumerated across all images, establishing putative equivalence between keypoints in different images, and the Jaccard similarity score was computed for each image pair based on the proportions of keypoint matches they shared.

\begin{figure}[ht]
	\centering
	\includegraphics[width=1\textwidth]{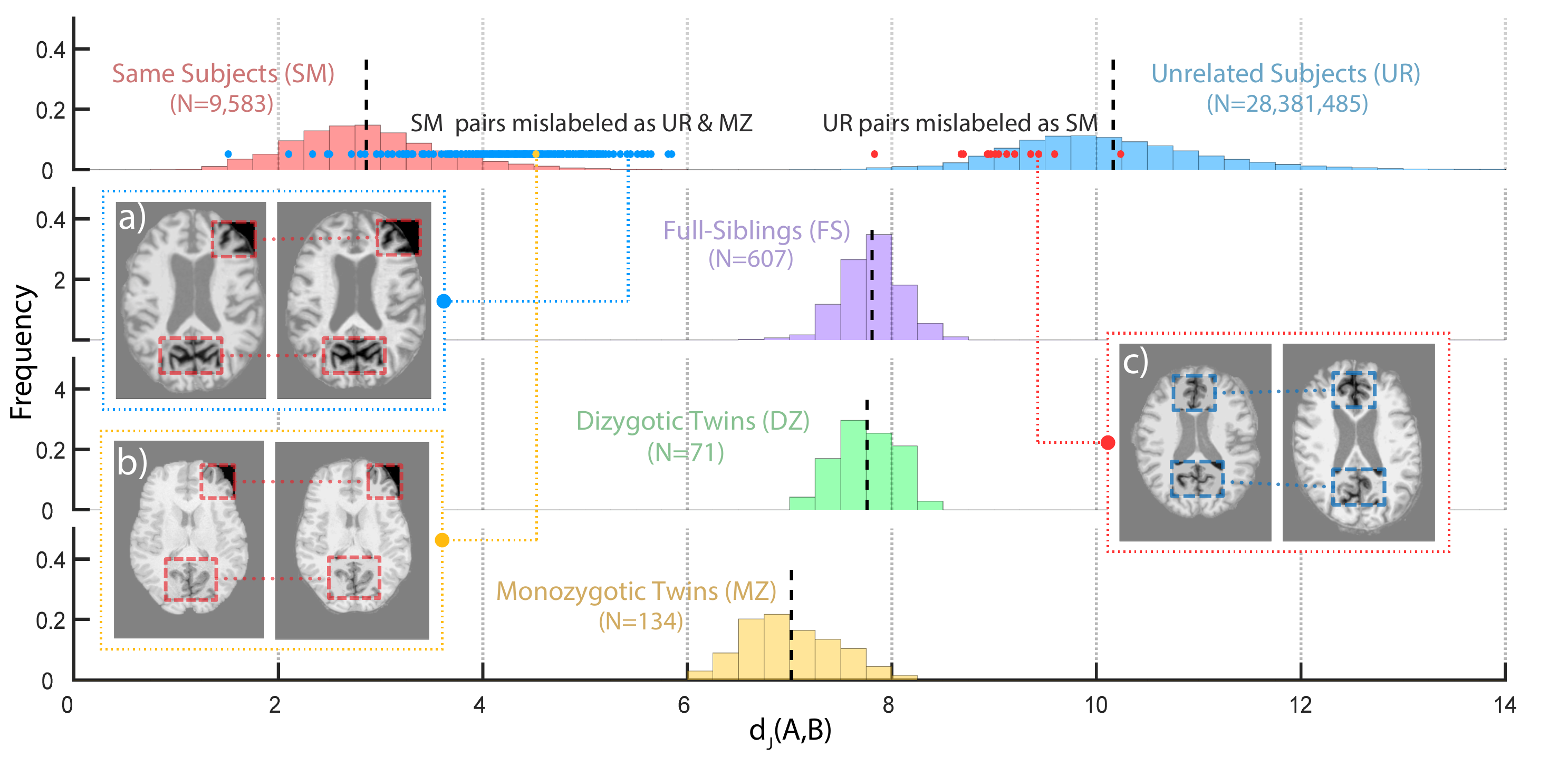}
	\caption{Distributions of the pairwise Jaccard distances conditional on relationship labels including: unrelated (UR, blue), same (SM, red), full siblings (FS, purple), dizygotic (DZ, green) and monozygotic (MZ, yellow) twin subject relationship labels, where black dashed lines indicate distribution means. Note the high degree of separation between SM and UR scores. Dots indicate data labeling inconsistencies automatically flagged by unexpected Jaccard distance. The correct relationship labels are evident upon visual inspection of cortical patterns (highlight), which are virtually identical in SM pairs mislabeled as (a) UR and (b) MZ, and highly different in a UR pair mislabelled as (c) SM
	}
\label{fig:jaccard_distributions}
\end{figure}

The Jaccard overlap $J(A,B)$ can be viewed as a whole-brain similarity measure ranging from $[0,1]$ for lowest to highest similarity. Equivalently, a monotonic transform such as the negative logarithm can be used to map $J(A,B)$ to a Jaccard distance measure $d_{J}(A,B) = -\log J(A,B)$ ranging from $[0,\infty]$. Figure~\ref{fig:jaccard_distributions} shows the empirical distributions of Jaccard distances, obtained for the five pair-wise relationships (indicated by color), where lower distance indicates a higher proportion of shared image content and neuroanatomic similarity. Distributions for each relationship label are unimodal and concentrated about a central tendency. The order of the mean distances for relationships (dashed vertical lines) is generally consistent with the degree of similarity in genetic and environmental developmental factors, i.e. increasing in order of SM, MZ, DZ/FS, UR. A number of outliers were identified (red and blue dots), and have been confirmed to arise from inconsistent image labels. These outliers will be discussed later.

The Jaccard distance distribution for same-subject (SM) pairs was highly unique, with no overlap with any other distribution including monozygotic twins (MZ). All SM pairs could thus be identified via a simple score threshold, supporting our hypothesis that keypoints capture highly unique aspects of individual neuroanatomy. Distributions for other relationships exhibit a degree of overlap, and the two-sample Kolmogorov-Smirnov test was used to evaluate the null hypothesis that samples arise from the same underlying distribution (Table~\ref{tab:kolmogorov-smirnov}). Most p-values were extremely low, allowing us to reject the null hypothesis with high confidence. The one exception was the case of FS and DZ pairs, where the p-value of 0.108 indicated no significant difference between Jaccard distance distributions for FS and DZ siblings.

\begin{table}[ht]
    \centering
    \captionsetup{width=0.8\textwidth}
    \caption{p-values for two-sample Kolmogorov-Smirnov tests between Jaccard distance distributions for (SM) Same Subject, (MZ) Monozygotic, (DZ) Dizygotic, (FS) Full-Sibling and (UR) Unrelated pairwise relationships}
    \label{tab:kolmogorov-smirnov}
    \setlength\extrarowheight{3pt}
        \begin{tabular}{|c|ccccc|}
            \hline
            & \textbf{SM} & \textbf{MZ} & \textbf{DZ} & \textbf{FS} & \textbf{UR}\\
            \hline
            \textbf{SM} & - & $\num{1.38e-233}$ & $\num{1.55e-125}$ & $0$ & $0$\\
            \textbf{MZ} & - & - & $\num{1.40e-41}$ & $\num{6.09e-97}$ & $\num{1.18e-228}$\\
            \textbf{DZ} & - & - & - & $\bm{0.108}$ & $\num{2.87e-120}$\\
            \textbf{FS} & - & - & - & - & $0$\\
            \textbf{UR} & - & - & - & - & -\\
            \hline
      \end{tabular}
\end{table}

\begin{figure}[ht]
    \centering
	\includegraphics[width=1\textwidth]{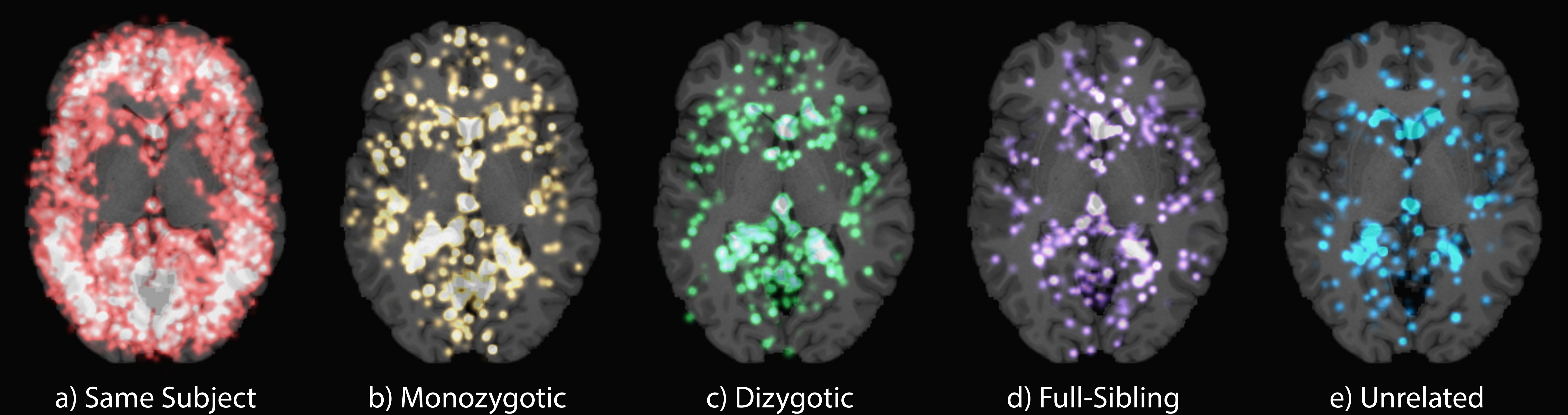}
	\caption{The spatial layout of keypoint matches for five pairwise relationship labels: a) SM, b) MZ, c) DZ, d) FS, e) UR. Heatmaps represent distributions of matching keypoints within the standard MNI305 neuroanatomic reference space and accumulated over 71 image pairs per label from HCP dataset}
	\label{fig:matches_visualization}
\end{figure}

\begin{figure}[ht]
	\centering
	\includegraphics[width=1\textwidth]{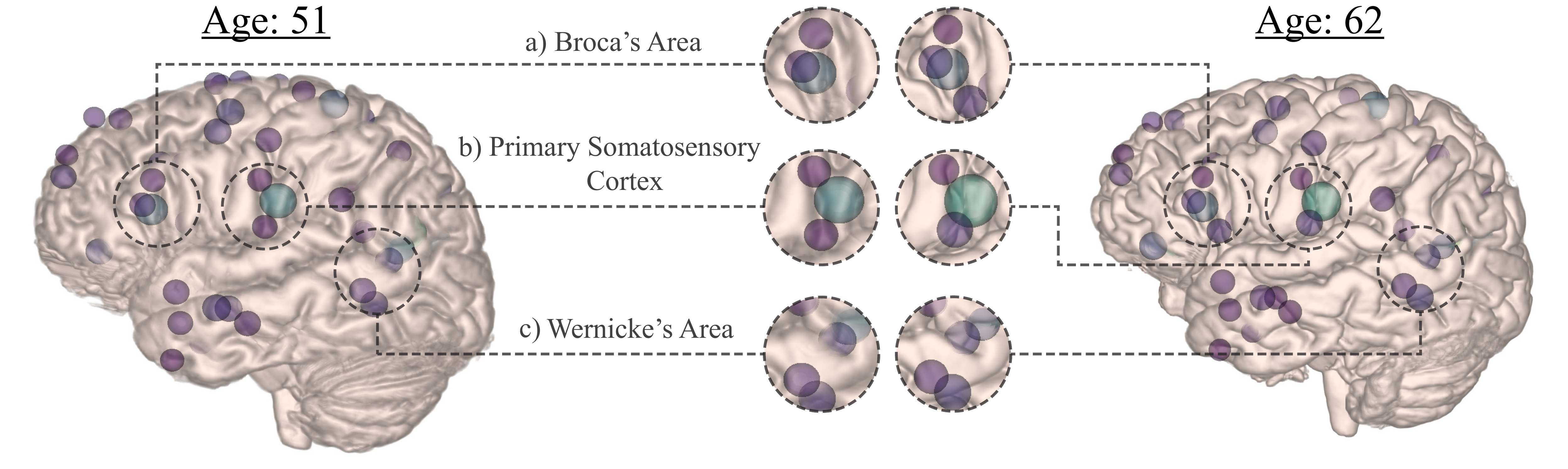}
	\caption{Keypoints matched between MRIs of the same individual acquired at 51 and 62 years of age. Matching keypoints (spheres, color indicating scale) represent patterns of local cortical structure that are highly unique to this individual, with notable concentrations in a) Broca's area, b) the primary somatosensory cortex and c) Wernicke's area. Note that keypoints have been slightly extruded from within the cortex for improved visualization. The visualization was generated using the 3D Slicer software~\citep{Fedorov20123DNetwork}}
	\label{fig:visualization_matching_keypoints}
\end{figure}

The Jaccard overlap is derived from image-to-image keypoint matches between unique neuroanatomic patterns identified in different images. Figure~\ref{fig:matches_visualization} illustrates the spatial distributions of matching keypoints as heatmaps within a common reference space for each relationship label. Keypoints were distributed similarly throughout the brain for all relationships, and generally concentrated in regions with significant intensity contrast variations, i.e. the interfaces between cortical sulci or sub-cortical structures and cerebral spinal fluid (CSF). The primary quantitative difference between relationship groups was the number of keypoint matches, which was much higher for same subject images, thus reflecting a higher degree of shared anatomic structure. Figure~\ref{fig:visualization_matching_keypoints} shows an example of keypoints matching between MRIs of the same individual acquired 11 years apart.

\begin{figure}[ht]
	\centering
	\includegraphics[width=1\textwidth]{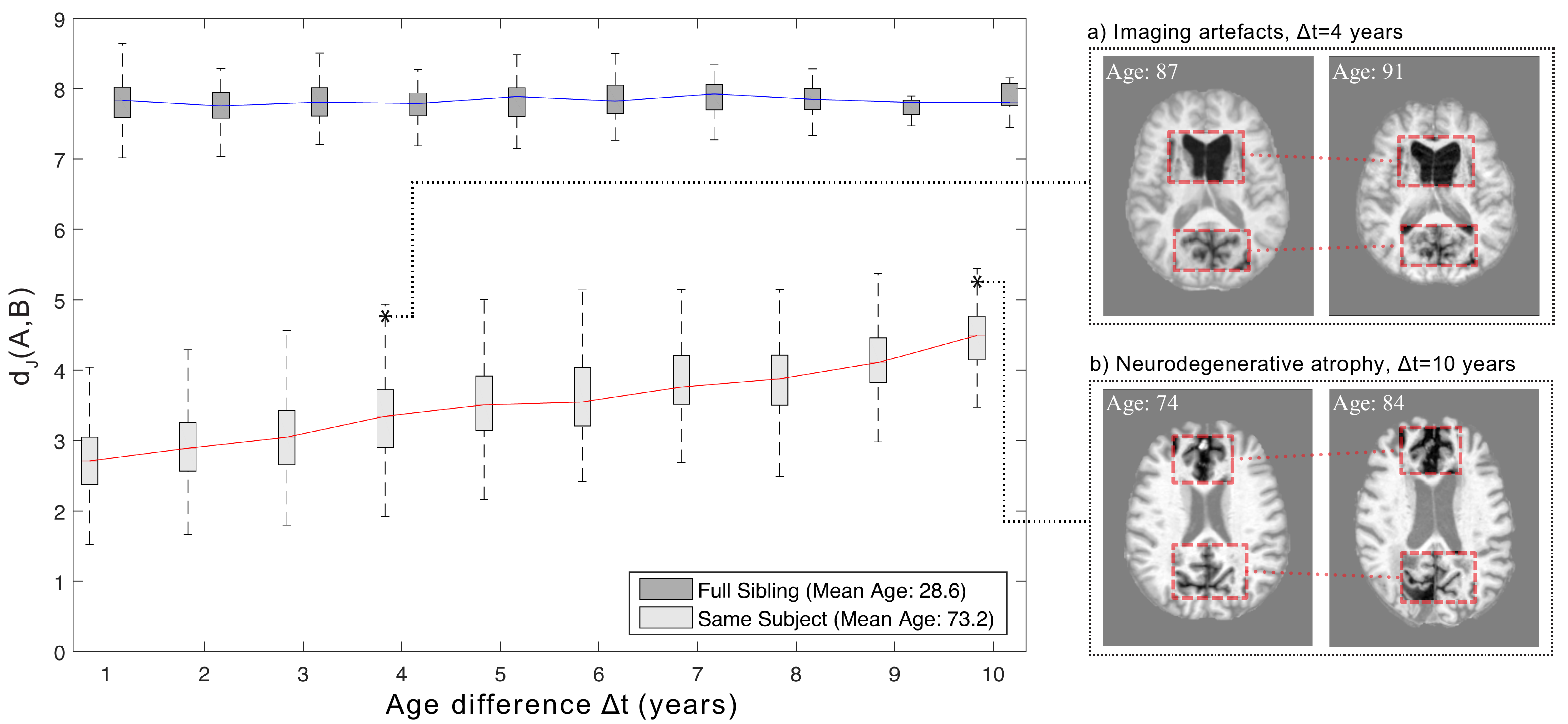}
	\captionsetup{width=\textwidth}
	\caption{Jaccard distance as a function of the age difference $\Delta t$ between scans for SM (N=9885) and FS (N=607) pairs. Lines and boxes represent the means and standard deviations of the Jaccard distance. a) and b) show examples of image pairs successfully identified as SM, despite a relatively high Jaccard distance due to aging, noticeable image artefacts in a) and neurodegenerative atrophy in b)}
	\label{fig:jaccard_over_time}
\end{figure}

The primary systematic confound was the age difference between image acquisitions, which was positively correlated with Jaccard distance as shown in Figure~\ref{fig:jaccard_over_time}. This likely reflected changes in brain morphology due to both natural aging and disease progression, primarily in SM pairs of older adults from the ADNI and OASIS datasets designed to study Alzheimer's disease (mean age 73 years). Data was unavailable to investigate the impact of age difference for younger SM subjects, however age separation between FS pairs (mean age 28 years) from the HCP dataset had no noticeable impact on the Jaccard distance, likely reflecting the relatively stable brain anatomy across the younger, healthy HCP cohort. By inspection, the highest Jaccard distances for SM pairs were typically associated with random confounds including image artifacts (e.g. due to MRI acquisition, pre-processing, etc.) or morphological changes in the brain (e.g. due to natural aging, neurodegenerative disease, etc.), see Figure~\ref{fig:jaccard_over_time} a) and b). \\
Other confounds including sex, race and age had no significant impact on Jaccard distances, and invariant keypoint matching is independent of the image (mis)alignment, an important benefit of our method.

A surprising result was the discovery of 184 outlier image pairs with Jaccard distances that were noticeably outside of the distributions associated with their relationship labels. These included pairs labeled as UR or MZ with Jaccard distances similar to same subjects (Figure~\ref{fig:jaccard_distributions}, blue dots), and pairs labeled as SM with Jaccard distances similar to those for unrelated subjects (Figure~\ref{fig:jaccard_distributions}, red dots). Upon visual inspection of the cortical folding patterns (see examples in Figure~\ref{fig:jaccard_distributions}), and with the help of the respective database administrators, we established that all were most likely labeled incorrectly. All individual datasets contained at least one case where images of the same subject were labelled as different subjects. Instances of the same subject were identified across the OASIS and ADNI datasets. Table~\ref{tab:duplicates_subjects} lists the numbers of subjects with inconsistent labels within and between datasets. Additionally, 11 cases of perfectly identical images were identified by unusually low Jaccard distance, these were labeled as images of the same subject acquired at different time points.
% Removed in preprint * These results have been communicated to the respective database administrators.

 An important algorithmic parameter is the number of NN descriptor matches $K$ used in estimating the Jaccard score. In the case of hard set equivalence~\citep{Toews2016HowFeatures,Kumar2018MultimodalFramework}, each keypoint match contributes 1 to the set union $|A\cap B|$, and an optimal $K$ depends on the number of relevant image samples, i.e. MRIs of the same subject or group, which is generally unknown and variable. In the case of our soft weighting approach however, each match contributes a weight proportional to a Gaussian density, and thus $K$ may be set large enough to include all relevant image samples.
 
\begin{table}[ht]
    \centering
    \captionsetup{width=0.8\textwidth}
    \caption{\normalsize Mislabeled subject relationships identified across and within databases. Most cases are subjects mislabeled as UR\\ \textsuperscript{*} A pair labeled as MZ was established to be SM and one dataset was subsequently removed from the HCP dataset, and a pair of scans labeled as UR were confirmed to be SM from the OASIS 1 dataset\\ \textsuperscript{**} OASIS 3 and OASIS 1 are known to share data from numerous subjects\\ \textsuperscript{***} OASIS 3 contained 2 pairs of SM subjects mislabeled as UR, and 3 pairs of UR subjects labeled as SM}
    \label{tab:duplicates_subjects}
    \setlength\extrarowheight{1.85pt}
    \setlength{\tabcolsep}{15pt}
        \begin{tabular}{|c|cccc|}
            \hline
            & \textbf{HCP Q4} & \textbf{ADNI 1} & \textbf{OASIS 1} & \textbf{OASIS 3} \\
            \hline
            \textbf{HCP Q4} & 1\textsuperscript{*} & 0 & 0 & 0 \\
            \textbf{ADNI 1} & - & 3 & 4 & 2 \\
            \textbf{OASIS 1} & - & - & 1\textsuperscript{*} & 79\textsuperscript{**} \\
            \textbf{OASIS 3} & - & - & - & 2 + 3\textsuperscript{***} \\
            \hline
      \end{tabular}
\end{table}  

\section{Discussion}
\label{sec4}

In this paper, we proposed to model neuroanatomy as a collection of distinctive image keypoints, hypothesizing that this would more accurately preserve aspects of anatomy unique to individuals or close family members, distinctive neuroanatomic signatures that might otherwise be averaged out by traditional parcellation or voxel-wise representations.\\
The whole-brain similarity of an image pair was assessed in terms of the proportion of keypoints they share using the Jaccard measure of set overlap, which can be computed from arbitrarily large datasets using a highly efficient keypoint matching procedure.

Experiments validated our hypothesis in the largest study of individual identification from MRI data to date, involving 7536 T1w MRIs of 3334 unique subjects pooled from four large, public neuroimaging datasets: ADNI, OASIS1, OASIS3 and HCP. Distributions of Jaccard distances for same vs. unrelated subject MRI pairs are separated by a wide margin, and a simple threshold on the Jaccard distance was sufficient to identify all same-subject pairs with 100\% accuracy\footnote{For completeness, there is the possibility of coincidental errors in both metadata labeling and automatic identification, however the probability of this is very low, given these events are unrelated, independent and individually highly improbable.}. In contrast, the largest previous study involved almost 700 subjects from the ADNI dataset alone, required multiple scans per subject as training data, and achieved less than perfect accuracy using features derived from a standard neuroanatomic segmentation~\citep{Wachinger2015BrainPrintMorphology}.

An important potential confound is within-dataset scan similarity; same-subject scans are typically found within the same dataset, whereas scans from the same dataset are known to generally exhibit similarity due to a number of commonalities including subject demographics, age, site, scan sequence, scanner artifacts, etc~\citep{Wachinger2019QuantifyingInference}. As expected, Jaccard distance distributions were generally lower within-dataset vs. across-datasets: for SM pairs $(2.12 \pm 0.65~\text{vs.}~4.46 \pm 0.66)$ and for UR pairs $(8.99 \pm 0.76~\text{vs.}~10.01 \pm 0.80)$, indicating a within-dataset similarity effect. However there was no overlap between distance distributions for SM and UR pairs, and we thus expect these to remain separable by a Jaccard distance threshold in new T1-weighted MRI data acquired and pre-processed (e.g. skull-stripped) with protocols similar to those used for the three datasets used here. Furthermore, the pattern of decreasing similarity in the order of SM, MZ, (DZ/FS) to UR pairs did not change when analysis was restricted to single datasets.

A surprising result was the discovery of previously unknown subject labeling inconsistencies, identified as clear outliers from their expected Jaccard distributions. These included MRIs of the same person labeled as unrelated, or MRIs of different people labelled as the same, and were identifed both within and across the datasets. Such inconsistencies may lead to bias in cross-validation studies, e.g. computer-assisted prediction~\citep{Desikan2009AutomatedDisease,Samper-Gonzalez2018ReproducibleData} where protocols assume independent training and testing data, and similar errors in a clinical context could potentially lead to errors in patient care. The ability to identify these in widely used, public datasets is noteworthy, and demonstrates the potential for the keypoint signature as a powerful tool in curating and validating large neuroimage datasets. % File IDs associated with labeling inconsistencies are provided in the Supplementary Tables document.

The Jaccard measure is driven by keypoint matches representing instances of unique neuroanatomic patterns shared between pairs of images of individual or siblings. The Jaccard overlap generally decreased with the degree of genetic and developmental separation in the pairwise relationship label, i.e. in the order of SM, MZ, DZ/FS, UR pairs, indicating decreasing proportions of unique anatomic structure shared by these groups as predicted. A notable exception was the case of dizygotic twin (DZ) and non-twin full sibling (FS) relationships, which showed no statistically significant difference in terms of their pairwise Jaccard distributions. Keypoint matches were generally distributed throughout the brain and across interfaces between adjacent tissues exhibiting high intensity contrast in MRI, e.g. cerebrospinal fluid, grey and white matter, in a manner unique to the specific image pair, rather than within regular loci defined by typical parcellation schemes. Combined with the high accuracy of identification experiments, this suggests that aspects of neuroanatomy most characteristic of individuals or close relatives may be highly idiosyncratic and not ultimately be tied to a fixed parcellation. Keypoint signatures provide a robust and efficient means of exploring these aspects across large datasets, and a combination of keypoint signatures and traditional segmentations or parcellations may ultimately prove most effective in understanding the variability of individuals and genetic families.

In terms of technology, the keypoint signature affords the capability of rapidly comparing a new image against a large dataset, e.g. identifying all keypoint sets with high Jaccard similarity in 0.35 seconds here. This is a memory-based learning approach, which requires no explicit training procedure. New data are easily incorporated, and it is limited only by the amount of memory available, a limit that is continually reduced by technological advancement. In contrast, representations based on traditional neuroanatomic parcellations generally require extensive pre-processing, including image registration and segmentation, and alternative machine learning approaches require training procedures with multiple MRIs per subject~\citep{Wachinger2015BrainPrintMorphology} which may be unavailable a priori. Keypoint detection here is based on a recursive Gaussian filtering process that is analogous to a highly efficient, unbiased convolutional neural network (CNN) used in deep learning~\citep{LeCun2015DeepLearning}. Machine learning could potentially be used to optimize filters, however this would require training procedures and data that might not be readily available (i.e. multiple labeled images of an individual), and would introduce bias related to a particular training set.

Our experiments here derived keypoint signatures from the ubiquitous T1w structural MRI modality, however keypoint detection can be performed from arbitrary scalar image modalities, e.g. fractional anisotropy derived from diffusion MRI (dMRI)~\citep{Kumar2018MultimodalFramework} or statistical parameter maps derived from fMRI data, and descriptors can be used to encode vector-valued data, e.g. histograms of diffusion gradient orientations in dMRI~\citep{Chauvin2018DiffusionAnalysis}. Keypoint matching across different modalities is generally non-trivial and an avenue for future investigation~\citep{Toews2013EfficientFeatures}. Our analysis focused on neuroanatomy and automatic skull stripping was used to remove image content associated with non-brain tissues. Extensive pre-processing is not generally required, as keypoints can be reliably detected despite variations in intensity or pose, and can be used to analyze non-brain image content. In fact, we found the Jaccard similarity between related subjects to be higher when non-brain structure is included. Nevertheless, processes for normalizing or correcting intensity values, e.g. correction of MRI inhomeneities using field maps, should generally improve the repeatability of keypoint detection, and the question of an optimal image pre-processing pipeline is left for future research. Experiments here were limited to sibling relationships and adult subjects ranging of 18-96 years of age, future investigations will consider younger age groups such as infants with the additional confound of rapid neurodevelopment and other relationships including parent-child or cousins with varying amounts of shared genetics. Finally, our work here does not investigate links between anatomical keypoints and subject abilities or behaviors. However, the keypoint representation has previously been used to interpret single anatomical scans according to group-wise clinical symptoms or labels, e.g. Alzheimer's disease classification~\citep{Toews2010FeaturebasedPatterns} and neonatal age prediction~\citep{Toews2012FeaturebasedMRI}, and similar keypoint analysis techniques could potentially be applied to other modalities including functional MRI data in future investigations.

\subsection{Code and Data Availability}
Complete C++ source code for the method is provided by the authors at \url{https://github.com/3dsift-rank/3DSIFT-Rank}. Keypoint data are available at \url{https://central.xnat.org/data/projects/SIFTFeatures}.

\section{Acknowledgements}
\noindent
\textbf{OASIS}. Data were provided in part by OASIS: Cross-Sectional: Principal Investigators: D. Marcus, R, Buckner, J, Csernansky J. Morris; P50 AG05681, P01 AG03991, P01 AG026276, R01 AG021910, P20 MH071616, U24 RR021382 and OASIS-3: Principal Investigators: T. Benzinger, D. Marcus, J. Morris; NIH P50AG00561, P30NS09857781, P01AG026276, P01AG003991, R01AG043434, UL1TR000448, R01EB009352. AV-45 doses were provided by Avid Radiopharmaceuticals, a wholly owned subsidiary of Eli Lilly.

\noindent
\textbf{HCP}. Data were provided in part by the Human Connectome Project, WU-Minn Consortium (Principal Investigators: David Van Essen and Kamil Ugurbil; 1U54MH091657) funded by the 16 NIH Institutes and Centers that support the NIH Blueprint for Neuroscience Research; and by the McDonnell Center for Systems Neuroscience at Washington University.

\noindent
\textbf{ADNI}. Data collection and sharing for this project was funded by the Alzheimer's Disease Neuroimaging Initiative (ADNI) (National Institutes of Health Grant U01 AG024904) and DOD ADNI (Department of Defense award number W81XWH-12-2-0012). ADNI is funded by the National Institute on Aging, the National Institute of Biomedical Imaging and Bioengineering, and through generous contributions from the following: AbbVie, Alzheimer's Association; Alzheimer's Drug Discovery Foundation; Araclon Biotech; BioClinica, Inc.; Biogen; Bristol-Myers Squibb Company; CereSpir, Inc.; Cogstate; Eisai Inc.; Elan Pharmaceuticals, Inc.; Eli Lilly and Company; EuroImmun; F. Hoffmann-La Roche Ltd and its affiliated company Genentech, Inc.; Fujirebio; GE Healthcare; IXICO Ltd.; Janssen Alzheimer Immunotherapy Research \& Development, LLC.; Johnson \& Johnson Pharmaceutical Research \& Development LLC.; Lumosity; Lundbeck; Merck \& Co., Inc.; Meso Scale Diagnostics, LLC.; NeuroRx Research; Neurotrack Technologies; Novartis Pharmaceuticals Corporation; Pfizer Inc.; Piramal Imaging; Servier; Takeda Pharmaceutical Company; and Transition Therapeutics. The Canadian Institutes of Health Research is providing funds to support ADNI clinical sites in Canada. Private sector contributions are facilitated by the Foundation for the National Institutes of Health (www.fnih.org). The grantee organization is the Northern California Institute for Research and Education, and the study is coordinated by the Alzheimer's Therapeutic Research Institute at the University of Southern California. ADNI data are disseminated by the Laboratory for Neuro Imaging at the University of Southern California.

\noindent
This work was supported by NIH grant P41EB015902, a Canadian National Sciences and Research Council (NSERC) Discovery Grant and the Canada Research Chair in 3D Imaging and Biomedical Engineering.

\chapter{Efficient Pairwise Neuroimage Analysis using the Soft Jaccard Index and 3D Keypoint Sets}
\label{chap4}
\chaptermark{Efficient Pairwise Neuroimage Analysis}

\articleAuthors{
{Laurent Chauvin\textsuperscript{a}}{Kuldeep Kumar\textsuperscript{b}}{Christian Desrosiers\textsuperscript{a}}{William Wells III\textsuperscript{c,d}}{and Matthew Toews\textsuperscript{a}}
}{
{\setstretch{1.2}
\textsuperscript{a} Département de Génie des Systemes, École de Technologie Supérieure, Canada%,\\
%1100 Notre-Dame Ouest, Montréal, Québec, Canada H3C 1K3
\\
\textsuperscript{b} CHU Sainte-Justine Research Center, University of Montreal, Canada%,\\ %3175 Chemin de la Côte-Sainte-Catherine, Montreal, Québec, Canada H3T 1C5
\\
\textsuperscript{c} Brigham and Women's Hospital, Harvard Medical School, USA%, \\
%25 Shattuck Street, Boston, MA 02115, USA
\\
\textsuperscript{d} Computer Science and Artificial Intelligence Lab, Massachusetts Institute of Technology, USA%, \\
%77 Massachusetts Avenue, Cambridge, MA 02139, USA
\\~\\
Paper published in \textit{IEEE Transactions on Medical Imaging}, September 2021}
}
\\
\textbf{Abstract}\\
We propose a novel pairwise distance measure between image keypoint sets, for the purpose of large-scale medical image indexing. Our measure generalizes the Jaccard index to account for soft set equivalence (SSE) between keypoint elements, via an adaptive kernel framework modeling uncertainty in keypoint appearance and geometry. A new kernel is proposed to quantify the variability of keypoint geometry in location and scale. Our distance measure may be estimated between $O(N^2)$ image pairs in $O(N~\log~N)$ operations via keypoint indexing. Experiments report the first results for the task of predicting family relationships from medical images, using 1010 T1-weighted MRI brain volumes of 434 families including monozygotic and dizygotic twins, siblings and half-siblings sharing 100\%-25\% of their polymorphic genes. Soft set equivalence and the keypoint geometry kernel improve upon standard hard set equivalence (HSE) and appearance kernels alone in predicting family relationships. Monozygotic twin identification is near 100\%, and three subjects with uncertain genotyping are automatically paired with their self-reported families, the first reported practical application of image-based family identification. Our distance measure can also be used to predict group categories, sex is predicted with an AUC=0.97. Software is provided for efficient fine-grained curation of large, generic image datasets.

\section{Introduction}
\label{sec:introduction}
Health treatment is increasingly personalized, where treatment decisions are conditioned on patient-specific information in addition to knowledge regarding the general population~\citep{Hamburg2010PathMedicine}. Modern genetic testing allows us, based on large libraries of human DNA samples, to cheaply predict patient-specific characteristics, including immediate family relationships, and also characteristics shared across the population including racial ancestry, sex, hereditary disease status, etc.~\citep{Annas201423andMeFDA,Auton2015GlobalVariation}. 
The brain is the center of cognition and a complex organ, tightly coupled to the genetic evolution of animals and in particular that of the human species. To what degree is the human brain phenotype coupled to the underlying genotype? How does the brain image manifold vary locally with genotype, i.e. immediate relatives sharing 25-100\% of their polymorphic genes, or between broad groups defined by subtle genetic factors such as racial ancestry or sex?

Large, publicly available databases allow us to investigate these questions from aggregated MRI and genetic data, e.g. 1000+ subjects~\citep{VanEssen2012HumanPerspective}. The shape of the brain has been modeled as lying on a smooth manifold in high dimensional MRI data space~\citep{Gerber2010ManifoldAnalysis,Brosch2013ManifoldMRIs}, where phenotype can be described as a smooth deformation conditioned on developmental factors including the environment. However the brain is also naturally described as a rich collection of spatially localized neuroanatomical structures, including common structures such as the basal ganglia shared across the population, but also highly specific patterns such as cortical folds that may only be observable in specific individuals or close family members~\citep{Ono1990AtlasSulci}.

\begin{figure}[!h]
  \centering
   \includegraphics[width=0.75\linewidth,draft=false]{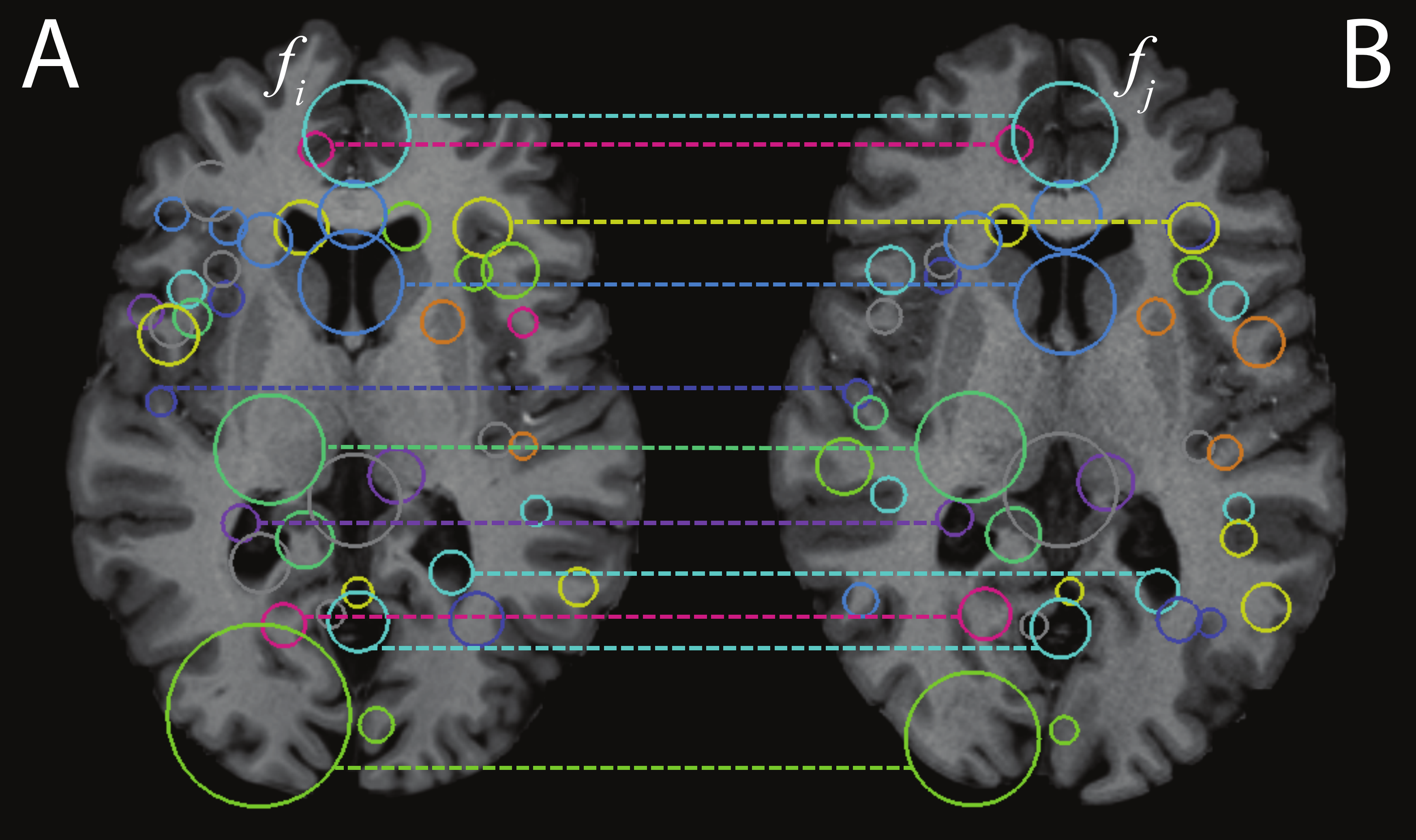}
   \captionsetup{width=0.75\textwidth}
   \caption{Illustrating two sets of 3D keypoints $A=\{f_i\}$ and $B=\{f_j\}$ extracted from MRI brain volumes of monozygotic twins. Colored circles represent the 3D locations $\bar{x} \in R^3$ and scales (radii) $\sigma \in R^1$ of keypoints in 2D axial slices, and links are correspondences identified via the SIFT-Rank~\citep{Toews2009SIFTRankCorrespondence} algorithm, note the degree of similarity between keypoint pairs}
  \label{fig:sift_visualization}
\end{figure}

The keypoint representation is an intuitively appealing means of modeling specific, localized phenomena, i.e. a set of descriptors automatically identified at salient image locations as shown in Figure~\ref{fig:sift_visualization}. A keypoint set can be viewed as an element of a high-dimensional manifold, and medical imaging  applications such as regression or classification can be formulated in terms of a suitable geodesic distance between sets. As keypoint sets are variable sized, typical metrics based on fixed-length vectors such as L-norms~\citep{Gerber2010ManifoldAnalysis,Brosch2013ManifoldMRIs} do not readily apply. Distances defined based on set overlap measures as the Jaccard index~\citep{Levandowsky1971DistanceSets} have proven to be effective in recent studies investigating genetics and brain MRI~\citep{Toews2016HowFeatures,Chauvin2020NeuroimageRelatives}. For example, by defining set equivalence in terms of nearest neighbor (NN) keypoint correspondences, the Jaccard distance may be used to predict pairwise relationships. Nevertheless, the assumption of binary or hard set equivalence (HSE) between keypoints is a crude approximation given probabilistic uncertainty inherent to natural image structure, and is ill-defined for variable sized datasets where the number of NN correspondences may be variable or unknown a-priori.

Our primary contribution is a new pairwise distance measure for 3D keypoint sets, generalizing the Jaccard index to account for soft set equivalence (SSE) between keypoint elements. SSE is achieved via a kernel framework for keypoint similarity, and we introduce a new kernel that normalizes pairwise keypoint displacement by keypoint scale, accounting for localization uncertainty in scale-space. Our work extends and compares against keypoint-based neuroimaging analysis methods~\citep{Toews2016HowFeatures,Kumar2018MultimodalFramework,Chauvin2020NeuroimageRelatives}, specifically the keypoint signatures approach~\citep{Chauvin2020NeuroimageRelatives}, currently the only method reporting perfect accuracy at identifying scans of the same individual in large public neuroimage MRI datasets, e.g. OASIS~\citep{Marcus2007OpenAdults}, ADNI~\citep{JackJr2008AlzheimerMethods} and HCP~\citep{VanEssen2013WUMinnOverview}. Here, following preliminary work~\citep{Chauvin2019AnalyzingManifold}, we demonstrate that the soft Jaccard distance is specific enough to predict family relationships from medical images, and we present the first results for this new task using 1010 MRIs of 434 families from the HCP~\citep{VanEssen2013WUMinnOverview} dataset. The entire method is based on highly efficient keypoint indexing, and scales to large datasets of volumetric image data.

% and instances of previously unknown subject labeling consistencies in widely used, public neuroimage MRI datasets (e.g. OASIS~\citep{marcus2007open}, ADNI~\citep{jack2008alzheimer}, HCP~\citep{van2013wu}). 

%This does not belong in the intro: It should be noted that the visualization of the results differ between~\citep{Chauvin2019AnalyzingManifold} and this work, for a better visualuation of outliers.

\section{Related Work}

Our work is motivated both by the study of the link between genotype and the human brain phenotype from large datasets~\citep{Sabuncu2016MorphometricityTrait}, and by practical applications such as maintaining accurate patient records in hospital Picture Archiving and Communication Systems (PACS).
We adopted a memory-based learning model where all data are stored in memory and accessed via highly-specific keypoint queries. Memory-based learning requires no explicit training procedure~\citep{Boiman2008DefenseClassification}, and approaches Bayes optimal performance as the number of data $N$ becomes large~\citep{Cover1967NearestClassification}. 3D SIFT keypoints~\citep{Toews2013EfficientFeatures} derived from Gaussian scale-space theory~\citep{Lowe2004DistinctiveKeypoints,Lindeberg1994ScalespaceScales} are invariant to global similarity transforms and contrast variations, and thus account for misalignment and scanner variability between images. Efficient indexing reduces the quadratic $O(N^2)$ complexity of nearest neighbor (NN) keypoint lookup to $O(N~\log~N)$, e.g. via $O(\log N)$ KD-tree indexing~\citep{Muja2014ScalableData}, ensuring that the method scales gracefully to large datasets, e.g. 7500 brain MRIs~\citep{Chauvin2020NeuroimageRelatives} or 20000 lung CTs~\citep{Toews2015FeatureBasedImages}.

Our method seeks a pairwise distance measure between 3D keypoint sets. As different images generally contain different numbers of elements, a natural choice for pairwise distance between variable sized sets are measures based on set intersection or overlap, e.g. the Jaccard distance metric~\citep{Levandowsky1971DistanceSets} first proposed in~\citep{Marczewski1958CertainFunctions} or the Tanimoto distance~\citep{Bajusz2015WhyCalculations,Rogers1960ComputerPlants}. In medical image analysis, the Jaccard distance is typically used to assess pixel-level segmentation accuracy~\citep{Yuan2017AutomaticDistance}, and has proven highly effective for assessing pairwise neuroimage similarity from 3D SIFT keypoint data~\citep{Toews2016HowFeatures,Kumar2018MultimodalFramework}, where hard set equivalence is determined by NN keypoint descriptor correspondences.
Set theory is predicated on binary equivalence between set elements, which is difficult to justify in the case of keypoints extracted from image data. Soft set theory was been developed to account for non-binary equivalence~\citep{Molodtsov1999SoftResults,Park2012PropertiesRelations,Gardner2014MeasuringSizes}, where set operations including intersection and union are defined in terms of non-binary equivalence ranging from [0,1]. Jaccard-like distance was used for retrieving near-duplicate 2D photos~\citep{Chum2008DuplicateWeighting}, where the ratio of soft intersection and union operators was used to account for  document frequency. Our work here extends the Jaccard distance proposed in~\citep{Chauvin2020NeuroimageRelatives} to include keypoint geometry in 3D Euclidean metric space coordinates, and compares to the HSE Jaccard measure used in~\citep{Toews2016HowFeatures,Kumar2018MultimodalFramework}, showing a significant improvement in predicting family relationships.

% Why use SIFT / Local descriptors
The keypoint methodology has a number of advantages, such as robustness to occlusion or missing correspondences, invariance to translation, scaling, rotation and intensity contrast variations between images. Our work adopts the 3D SIFT-Rank~\citep{Toews2013EfficientFeatures} representation, a robust, general tool used for a variety of imaging tasks, including registration~\citep{Toews2013EfficientFeatures,Machado2018NonrigidMatching}, whole body segmentation~\citep{Wachinger2018KeypointSegmentation}, classification~\citep{Toews2015FeatureBasedImages,Toews2010FeaturebasedPatterns}, regression~\citep{Toews2012FeaturebasedMRI}, without the need for context-specific training procedures or data. As in the original 2D SIFT approach, keypoint geometry is represented as a location $\bar{x}$ and scale $\sigma$, these are detected as extrema of a difference-of-Gaussian scale-space~\citep{Lowe2004DistinctiveKeypoints}, as $\argmax_{\bar{x},\sigma} \ |I(\bar{x},\sigma) \, - \, I(\bar{x},\kappa\sigma)|$, approximating the Laplacian-of-Gaussian~\citep{Lindeberg1998FeatureSelection},  where $I(\bar{x},\sigma)=I(\bar{x})*\mathcal{N}(\bar{x},\sigma^2)$ represents the image convolved with the Gaussian kernel $\mathcal{N}(\bar{x},\sigma^2)$. The geometry of a keypoint may thus be characterised as an isotropic Gaussian density centered on $\bar{x}$ with variance $\sigma^2$ representing spatial extent $\sigma$ in 3D Euclidean space. Rotationally symmetric Gaussian filtering and uniformly sampled derivative operators lead to scale and rotation invariance. The 3D SIFT-Rank descriptor is a 64-element histogram of rank-ordered oriented gradients (HOG)~\citep{Toews2013EfficientFeatures}, sampled about a scale-normalized reference frame centered on $\bar{x}$ and quantized uniformly into 2x2x2=8 spatial locations x 8 orientations. Rank transformation provides invariance to arbitrary monotonic variations of descriptor element values~\citep{Toews2009SIFTRankCorrespondence}.

The 3D SIFT-Rank keypoint method we use is a robust and well-established baseline, however our distance measure may generally be used with any keypoint representation. The search for new keypoint detectors and descriptors remains an active research focus, primarily in the context of 2D computer vision. We refer the reader to an extensive literature review of traditional hand-crafted solutions~\citep{Mukherjee2015ComparativeDescriptors} and a recent discussion of 2D SIFT matching technology~\citep{Bellavia2020ThereMatching}. Deep learning keypoint methods have been investigated, examples include LIFT~\citep{Yi2016LIFTTransform}, DISK~\citep{Tyszkiewicz2020DISKGradient}, LF-Net~\citep{Ono2018LFNetImages}, SuperPoint~\citep{Detone2018SuperPointDescription}. Nevertheless, variants of the original SIFT histogram descriptor, including SIFT-Rank~\citep{Toews2009SIFTRankCorrespondence}, DSP-SIFT~\citep{Dalal2005HistogramsDetection} or RootSIFT~\citep{Arandjelovic2012ThreeRetrieval} remain competitive with descriptors derived from training~\citep{Balntas2017HPatchesDescriptors, Schonberger2017ComparativeFeatures}, particularly for non-planar objects~\citep{Bellavia2020ThereMatching}, while requiring no training and few hyperparameters. Learning-based approaches often rely on hand-crafted keypoint detectors to generate training data~\citep{Yi2016LIFTTransform,Detone2018SuperPointDescription}. The challenges of developing deep keypoint architectures include training of individual components (i.e. keypoint detector, orientation estimator, descriptor), achieving invariance to rotation or scale changes, the need to retrain for different imaging modalities and body parts, and avoiding bias due to training data procedures~\citep{Geirhos2019ImageNettrainedRobustness}.  

% Neuro fingerprinting

In terms of neuroimage analysis, brain fingerprinting methods have been used to investigate the variability of individuals based on neuroimage-specific preprocessing and data, e.g. brain segmentations from structural MRI~\citep{Wachinger2015BrainPrintMorphology,Valizadeh2018IdentificationFeatures}, neuroanatomic parcellations and functional MRI~\citep{Finn2015FunctionalConnectivity}, fiber tracts from diffusion MRI~\citep{Kumar2017FiberprintAnalysis}. None of these methods have been sufficiently specific to achieve perfect accuracy at individual identification, and none have been used for the more difficult task of predicting family members. Heritability studies have investigated correlations in MRIs of siblings, including structural~\citep{Thompson2001GeneticStructure,VanDerLee2017GrayMorphometry} and functional MRI~\citep{Colclough2017HeritabilityActivity}, however these are not designed for prediction. In contrast, the keypoint methodology we adopt may be applied to generic imaging data with no preprocessing, representing distinctive patterns potentially present only in images of specific individuals or immediate family members. Our work here extends and compares against the keypoint-based neuroimage signatures method~\citep{Chauvin2020NeuroimageRelatives}, the only method reporting perfect accuracy at same-subject image prediction from standard neuroimage datasets (e.g. OASIS~\citep{Marcus2007OpenAdults}, ADNI~\citep{JackJr2008AlzheimerMethods}, HCP~\citep{VanEssen2012HumanPerspective}).

Our results here are the first in the literature for the task of family member prediction from volumetric medical image data. Family member prediction has been investigated in the context of 2D face photographs, based on clustering~\citep{Kiley2020WhoLearning}, conditional random fields~\citep{Dai2015FamilyCollections}, deep neural networks~\citep{Ahmad2019DeepTwins}, discriminant analysis~\citep{Juefei-Xu2013AugmentedFeatures}. These solutions are not directly related to our work, as they generally address aspects and challenges specific to the context of 2D photography, e.g. facial expression and lighting conditions~\citep{Paone2014DoubleRecognition}, multiple images per person, group photos of relatively few families (e.g. 16)~\citep{Dai2015FamilyCollections}. Accuracy is not generally available for siblings based on genotyping, and near 100\% accuracy for monozygotic twin identification from face photographs has not been reported.

\section{Method}

We seek a pairwise distance measure between two sets $A = \{f_i\}$ and $B = \{ f_j\}$ that can be used to estimate proximity and thus genetic relationships between subjects from image data. We first describe our method in terms of general set theory, then later include details pertaining to set elements $f_i$ defined as 3D image keypoints. Our measure begins with the Jaccard index or intersection-over-union $J_{HSE}(A,B)$ defined as
\begin{align}
    J_{HSE}(A,B) &= \frac{|A \cap B|}{|A| + |B| - |A \cap B|},
    \label{eq:tmi-jaccard}
\end{align}
where in Equation~\eqref{eq:tmi-jaccard}, $J_{HSE}(A,B)$ is defined by binary or hard set equivalence (HSE) between set elements. $A \cap B$ is the intersection operator between sets $A$ and $B$, and $|A \cap B|$, $|A|=|A \cap A|$, $|B|=|B \cap B|$ represent set cardinality operators that count the numbers of elements present in each set, where $|A \cap B| \le \min\{ |A|, |B|\}$. The Jaccard index is a similarity measure, and may be used to define various distance measures including the Jaccard distance metric~\citep{Levandowsky1971DistanceSets} $1-J_{HSE}(A,B)$ or the Tanimoto distance $-\log J_{HSE}(A,B)$. 

In the case of sets of real data samples, for example keypoint descriptors, binary equivalence may be difficult to establish due to noise or uncertainty in the measurement process, and we seek to redefine $J_{HSE}(A,B)$ in Equation~\eqref{eq:tmi-jaccard} such that it more accurately accounts for non-binary equivalence between set elements. This may generally be accomplished by redefining the standard set intersection cardinality operator $|A \cap B|$ in Equation~\eqref{eq:tmi-jaccard} by an analogous soft cardinality operator $\mu(A \cap B)$, leading to the following expression for our proposed Jaccard index based on soft set equivalence $J_{SSE}(A,B)$
\begin{equation}
	J_{SSE}(A,B) = \frac{\mu(A \cap B)}{\mu(A) + \mu(B) - \mu(A \cap B)} \label{eq:soft_jaccard}.
\end{equation}
In Equation~\eqref{eq:soft_jaccard}, $J_{SSE}(A,B)$ is defined by the soft cardinality operator $\mu(A \cap B)$ as described in the following section, including cardinality operators $\mu(A)=\mu(A \cap A)$, $\mu(B)=\mu(B\cap B)$. In order to ensure that $J_{SSE}$ remains bounded to the range $[0,1]$, it is important that $\mu(A \cap B)$ be upper bounded by the minimum of the individual soft set cardinalities $\mu(A \cap B) \le \min \left\{ \mu(A), \mu(B)\right\}$.  \\

\noindent {\bf Defining $\boldsymbol{\mu(A \cap B)}$:}
The cardinality of the soft set intersection $\mu(A \cap B)$ is intended to reflect the uncertainty in equivalence between set elements $f_i \in A$ and $f_j \in B$. We define soft equivalence between elements $f_i \in A$ and $f_j \in B$ as a similarity function or kernel $\mathcal{K}(f_i,f_j)$ ranging from $[0,1]$, where $\mathcal{K}(f_i,f_j)=1$ indicates exact equivalence and $\mathcal{K}(f_i,f_j) = 0$ represents the absence of equivalence. We then define a measure $\mu(A \rightarrow B)$ over a mapping  $A \rightarrow B$ from set $A$ to set $B$ as:
\begin{align}
\mu(A \rightarrow B) 
&= \sum_i^{|A|}\max_{f_j \in B}\mathcal{K}(f_i,f_j)
\label{eq:gen2}
\end{align}
The similarity function $\mathcal{K}(f_i,f_j)$ in Equation~\eqref{eq:gen2} may generally be defined according to the specific representation of set elements $f_i$ and $f_j$. The maximum operator $\max_{f_j \in B}\mathcal{K}(f_i,f_j)$ ensures the bound $\mu(A \rightarrow B) \leq \mu(A)$ via a partial injective mapping from A to B, embodying the intuition that each element $f_i \in A$ has at most one counterpart $f_j \in B$. Note also that max operator may be computed via computationally efficient NN indexing methods. The soft intersection cardinality $\mu(A \cap B)$ may be defined as a symmetrized version of $\mu(A \rightarrow B)$:
\begin{align}
\mu(A \cap B) &= \min \left\{~\mu(A \rightarrow  B),~ \mu(B \rightarrow  A)~\right\} \notag \\
&= \min \left\{ \sum_i^{|A|}\max_{f_j \in B}\mathcal{K}(f_i,f_j), \sum_j^{|B|}\max_{f_i \in A}\mathcal{K}(f_j,f_i) \right\}
\label{eq:gen3}
\end{align}
Note that the Jaccard distance measure derived using the cardinality operator $\mu(A \cap B)$ as defined by Equations~\eqref{eq:gen2} and~\eqref{eq:gen3} is symmetric and follows the identity of indiscernibles, and thus may be considered to be a semi-metric. The triangle inequality holds in the case of binary equivalence~\citep{Levandowsky1971DistanceSets} but not generally for soft equivalence relationships. Equation~\eqref{eq:gen3} may be used to define the cardinality of a single set $A$ as $\mu(A)=\mu(A \cap A)=|A|$ both for hard and soft equivalence since, since for a given element $f_i \in A$, $\max_{f_j \in A}\mathcal{K}(f_i,f_j) = \mathcal{K}(f_i,f_i) = 1$. The standard Jaccard index in Equation~\eqref{eq:tmi-jaccard} is thus a special case of the soft Jaccard index in the case of a binary-valued kernel $\mathcal{K}(f_i,f_j) \in \{0,1\}$.

\noindent {\bf Defining $\boldsymbol{\mathcal{K}(f_i,f_j)}$:} The soft set intersection cardinality $\mu(A \cap B)$ may be adapted to a specific task by defining $\mathcal{K}(f_{i},f_{j})$ according to the representation of elements $f_{i}$. In the context of this article, a set element $f_{i}$ is a 3D scale-invariant keypoint $f_{i} = \{ \bar{a_{i}}, \bar{g_{i}} \}$ as described in~\citep{Toews2013EfficientFeatures}, where $\bar{g_{i}}$ and $\bar{a_{i}}$ are descriptors of local keypoint geometry and appearance, respectively. Keypoint geometry $\bar{g_{i}} = \{ \bar{x_{i}},\sigma_{i} \}$ consists of 3D location $\bar{x_{i}}$ and scale $\sigma_{i}$, and appearance $\bar{a_{i}}$ is a vector of local image information, i.e. a rank-ordered histogram of oriented gradients (HOG)~\citep{Toews2009SIFTRankCorrespondence}.

%In the limiting case of a kernel with binary support $\mathcal{K}(f_i,f_j) \in \{0,1\}$ such as the Iverson bracket $\mathcal{K}(f_i,f_j)=[f_i=f_j]$, the expression in Equation~\eqref{eq:soft_jaccard} is equivalent to the standard Jaccard with hard set equivalence. 

The kernel $\mathcal{K}(f_{i},f_{j})$ operates on keypoint elements $f_{i} = \{ \bar{a_{i}}, \bar{g_{i}} \} \in A$ and $f_{j} = \{ \bar{a_{j}}, \bar{g_{j}} \} \in B$. Here we relax the assumption of hard equivalence using squared exponential kernels with non-zero support, factored into kernels operating separately on local keypoint appearance $\mathcal{K}(\bar{a_{i}},\bar{a_{j}})$ and geometry $\mathcal{K}(\bar{g_{i}},\bar{g_{j}})$ variables:
\begin{align}
    \mathcal{K}(f_{i},f_{j}) &= \mathcal{K}(\bar{a_{i}},\bar{a_{j}})\mathcal{K}(\bar{g_{i}},\bar{g_{j}})
    \label{eq:product_weights}
\end{align}
The factorization in Equation~\eqref{eq:product_weights} is due to the use of descriptors $\bar{a_{i}}$ that are invariant to 7-parameter similarity transforms of the 3D image coordinate system from which geometry $\bar{g_{i}}$ is derived. The two kernels in Equation~\eqref{eq:product_weights} are defined as squared exponential functions as follows. 

The appearance kernel $\mathcal{K}(\bar{a_{i}},\bar{a_{j}})$ is defined by the squared Euclidean distance $\lVert \bar{a_{i}} - \bar{a_{j}} \rVert^{2}_{2}$ between appearance vectors $\bar{a_{i}}$ and $ \bar{a_{j}}$:
\begin{align}
	\mathcal{K}(\bar{a_{i}},\bar{a_{j}}) &= \exp \left( -\frac{\lVert \bar{a_{i}} - \bar{a_{j}} \rVert^{2}_{2}}{\alpha^{2}} \right) \label{eq:appearance_weight}
\end{align}
where in Equation~\eqref{eq:appearance_weight}, $\alpha$ is a bandwidth parameter that may be estimated adaptively as
\begin{equation}
	\alpha = \min_{f_{j} \in \Omega}\lVert \bar{a_{i}} - \bar{a_{j}} \rVert^{2}_{2},~s.t.~\lVert \bar{a_{i}} - \bar{a_{j}} \rVert^{2}_{2} > 0
\end{equation}
the minimum Euclidean distance between appearance descriptor $\bar{a_{i}} \in A$ and the nearest descriptor $\bar{a_{j}} \in \Omega \setminus A$ within the entire available dataset $\Omega$ excluding $A$. Note this choice of estimator is not strictly symmetric, however it allows the kernel to adapt to arbitrary dataset sizes, shrinking the resolution of prediction as the number of data grows large, and does not affect the symmetry of Equation~\eqref{eq:gen3}.

\begin{figure}[!h]
  \centering
  \includegraphics[width=0.75\textwidth,draft=false]{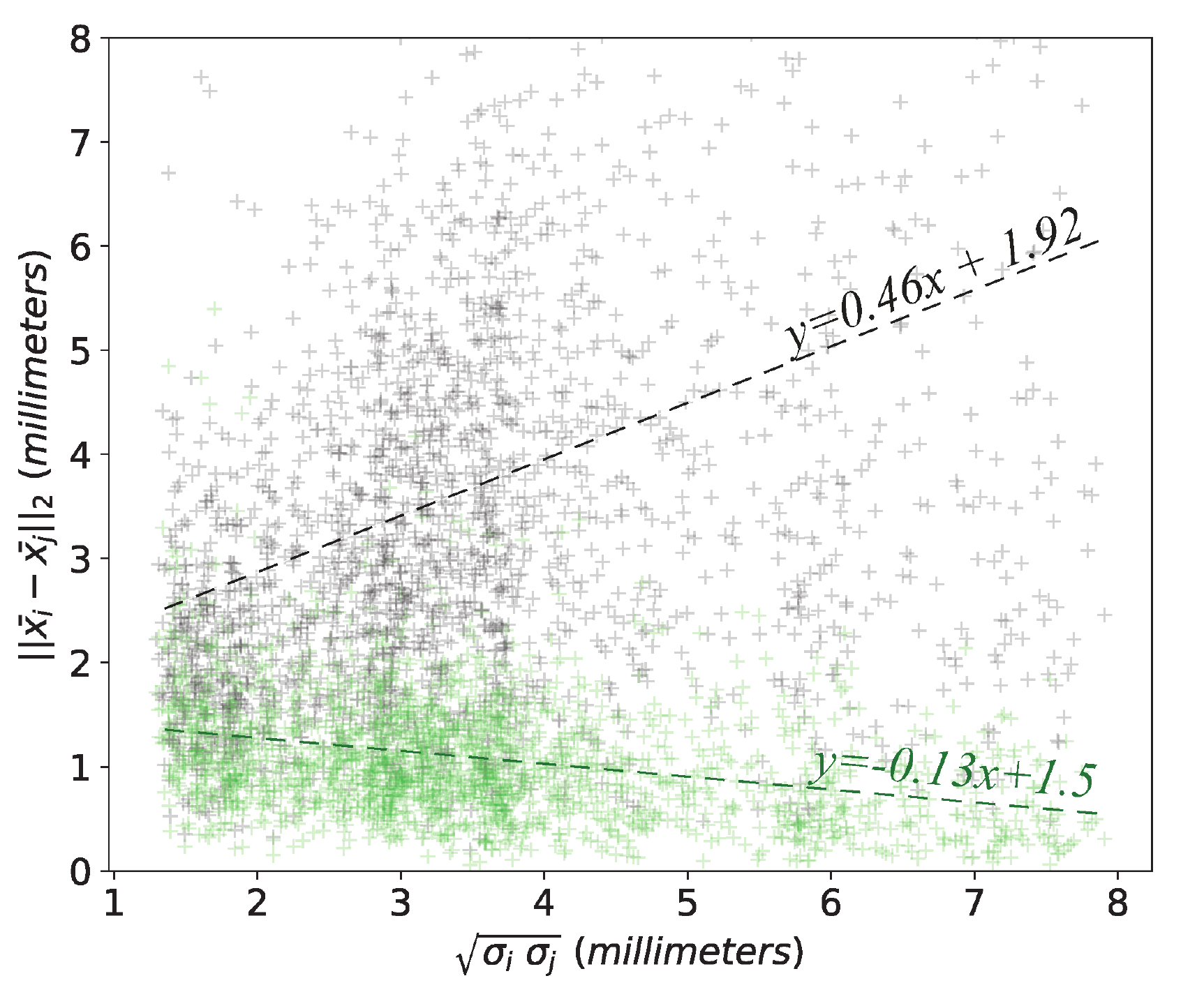}
  \captionsetup{width=0.75\textwidth}
  \caption{Graph of spatial localization error $\lVert \bar{x_i}-\bar{x_j} \rVert_{2}$ vs. geometric mean scale $\sqrt{\sigma_i~\sigma_j}$. Lines represent the linear regression based on each point set. Note how location error increases with scale (black) due to uncertainty, however this effect is reduced by scale normalization (green) as in Equation~\eqref{eq:geometrical_weight}. This visualization is based on keypoint correspondences between 10 MRI aligned volumes}
  \label{fig:scale_related_error}
\end{figure}

The geometry kernel $\mathcal{K}(\bar{g_{i}},\bar{g_{j}})$ is novel to this work, and is defined as the product of two kernels, one modeling keypoint location conditional on keypoint scale $\mathcal{K}(\bar{x_{i}},\bar{x_{j}};\sigma_{i},\sigma_{j})$ and the $\mathcal{K}(\sigma_{i},\sigma_{j})$ modeling scale alone  
\begin{align}
    \mathcal{K}(\bar{g_{i}},\bar{g_{j}})=
    \mathcal{K}(\bar{x_{i}},\bar{x_{j}};\sigma_{i},\sigma_{j})\mathcal{K}(\sigma_{i},\sigma_{j})
 \label{eq:geometry_kernel}
\end{align}
These kernels are defined as follows
\begin{align}
    \mathcal{K}(\bar{x_{i}},\bar{x_{j}};\sigma_{i},\sigma_{j}) &= \exp \left( -\frac{\lVert \bar{x_{i}} - \bar{x_{j}} \rVert^{2}_{2}}{\sigma_{i}\sigma_{j}} \right) \label{eq:geometrical_weight} \\
    \mathcal{K}(\sigma_{i},\sigma_{j}) &=  \exp \left( - \log^{2}\left( \frac{\sigma_{i}}{\sigma_{j}} \right) \right) 
    \label{eq:scale_weight}
\end{align}
Kernel $\mathcal{K}(\bar{x_{i}},\bar{x_{j}};\sigma_{i},\sigma_{j})$ in Equation~\eqref{eq:geometrical_weight} penalizes the squared distance between keypoint coordinates within a local reference frame, normalized by a variance proportional to the product of keypoint scales $\sigma_{i}\sigma_{j}$. This variance embodies uncertainty in keypoint location due to scale, and has a computational form that is reminiscent of mass in Newton's law of gravitation or electric charge magnitude in Coulomb's law. Figure~\ref{fig:scale_related_error} demonstrates how this kernel variance normalizes higher localization error associated with keypoints of larger scale. Kernel $\mathcal{K}(\sigma_{i},\sigma_{j})$ in Equation~\eqref{eq:scale_weight} penalizes multiplicative difference between keypoint scales $(\sigma_i,\sigma_j)$.

Note that while the appearance kernel in Equation~\eqref{eq:appearance_weight} is invariant to global similarity transforms due to descriptor invariance, the geometry kernel in Equation~\eqref{eq:geometry_kernel} measures zero-mean keypoint displacement and thus requires data to be aligned within a common spatial reference frame. Alignment may be established via standard subject-to-atlas registration prior to keypoint extraction or from keypoint correspondences after extraction using feature-based alignment~\citep{Toews2013EfficientFeatures}. The following workflow is used to compute pairwise distances for a set of $N$ images:
\begin{enumerate}
    \item Extract keypoints from each image
    \item Align keypoints to an atlas template (optional for pre-aligned images) 
    \item Identify k-NN correspondences $(\bar{a_{i}},\bar{a_{j}})$ minimizing the Euclidean distance between descriptors $\| \bar{a_{i}}-\bar{a_{j}} \|$
    \item Compute pairwise image distances by evaluating kernels from k-NN correspondences
\end{enumerate}

Figure~\ref{fig:workflow} shows the effect of our proposed geometry kernel in regularizing and favoring geometrically plausible correspondences for an example twin sibling pair.

\begin{figure}[!h]
  \centering
  \includegraphics[width=0.75\textwidth,draft=false]{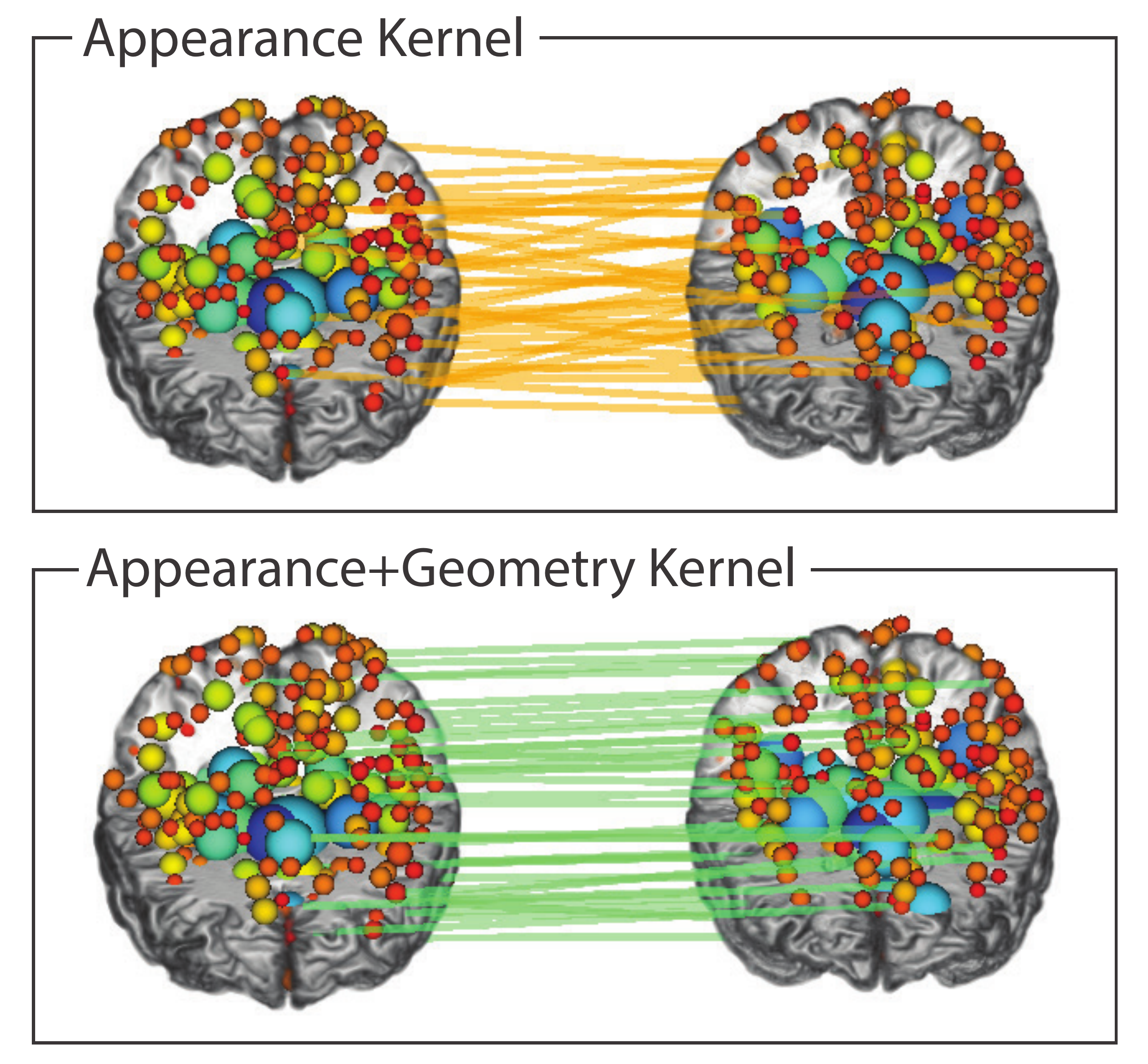}
  \captionsetup{width=0.75\textwidth}
  \caption{Visualizing the keypoint correspondences driving classification between a pair of identical twin brains, showing the 100 correspondences with the highest SSE kernel values. The appearance kernel alone (upper, Equation~\eqref{eq:appearance_weight}) may include geometrically inconsistent matches (diagonal lines), however these are down weighted when combined with the geometry kernel (lower, Equations~\eqref{eq:geometrical_weight} and~\eqref{eq:scale_weight}), leading to geometrically consistent matches (horizontal lines) and improved prediction performance}
  %Step I consists of the SIFT feature extraction for each image (via GPU or CPU implementation). In Step II and III, lines represents the 100 most similar features based respectively on kernels defined in equation~\eqref{eq:appearance_weight}, and the combination of kernels from equation~\eqref{eq:appearance_weight},~\eqref{eq:geometrical_weight} and~\eqref{eq:scale_weight}. Step II is achieved in $O(log N)$ complexity  via a k-Nearest Neighbor algorithm. It should be noted that the constraint introduced by our newly defined kernels in equations~\eqref{eq:geometrical_weight} and~\eqref{eq:scale_weight} in Step III results in a more accurate feature matching.}
  \label{fig:workflow}
\end{figure}

\section{Experiments}

Experiments investigate the ability of the Jaccard distance to predict pairwise relationships between whole-brain MRI scans, where relationships include 1) close genetic links between siblings sharing 25-100\% of their polymorphic genes and 2) broad genetic links between non-siblings sharing nominal genetic information due to sex and common racial ancestry. We hypothesize that closer genetic proximity will be reflected in higher pairwise similarity and thus lower Jaccard distance. We expect that pairwise distance based on our proposed soft set equivalence (SSE) $d_{J}(A,B) = -\log J_{SSE}(A,B)$ as defined in Equations~\eqref{eq:soft_jaccard} and~\eqref{eq:gen3} and our geometry kernel as defined by Equations~\eqref{eq:geometrical_weight} and~\eqref{eq:scale_weight} will lead to improved identification of pairwise relationship labels.

Prediction is based on pairwise Jaccard distance and a leave-one-out protocol, analogous to querying a subject MRI in a hospital PACS. Note that this is a deterministic procedure with no explicit training stage, where the primary hyperparameter is the number of nearest neighbors $k$ per keypoint. Our computational workflow, as previously described, follows three steps 1) 3D SIFT-Rank keypoints are extracted from individual pre-processed images, 2) kNN correspondences are identified between keypoints from all $N$ images, and 3) Jaccard distances are evaluated between all image pairs from kernels and kNN correspondences.

\subsection{Data and Computational Details}
Our data set consists of MRI scans of $N=1010$ unique subjects from the Human Connectome Project (HCP) Q4 release~\citep{VanEssen2012HumanPerspective}, aged 22-36 years (mean 29 years), acquired from a diverse population including 468 males and 542 females, and 434 unique families. The MRI data are provided as T1w volumes at isotropic 0.7mm voxel resolution and preprocessed via a standard neuroimaging pipeline, including rigid subject-to-atlas registration and skull-stripping. Keypoint extraction requires approximately 3~sec.~/~per image and results in an average of 1,400 keypoints per image for a total of 1,488,065 keypoints. We use a GPU implementation~\citep{Pepin2020LargeScaleMasking} of the 3D SIFT-Rank keypoint algorithm~\citep{Toews2013EfficientFeatures} that produces identical keypoints at a 7$\times$ speedup. Approximate kNN correspondences between appearance descriptors are identified across the entire database using efficient KD-tree indexing~\citep{Muja2014ScalableData}, where lookup requires 0.8~sec.~/~subject for k=200 nearest neighbors on an i7-5600@2.60Ghz machine with 16 GB RAM (1.64 GB used).

Given N=1010 images, there are a total of $N(N-1)/2=509545$ pairwise relationships to be evaluated. Each subject pair is assigned by one of five possible relationship labels $L = \{MZ, DZ, FS, HS, UR\}$ for monozygotic twins (MZ), dizygotic twins (DZ), full non-twin siblings (FS), half-siblings (HS) and unrelated non-siblings (UR), for totals of $\{134, 71, 607, 44, 508689\}$ relationships per label. Family relationship labels are based on mother and father identity and zygoticy (for twins), confirmed via genome-wide single nucleotide polymorphism (SNP) genotyping~\citep{1200Connectome}. Note the sparse structure of family relationship labels, where each sibling label (MZ, DZ, FS, HS) is unique to one of 434 families and the vast majority of pairs are UR ($0.998=508689/509545$). The nominal probability of randomly guessing the correct family is approximately $0.0023=1/434$, and the challenge is thus to efficiently and accurately identify the small number of family relationships. Note that evaluating the similarity of $N(N-1)/2$ pairs via brute force image matching or registration becomes computationally intractable for large $N$.

Image alignment is fundamental to the functioning of our proposed geometry kernel, which measures zero-mean deviations in keypoint location within a common spatial reference frame. We thus evaluated prediction following two different subject-to-atlas registration methods: the original 3D rigid transform provided by the HCP mapping images to the standard MNI atlas using the FLIRT algorithm~\citep{Glasser2013MinimalProject,Jenkinson2002ImprovedImages}, and feature-based alignment (FBA)~\citep{Toews2013EfficientFeatures} using keypoint correspondences to estimate a 3D similarity transform mapping images to a subject atlas, for 20 different randomly selected atlases. All alignment solutions were very close to the reference MNI alignment (rotation differed by $4.31^{\circ}\pm 2.84$, translation by $3 mm \pm 3.5$), reflecting a degree of inter-subject variability given different atlases. Most notably, however, was that AUC values were virtually identical for all prediction trials, with a maximum standard deviation of $\epsilon = 0.0017$ (see Table~\ref{tab:hard_soft_ROC}), indicating that prediction was not sensitive to the alignment solution used. The results we report are thus consistent with the reference MNI alignment provided with the HCP data.

\subsection{Close genetic proximity: Siblings}

\begin{figure*}[!h]
  \centering
  \includegraphics[width=\textwidth,draft=false]{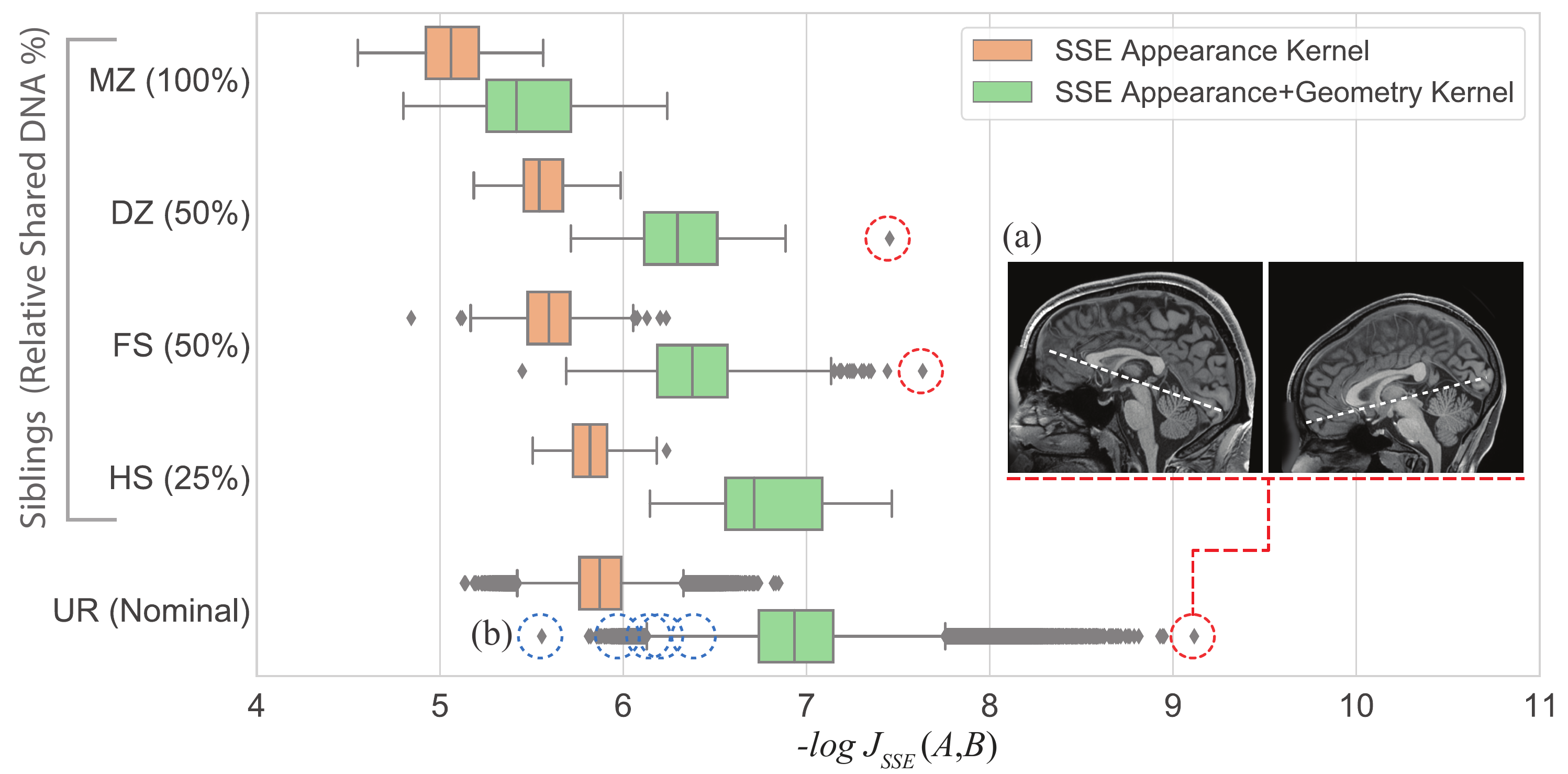}
  \caption{Distributions of Jaccard distances $-\log J_{SSE}(A,B)$ conditioned on pairwise labels $L = \{MZ, DZ, FS, HS, UR\}$, comparing kernels with and without geometry (green vs. orange). Note that the distance generally increases with genetic separation, and that geometry kernels increase separation between sibling relationship labels. Outliers, i.e. pairs outside of their expected relationship distance distributions, can be easily identified and inspected for irregularity, e.g. (a) unusually high distances may indicate spatial misalignment (red circles) and (b) unusually low UR distances may indicate cases of incorrect family labels (blue circles)
  %like the blue circles, indicating pairs of self-reported family members flagged as unrelated due uncertain genotyping~\citep{wu20171200}.
  } 
  \label{fig:rel_distances}
\end{figure*}

Figure~\ref{fig:rel_distances} shows distributions of pairwise Jaccard distance conditioned on pairwise relationships. Our proposed kernel based on appearance and geometry (Figure~\ref{fig:rel_distances}, green) generally increases separation between different sibling relationships in comparison to appearance only (Figure~\ref{fig:rel_distances}, orange) as in~\citep{Chauvin2020NeuroimageRelatives}. A two-tailed Kolmogorov-Smirnoff (KS) test shows all distributions to be significantly different ($p-value < 1e-10$) except those of DZ and FS siblings sharing approximately 50\% of their genes ($p-value = 0.0199$). Furthermore, pairwise KS statistics were all lower for appearance+geometry kernels compared to appearance only, indicating that the geometry kernels lead to increased separation of relationship categories. The Jaccard distance may be rapidly evaluated, and pairs which are outliers from their expected distributions may be easily flagged and inspected for irregularities. For example, the highest Jaccard distance outliers indicated noticeable spatial misalignment between a small number of subject pairs (e.g. Figure~\ref{fig:rel_distances}a. The lowest distance outliers for UR pairs may indicate potential sibling relationships. The blue circles near Figure~\ref{fig:rel_distances}b are pairs involving 3 subjects flagged as UR due to genotyping errors (including a MZ twin)~\citep{1200Connectome}. Here, nearest neighbor Jaccard distance was used to correctly predict the self-reported families for all three cases, a result confirmed by the HCP.

\begin{figure*}
  \centering
  \includegraphics[width=1\textwidth,draft=false]{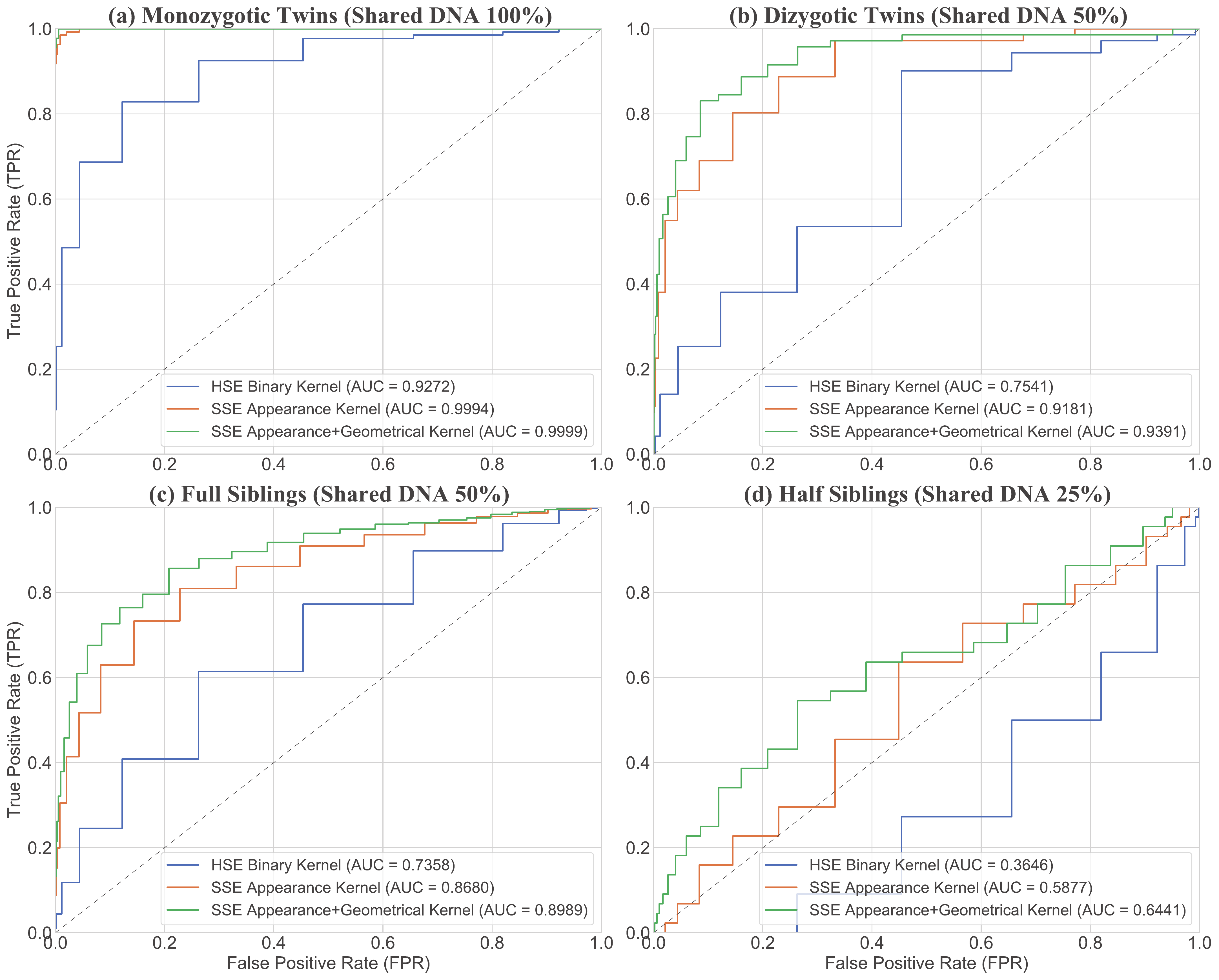}
  \caption{ROC curves for sibling identification based on Jaccard distance for (a) MZ, (b) DZ, (c) FS and (d) HS pairs, comparing three kernels for k=200 NNs. Note that the prediction AUC is always highest for our proposed kernel combining appearance and geometry (green), and decreases as expected with the degree genetic and developmental difference, e.g. in the order MZ, DZ, FS, HS}
  \label{fig:roc_curves}
\end{figure*}

As sibling pairs exhibit significantly lower distance than UR pairs, we investigate the degree to which they can be distinguished from unrelated pairs based on a simple distance threshold. Figure~\ref{fig:roc_curves} shows the Receiver Operating Characteristic (ROC) curves for sibling relationships based on distance, comparing our SSE kernels for appearance and geometry, appearance only and the binary HSE kernel. Table~\ref{tab:hard_soft_ROC} quantifies the improvement of SSE vs HSE, for various numbers of keypoint nearest neighbors (20, 100, 200), where the highest area-under-curve (AUC) values are obtained for SSE with k=200. The performance of SSE kernels generally increases with the number of NNs, and is always superior for combined appearance and geometry kernels. Conversely, HSE classification performance decreases, even falling below the diagonal in the most challenging case of HS pairs in Figure~\ref{fig:roc_curves}d, as additional hard correspondences appear to accumulate systematically on a subset of typical but unrelated subjects. For completeness, the highest UR prediction result was AUC$=0.951$ for SSE, appearance and geometry kernels, k=200.
 
\begin{table}
\centering
\caption{Area Under the Curve (AUC) values for sibling classification, comparing kernels and numbers of nearest neighbors (k). $\epsilon = 0.0017$ is the maximum AUC standard deviation observed from 21 different subject-to-atlas registration trials. Note $SSE_{App+Geo}$ kernels significantly outperform $SSE_{App}$ and $HSE$ for the more difficult DZ and FS pairs}
\label{tab:hard_soft_ROC}
\bgroup
\def\arraystretch{1.5}
\resizebox{\columnwidth}{!}{
\begin{tabular}{|cc|ccc|ccc|ccc}
    \hline
    \multicolumn{2}{|P{2cm}|}{\multirow{2}{*}{\backslashbox[2.4cm]{Label}{Kernel}}}&\multicolumn{3}{|c|}{\textit{k = 20}}&\multicolumn{3}{|c|}{\textit{k = 100}}&\multicolumn{3}{|c|}{\textit{k = 200}}\\ 
    \cline{3-11}
    \multicolumn{2}{|P{2cm}|}{}&\multicolumn{1}{|P{1.2cm}|}{\textit{HSE}}&\multicolumn{1}{|P{1.2cm}|}{\textit{SSE\textsubscript{A}}}&\multicolumn{1}{P{1.5cm}|}{\textit{SSE\textsubscript{A+G}}}&\multicolumn{1}{|P{1.2cm}|}{\textit{HSE}}&\multicolumn{1}{|P{1.2cm}|}{\textit{SSE\textsubscript{A}}}&\multicolumn{1}{P{1.5cm}|}{\textit{SSE\textsubscript{A+G}}}&\multicolumn{1}{|P{1.2cm}|}{\textit{HSE}}&\multicolumn{1}{|P{1.2cm}|}{\textit{SSE\textsubscript{A}}}&\multicolumn{1}{P{1.5cm}|}{\textit{SSE\textsubscript{A+G}}}\\
    \thickhline
    \multicolumn{2}{|P{2cm}|}{\textbf{MZ}}&\multicolumn{1}{|P{1.2cm}|}{0.9983}&\multicolumn{1}{P{1.2cm}|}{0.9993}&\multicolumn{1}{P{1.5cm}|}{\textbf{0.9996 $\bm{\pm~\epsilon}$}}&\multicolumn{1}{|P{1.2cm}|}{0.9544}&\multicolumn{1}{P{1.2cm}|}{0.9995}&\multicolumn{1}{P{1.5cm}|}{\textbf{0.9998 $\bm{\pm~\epsilon}$}}&\multicolumn{1}{|P{1.2cm}|}{0.9272}&\multicolumn{1}{P{1.2cm}|}{0.9994}&\multicolumn{1}{P{1.5cm}|}{\textbf{0.9999 $\bm{\pm~\epsilon}$}}\\
    \hline
    \multicolumn{2}{|P{2cm}|}{\textbf{DZ}}&\multicolumn{1}{|P{1.2cm}|}{0.8922}&\multicolumn{1}{P{1.2cm}|}{0.8825}&\multicolumn{1}{P{1.5cm}|}{\textbf{0.9044 $\bm{\pm~\epsilon}$}}&\multicolumn{1}{|P{1.2cm}|}{0.8018}&\multicolumn{1}{P{1.2cm}|}{0.9034}&\multicolumn{1}{P{1.5cm}|}{\textbf{0.9250 $\bm{\pm~\epsilon}$}}&\multicolumn{1}{|P{1.2cm}|}{0.7541}&\multicolumn{1}{P{1.2cm}|}{0.9181}&\multicolumn{1}{P{1.5cm}|}{\textbf{0.9391 $\bm{\pm~\epsilon}$}}\\
    \hline
    \multicolumn{2}{|P{2cm}|}{\textbf{FS}}&\multicolumn{1}{|P{1.2cm}|}{0.8423}&\multicolumn{1}{P{1.2cm}|}{0.8433}&\multicolumn{1}{P{1.5cm}|}{\textbf{0.8753 $\bm{\pm~\epsilon}$}}&\multicolumn{1}{|P{1.2cm}|}{0.7611}&\multicolumn{1}{P{1.2cm}|}{0.8569}&\multicolumn{1}{P{1.5cm}|}{\textbf{0.8888 $\bm{\pm~\epsilon}$}}&\multicolumn{1}{|P{1.2cm}|}{0.7358}&\multicolumn{1}{P{1.2cm}|}{0.8680}&\multicolumn{1}{P{1.5cm}|}{\textbf{0.8989 $\bm{\pm~\epsilon}$}}\\
    \hline
\end{tabular}
}
\egroup
\end{table}

\subsection{Distant genetic proximity: unrelated subjects}

\begin{figure*}[!h]
  \centering
  \includegraphics[width=1\textwidth,draft=false]{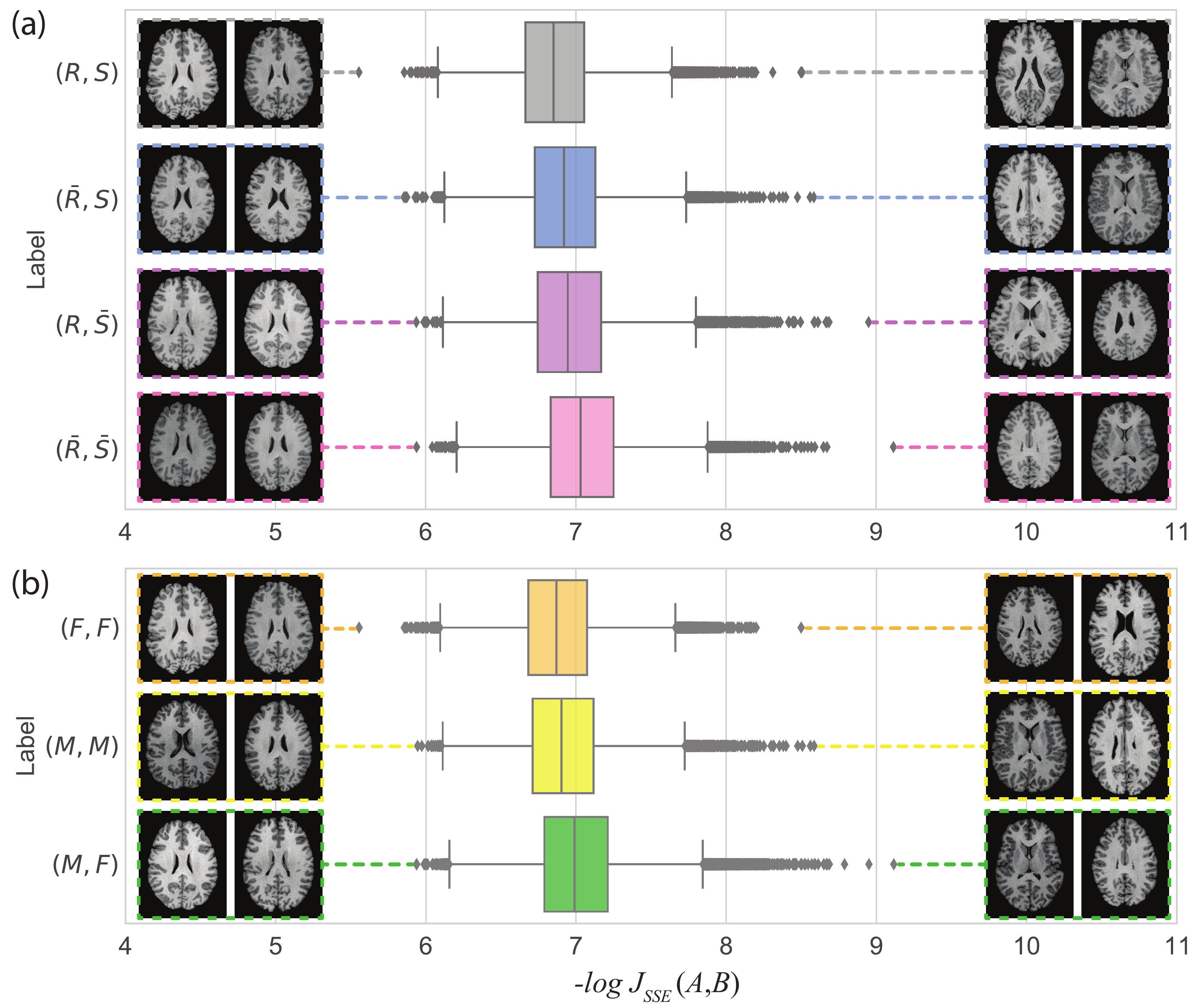}
  \caption{Jaccard distance $-\log J_{SSE}(A,B)$ distributions between unrelated pairs conditioned on shared demographic information. \textbf{(a)} same race, same sex $(R,S)$; different race, same sex $(\bar{R},S)$; same race, different sex $(R,\bar{S})$, and different race, different sex $(\bar{R},\bar{S})$ and \textbf{(b)} Female-Female $(F,F)$, Male-Male $(M,M)$, Male-Female $(M,F)$. Image pairs corresponding to minimum and maximum distances for each distribution are shown for visualization}
  \label{fig:gender_race}
\end{figure*}

Unrelated (UR) subject pairs share nominal amounts of genetic information, with subtle similarities due to demographic factors such as common sex and ancestral race. We thus expect whole-brain distance to be lowest for pairs of the same race and sex (R,S), highest for different race and sex ($\overline{\mbox{R}}$,$\overline{\mbox{S}}$), and intermediate for either same race (R,$\overline{\mbox{S}}$) or sex ($\overline{\mbox{R}}$,S). While the mean distances for conditional distributions in Figure~\ref{fig:gender_race} generally increase with differences in demographic labels, there are many exceptions where pairs with different labels exhibit lower distance than those with the same labels. We note this is consistent with pairwise genetic differences, where pairs of unrelated individuals from different populations may often exhibit higher genetic similarity than those from the same population~\citep{Witherspoon2007GeneticPopulations}.

\begin{figure}[!h]
  \centering
  \includegraphics[width=0.75\textwidth,draft=false]{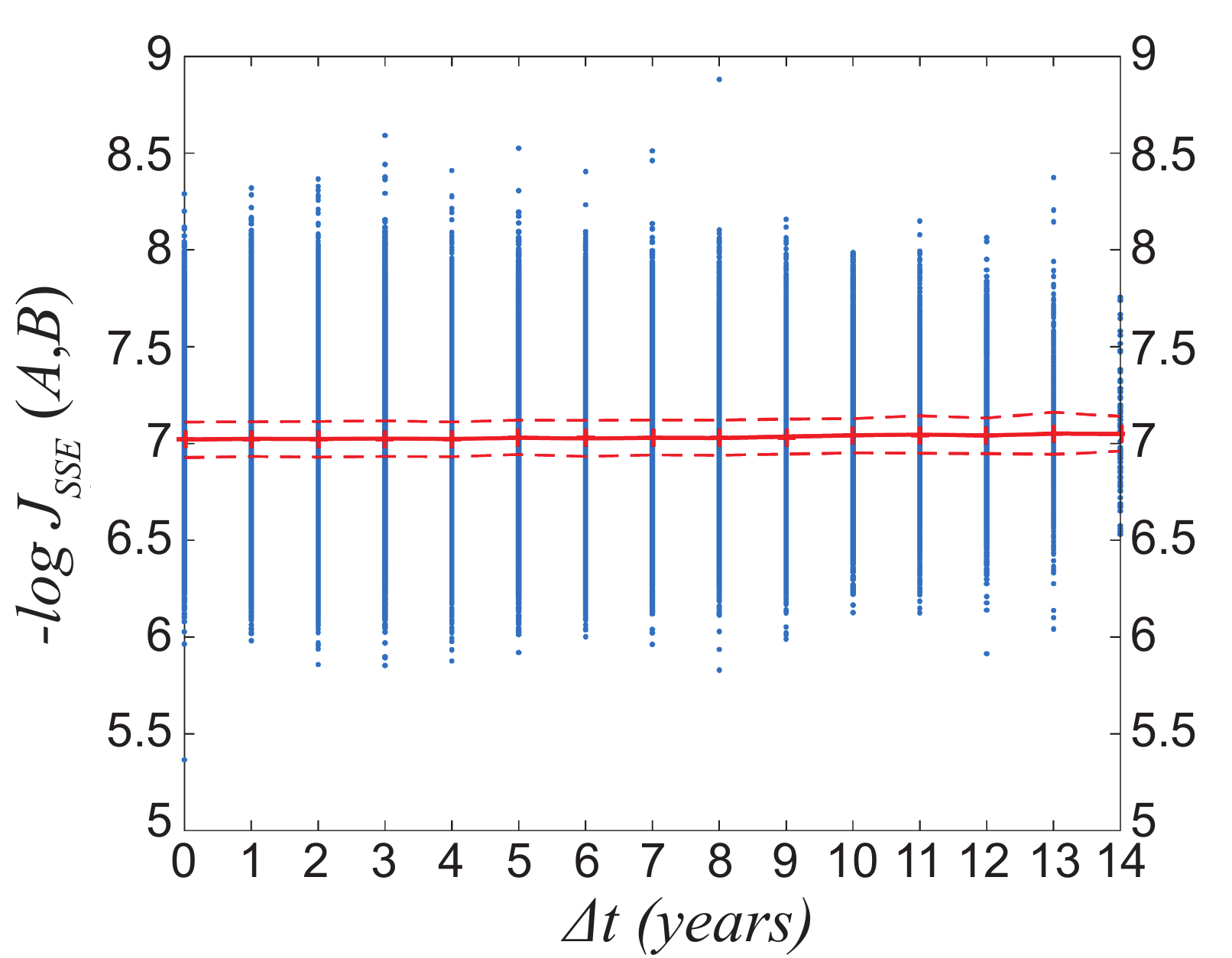}
  \captionsetup{width=0.75\textwidth}
  \caption{Distributions of Jaccard distance $-\log J_{SSE}(A,B)$ conditioned on age difference $\Delta t$ between unrelated pairs $(A,B)$ conditioned on age difference $\Delta t$. The mean (solid red line) and standard deviation (dashed red lines) are plotted for each $\Delta t$. Age difference has no significant impact on the distance for unrelated healthy young adult brains (age 22-36 years), eliminating a possible confounding factor}
  \label{fig:age_difference}
\end{figure}

Age difference between subjects is a potential confounding factor in whole brain Jaccard distance, and we plotted the variation of distance vs. age difference in Figure~\ref{fig:age_difference}. Distance distributions are virtually identical across the HCP subject age range spanning 22-36 years of age, indicating that age difference is not a major confound in this relatively young and healthy HCP cohort where brain morphology is relatively stable. Note that the Jaccard distance was found to increase with age difference in older subjects due to natural aging and neurodegenerative disease in~\citep{Chauvin2020NeuroimageRelatives}.
%similarly in rapid neurodevelopment over the infant range in~\citep{Toews2012Feature-basedMRI}.  

\subsection{Group Prediction: Sex}

While the pairwise Jaccard distance is highly informative regarding sibling relationships, it is insufficient for predicting group labels such as sex, age or disease. A simple modification can be used to predict group labels, by evaluating the distance between a single keypoint set A and supersets formed by the union of group members (excluding subject A), e.g. keypoints for {\em all Males} $d_{J}(A,\mathcal{M})$ and {\em all Females} $d_{J}(A,\mathcal{F})$. Supersets $\mathcal{M}$ and $\mathcal{F}$ are solely composed of unrelated subjects to avoid potential biases due to family relationships. These distances are combined in a basic linear classifier with a single threshold parameter $\tau$ to adjust for differences in the numbers of keypoints per group based on the following equation:
\begin{align}
   Class(A) = &
   \begin{cases}
    Female& \text{if } d_{J}(A,\mathcal{F}) - d_{J}(A,\mathcal{M}) + \tau > 0\\
    Male& \text{otherwise.}
   \end{cases}
   \label{eq:classification}
\end{align}

\begin{figure}[!h]
  \centering
  \includegraphics[width=0.75\textwidth,draft=false]{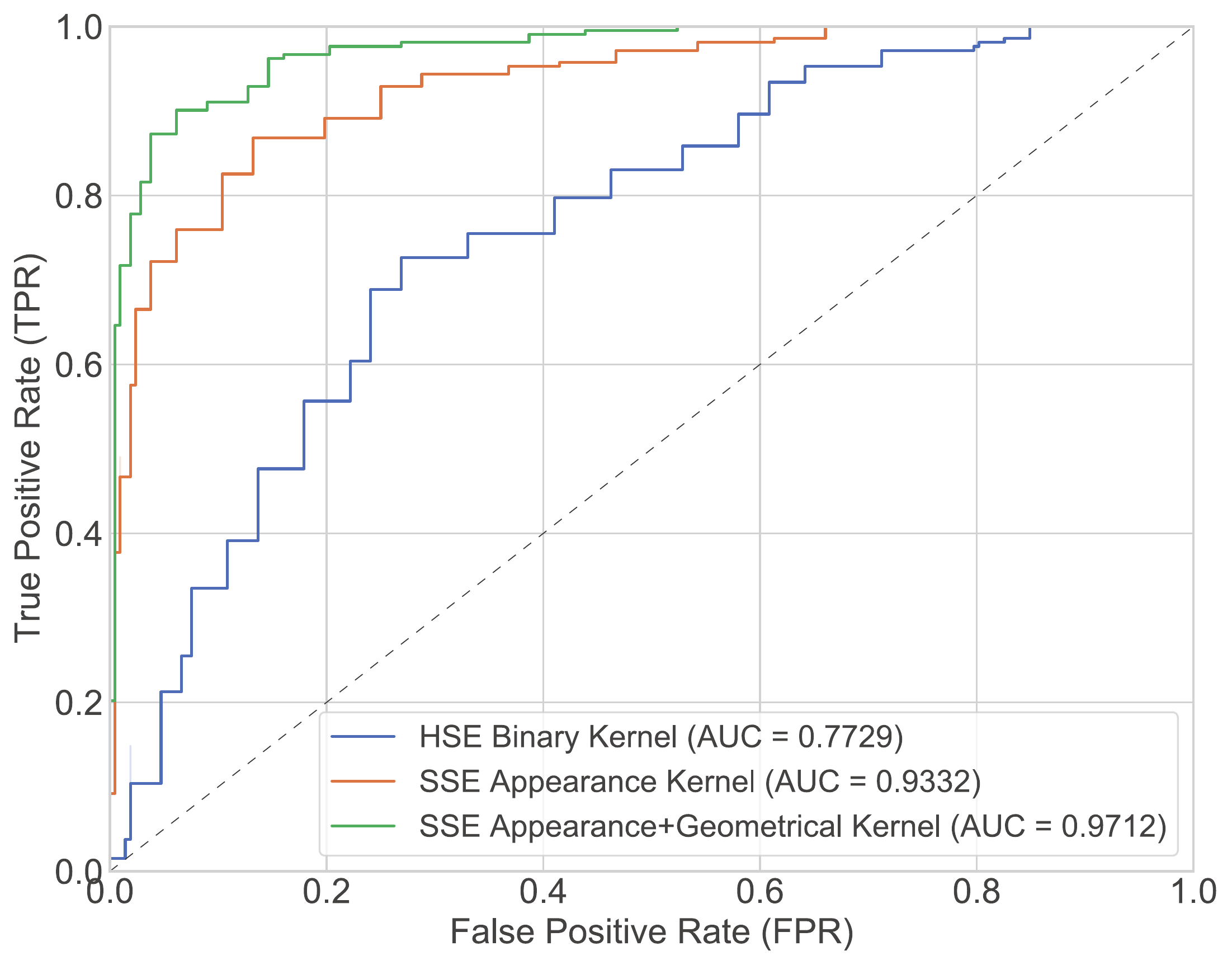}
  \captionsetup{width=0.75\textwidth}
  \caption{ROC curves and AUC for sex prediction based on individual-to-group Jaccard distance, comparing binary, appearance only, and combined appearance and geometry kernels.}
  \label{fig:ROC_male_female}
\end{figure}

Figure~\ref{fig:ROC_male_female} shows ROCs curves for sex prediction obtained by varying $\tau$ over the range $(-\infty,\infty)$, comparing Jaccard distance computed with binary, appearance only and combined appearance and geometry kernels. A possible confound is brain size, which is on average slightly larger for males than females~\citep{Eliot2021DumpSize}. As the appearance kernel is invariant to image scale, the AUC=0.93 reflects prediction accuracy independently of image size. Our proposed combined approach achieves the highest AUC=0.97, again outperforming other options.

\section{Discussion}

In this paper we propose a novel, highly specific distance measure between volumetric images represented as invariant keypoints. Our measure generalizes the Jaccard index to account for soft equivalence between keypoints, and a novel kernel estimator is proposed to model keypoint geometry in terms of location and scale within normalized image space. The soft Jaccard index is used as a distance measure to predict pairwise relationships between MRI brain scans of 1010 subjects from 434 families, including siblings and twins, and significantly improves upon previous work based on binary equivalence and appearance descriptors~\citep{Chauvin2020NeuroimageRelatives,Kumar2018MultimodalFramework}. We report the first results for predicting relationships from medical images, a new task, where monozygotic twins can be identified with virtually perfect accuracy. A minor modification allows the Jaccard distance to predict group labels such as sex with high accuracy. Our geometry kernel requires spatial normalization, however trials involving various linear subject-to-atlas alignment solutions including robust keypoint-based alignment~\citep{Toews2013EfficientFeatures} show that the prediction performance to be insensitive to the specific alignment solution used. The improved prediction afforded by our geometry kernel thus suggests that family members tend to align in a consistent manner regardless of the solution used.

Our method promises to be a useful tool for curating large medical image datasets for precision medicine and research purposes. A memory-based model using efficient and robust algorithms for 3D keypoint extraction and indexing~\citep{Toews2013EfficientFeatures,Toews2013FeaturebasedImages} allows for fine-grained comparisons between $O(N^2)$ image pairs in $O(N~\log~N)$ computational complexity. The Jaccard distance may thus be used to rapidly validate relationship labels, e.g. unexpectedly low distance may indicate related individuals and unexpectedly high distance may indicate unrelated individuals or spatially misaligned pairs. Previous work identified mislabelled scans of individuals in large neuroimage datasets~\citep{Chauvin2020NeuroimageRelatives}, here our improved method allowed us predict the self-reported families of three subjects labelled as unrelated due to inconclusive genotyping~\citep{1200Connectome}, on the basis of nearest neighbor soft Jaccard distance. This exceptional result was confirmed by the Human Connectome Project, and serves as a concrete example as to how our method can be used to validate labels associated with medical imaging data.

The task of pairwise relationship prediction may be viewed as the finest grain of categorization, where the number of unique pairwise relationship labels is linear $O(N)$ in the number of data $N$, e.g. 434 families from N=1010 images, as opposed to typical classification where all data are associated with a small number of labels (e.g. male, female). Pairwise prediction thus represents a challenge for ubiquitous deep neural network methods, which generally require large numbers of training data per category, and training for specific modalities and body parts. In contrast, generic 3D SIFT keypoints may be used as-is with arbitrary imaging modalities and contexts with no training. In future work, generic keypoint correspondences here could potentially be used to train domain-specific keypoint models~\citep{Yi2016LIFTTransform,Detone2018SuperPointDescription}. Auto-encoders used for anomaly detection could potentially be adapted to generate subject-specific codes between pairs of images~\citep{Baur2021AutoencodersStudy}. Analysis beyond pairs of subjects could be generalized via graph theory to model the clique structure of families~\citep{Blair1993IntroductionTrees}. All code required to reproduce our results may be obtained at https://github.com/3dsift-rank.

%\bibliographystyle{IEEEtran}
%\bibliography{IEEEabrv,TMI19,mendeley}

% \section{Table layout tests}

% Tables have the same constraints than the figures, except for the caption that has to be on top.

% \begin{table}
% \centering
% 		\parbox{0.65\textwidth}{\caption{Test of a long table caption, with linebreak}} % Contrainte manuelle de la largeur de la légende
		
% 		\begin{tabular}{|c|c|c|c|c|c|c|c|}
% 		\hline
% 			{\bf titre} & {\bf titre} & {\bf titre} & {\bf titre} & {\bf titre} & {\bf titre} & {\bf titre} & {\bf titre} \\
% 	  \hline
% 			blá & blá & blá & blá & blá & blá & blá & blá \\
% 	  \hline
% 			blá & blá & blá & blá & blá & blá & blá & blá \\
% 	  \hline
% 			blá & blá & blá & blá & blá & blá & blá & blá \\
% 	  \hline
% 			blá & blá & blá & blá & blá & blá & blá & blá \\
% 	  \hline
% 			blá & blá & blá & blá & blá & blá & blá & blá \\
% 	  \hline
% 			blá & blá & blá & blá & blá & blá & blá & blá \\
% 	  \hline
% 		\end{tabular}
% \end{table}

% \section{References test}

% \subsection{References to the bibliography}

% \subsection{References to the list of references "refs"}

% References from the list of references "refs", declared at the beginning of the document \citerefs{Test}.

% \subsection{References to a label of the document}

% Reference to a Figure associated to a label: Figure \ref{fig:vueEts}.

% \subsection{URL references}

% \subsubsection{Test of "href"}

% Href is used to integrate a link to a text:
% \href{http://www.etsmtl.ca/Etudiants-actuels/Cycles-sup/Realisation-etudes/Guides-gabarits}{Link to the template page.}.

% \subsubsection{Test de url}

% Url is used to format a clickable link:
% \url{http://www.etsmtl.ca/Etudiants-actuels/Cycles-sup/Realisation-etudes/Guides-gabarits}.

%%- Conclusion -%%
\begin{conclusion}
%\lipsum[1] % Text filling, to have an example of the layout

In this thesis, we presented an atom-like model of SIFT features derived from fundamental operators such as the Laplacian and including discrete orientation and sign to address earlier models' drawbacks such as orientation ambiguity and challenges with contrast inversion in multi-modal images. Our approach subsequently proposes evaluating the likelihood of two particles emerging from the same distribution based on kernel density estimates of feature appearance and geometry, and ultimately extending it to volumetric images allowing for fast image registration and quantification of image similarity based on a Jaccard-like measure of set overlap.
Our model has already proven useful for the neuroimaging community, by identifying mislabeled data in widely used large datasets or by efficiently registering 3D images in the case of stroke detection.
It has been cited as "incredibly powerful in capturing a keypoint signature ‘brainprint’ [...] despite ageing and neurodegenerative disease progression" in~\citep{Fetit2020TrainingBrain} and as: "Impressively, by identifying keypoint signatures robust to scan-to-scan variation, the algorithm matched MR images from an individual collected as many as 11 years apart" in~\citep{Finn2021FingerprintingConnectomes}.
A wide range of studies could stem from this research, assessing for example the impact of age, gender or diseases in large scale analyses, as in~\citep{Rokooie2020VisualizingSIFT}.

% Scale-space distance image to same image feature graph (geometry kernel) with constraints (see Karthik)
Although the work presented in this thesis focus on evaluating similarity between pairs of features from distinct images, the geometrical kernel $\mathcal{K}(g_i,g_j)$ could also serve, for example, to establish a fingerprint of cortical regions with pairs of features from the same image, via a distance graph between features on the scale-space, accounting for feature scale (see Figure~\ref{fig:sift_graph}). Graph constraints could also be added like in~\citep{Gopinath2022LearnableAnalysis} to learn surface data of cortical structures through graph convolution networks~\citep{Gopinath2021GeometricalData}.

\begin{figure}
    \centering
    \includegraphics[width=1\linewidth]{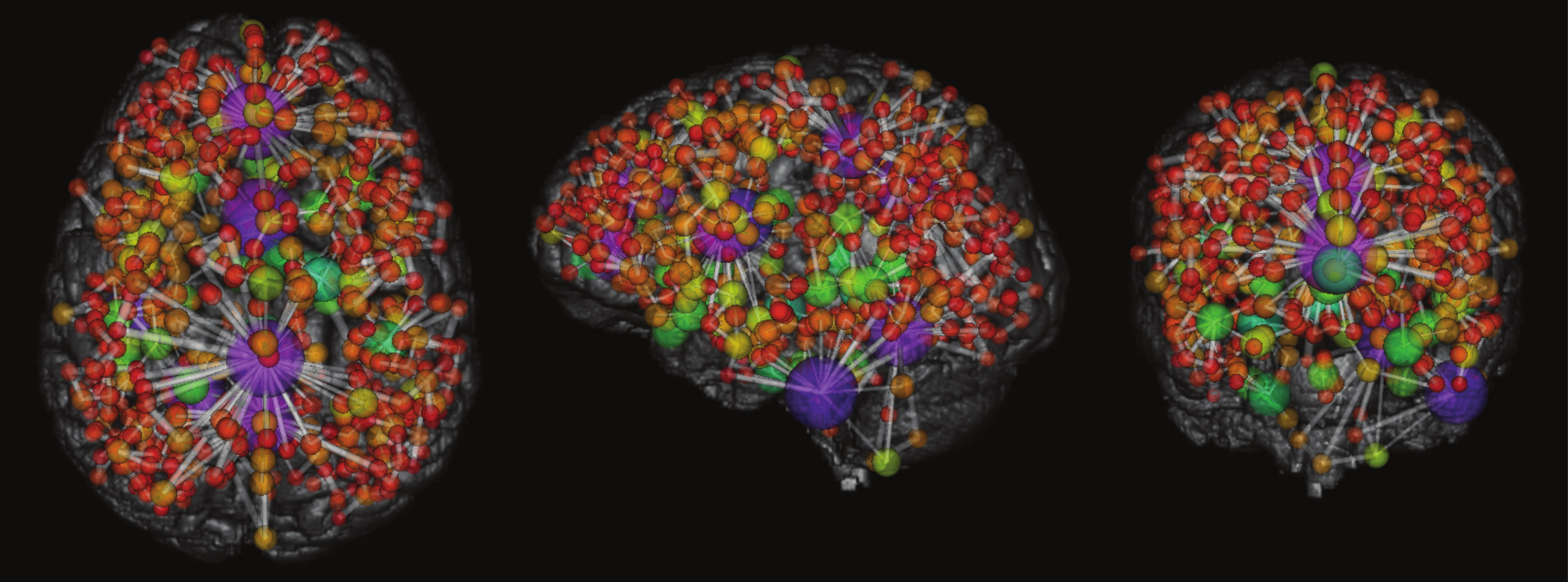}
    \captionsetup{width=0.9\textwidth}
    \caption{Illustration of first nearest neighbor graph between features in the scale-space based on the geometrical kernel $\mathcal{K}(g_i,g_j)$. Feature scales are represented by colors, from red for smaller ones to blue for larger ones.}
    \label{fig:sift_graph}
\end{figure}

% Feature fusion
Another conceivable future study route would be the combination of learn and memory-based models to accomplish some coarse to fine hierarchical classification, comparable to how the human brain recognises a face before identifying a specific individual. Deep learning algorithms might then be trained using local features to determine if they are more unique to their class (a general feature) or to their instance (a specific feature).

\end{conclusion}

%%%%%%%%%%%%%%%%%%%%%%%%%%%%%%%%%%%%%%%%%%%%%%%%%%%
%  Appendix example:
%%%%%%%%%%%%%%%%%%%%%%%%%%%%%%%%%%%%%%%%%%%%%%%%%%%
\appendix
%% To use more than one appendix
%\multiannexe

%% Appendix from an external file
% \include{extApp}

\chapter{Diffusion Orientations Histograms (DOH) for Diffusion Weighted Image Analysis}
\chaptermark{Diffusion Orientations Histograms (DOH)}
\label{app:DOH}

\articleAuthors{
{Laurent Chauvin\textsuperscript{a,b}}{Kuldeep Kumar\textsuperscript{a}}{Christian Desrosiers\textsuperscript{a}}{Jacques De Guise\textsuperscript{b}}{and Matthew Toews\textsuperscript{a}}
}{
{\setstretch{1.2}
\textsuperscript{a} Laboratory for Imagery, Vision and Artificial Intelligence (LIVIA), \\ \'Ecole de Technologie Sup\'erieure, Montreal, Canada%,\\
%1100 Notre-Dame Ouest, Montréal, Québec, Canada H3C 1K3
\\
\textsuperscript{b} Laboratoire de recherche en imagerie et orthop\'edie (LIO), \\ \'Ecole de Technologie Sup\'erieure, Montreal, Canada%,\\ %3175 Chemin de la Côte-Sainte-Catherine, Montreal, Québec, Canada H3T 1C5
\\~\\
Presented as poster in \textit{Computational Diffusion MRI}, MICCAI 2017}
}
\\
\textbf{Abstract}\\
This paper proposes a novel keypoint descriptor for Diffusion Weighted Image \index{Diffusion Weighted Image}(DWI) analysis, the Diffusion Orientation Histogram \index{Diffusion Orientation Histogram} (DOH). The DOH descriptor quantizes local diffusion gradients into histograms over spatial location and orientation, in a manner analogous to the quantization of image gradients in the widely used Histogram of Oriented Gradients \index{Histogram of Oriented Gradients} (HOG) technique. Diffusion gradient symmetry allows representing half of the orientation space at double the angular resolution, leading to a compact but highly informative descriptor. Quantitative preliminary experiments evaluate descriptors for the task of automatically identifying familial links (twins, non-twin siblings) from DWI keypoint correspondences. The DOH descriptor is found to be complementary to traditional HOG descriptors computed from scalar fractional anisotropy (FA) images, where concatenated DOH and HOG descriptors result in the highest rates of correct family member identification. Twin-twin descriptor correspondences are generally more concentrated about major white matter tracts, e.g. the internal capsule, in comparison to twin/non-twin sibling correspondences.

\section{Introduction}
Diffusion magnetic resonance imaging (dMRI) \index{dMRI} data offers the ability to observe patterns of brain connectivity in-vivo. dMRI data, particularly when sampled at high angular resolutions, is large and unwieldy, as diffusion information may be sampled over a large set of orientation directions. Large-scale Diffusion Weighted Images (DWI) analysis is thus typically performed based on reduced data representations for DWI, e.g. diffusion tensors~\citep{LeBihan2001DiffusionApplications,Basser2000VivoData}, tractography~\citep{Jahanshad2011SexTwins}, etc. 

In this paper we consider a potential alternative, the keypoint representation, where analysis focuses on a subset of salient, informative image local patches or features. Local feature information can be encoded into informative descriptors, and subsequently used in families of highly efficient analysis algorithms based on approximate nearest neighbors (NN) \index{Approximate Nearest Neighbors}, e.g. kernel density estimation (KDE). Approximate search routines can compute NN correspondences in $O(\log~N )$ complexity in memory and computation time given $N$ descriptors~\citep{Muja2009FastConfiguration}, and thus scale gracefully to arbitrarily large data sets, as opposed to $O(N)$ for brute force/naive correspondence methods. Furthermore, NN methods come with important theoretical guarantees, e.g. they approach near optimal Bayes error rates as $N\rightarrow\infty$~\citep{Cover1967NearestClassification}.

Assuming salient DWI keypoints can be identified, how can DWI information be most effectively encoded? The computer vision literature shown descriptors based on histograms of orientated image gradients (HOG), e.g. SIFT~\citep{Lowe2004DistinctiveKeypoints} \index{SIFT}, HOG~\citep{Dalal2005HistogramsDetection}, BRIEF~\citep{Calonder2010BriefFeatures} to be among the most effective at achieving high correct correspondence rates. The HOG encoding is highly competitive with alternative encodings based on recent convolutional neural networks (CNN)~\citep{Dong2015DomainsizeDSPSIFT}~\citep{Zheng2017SIFTRetrieval}.

In this work, we hypothesize that keypoints representing informative, local diffusion patterns can be identified and encoded for efficient DWI analysis, e.g. for tasks including registration, classification, etc. We propose a novel descriptor, the diffusion orientation histogram (DOH), that naturally extends the HOG descriptor to encode {\em diffusion gradient information}, as opposed to traditional image gradient information. We propose novel techniques for encoding diffusion gradients, accounting for expected diffusion distance and diffusion orientation symmetry. Experiments demonstrate that DOH descriptors are effective and offer important, complementary information, in comparison to standard HOG descriptors encoded from scalar fractional anisotropy (FA) images.

\section{Related Work}
The DOH descriptor bridges two major fields of research, local keypoint methods and DWI analysis. Keypoint methods reduce large images into collections of local salient keypoints, that can be efficiently encoded for NN methods. The most effective descriptors have been based on local image gradient information~\citep{Lowe2004DistinctiveKeypoints,Calonder2010BriefFeatures,Dalal2005HistogramsDetection}. They have been shown to be among the most effective for image correspondence tasks~\citep{Mikolajczyk2005PerformanceDescriptors}. They are the basis for the widely used SIFT descriptor~\citep{Lowe2004DistinctiveKeypoints}, which models image content in terms of multi-scale Gaussian derivatives~\citep{Lindeberg1994ScalespaceScales}. HOG-like filters, i.e. localized oriented gradient filters, result from machine learning approaches such as independent component analysis (ICA)~\citep{Bell1997IndependentFilters} or CNNs~\citep{Krizhevsky2012ImageNetNetworks} when applied to natural images, and have been identified in the mammalian visual system, i.e. orientation-sensitive hypercolumns~\citep{Hubel1962ReceptiveCortex}.

In medical imaging, dMRI allows observation of water molecule diffusion in-vivo, locally and throughout the brain, and thus inference of neural connectivity patterns \index{Neural Connectivity Patterns}, as water tends to diffuse preferentially in the direction of axonal fibers~\citep{LeBihan1986MRDisorders}. dMRI is thus a critical tool for in-vivo study of white matter geometry throughout the brain. A major challenge in large-scale DWI analysis, e.g. twin studies~\citep{VanEssen2012HumanPerspective,Jahanshad2011SexTwins}, is coping with the sheer amount of directional data. Traditionally dMRI data are processed in terms of reduced data representations such scalar images (e.g. Fractional Anisotropy (FA), Mean Diffusivity) fiber tracts~\citep{Jahanshad2011SexTwins}, diffusion tensors (DTI)~\citep{LeBihan2001DiffusionApplications,Basser2000VivoData}, etc. 

\section{Diffusion Orientation Histogram Descriptors}

We propose encoding dMRI data in terms of Diffusion Orientation Histograms, a model-free natural extension of the ubiquitous HOG descriptor from computer vision. In this section we describe the computational framework for DWI and then the DOH encoding.

{\bf Diffusion Framework:} Consider random variables of 3D spatial location $x \in R^3$, time $t \in R$, with corresponding displacement vectors $\Delta x$ and $\Delta t$. Assuming temporally stationary diffusion and a constant diffusion time $\Delta t$, the diffusion of a water molecule can be modeled as a posterior density over a random displacement vector $\Delta x$ conditioned on location $x$ a time period $\Delta t$:
\begin{equation}
p(\Delta x, \Delta t | t, x) \propto p(\Delta x | \Delta t, x)
\label{eq:odf}
\end{equation}
Equation~\eqref{eq:odf} is referred to as the Ensemble Average Propagator (EAP)~\citep{Descoteaux2009DiffusionPropagator} \index{Ensemble Average Propagator} (or true diffusion propagator~\citep{Cory1990MeasurementCompartmentation} under the \textit{narrow pulse approximation}). Representing net 3D displacement vector $\Delta x = \{r,\angle \Theta\}$ in polar coordinates of magnitude $r=\|\Delta x\|$ and 2D orientation $\angle\Theta = \{\theta,\phi\}$, the EAP may be expressed using Bayes rule as:
\begin{equation}
p(\Delta x | \Delta t, x) = p(r, \angle\Theta | \Delta t, x) = p(r| \angle\Theta, \Delta t, x)p(\angle\Theta|\Delta t, x)
\label{eq:odf2}
\end{equation}
where in Equation~\eqref{eq:odf2}, the EAP is factored into conditional densities over orientation density $p(\angle\Theta|\Delta t, x)$ and magnitude $p(r|\angle\Theta, \Delta t, x)$ conditional on orientation $\angle\Theta$.

Without loss of generality, orientation density  $p(\angle\Theta|\Delta t, x)$ is taken to be uniform, and the EAP is thus characterized by the conditional density of diffusion magnitude $p(r| \angle\Theta, \Delta t, x)$ along direction $\angle\Theta$. Adopting the Brownian motion model of diffusion~\citep{Einstein1905UberTeilchen}, this may be expressed as a unidimensional Gaussian density of the form
\begin{equation}
p(r|\angle\Theta, \Delta t, x) \propto exp -\{r^2 / 4\Delta t ADC_{\angle\Theta}\}
\label{eq:odf3}
\end{equation}
centered on mean $x$ with variance $2\Delta t~ADC_{\angle\Theta}$~\citep{Alexander2002DetectionData}, where $ADC_{\angle\Theta}$ is the apparent diffusion coefficient (ADC) measured along orientation $\angle\Theta$ in DWI. It is this density which we estimate and use to encode DWI gradient orientation information.

dMRI data comes in the form of images sampled at a set of diffusion-sensitizing gradient angles $\angle\Theta$. The DWI voxel intensity $S_{\angle\Theta}$ is proportional to the displacement of water molecules in the gradient direction $\angle\Theta$ applied during the acquisition, and the $ADC_{\angle\Theta}$ is estimated as
\begin{equation}
\label{eq:VoxIntensity}
ADC_{\angle\Theta} = -\frac{1}{b}ln\left(\frac{S_{\angle\Theta}}{S_{0}}\right)
\end{equation}
where $S_{0}$ is the baseline signal without diffusion gradient and $b$ is the $\textit{b-value}$ defined by the Stejskal-Tanner equation~\citep{Stejskal1965SpinGradient} as $b=\gamma^{2}G^{2}\delta^{2}(\Delta t - \frac{\delta}{3})$, with the gyromagnetic ratio $\gamma$, the sensitizing gradient pulse amplitude $G$, the gradient duration $\delta$, and the diffusion time $\Delta t$ between gradient pulses.

{\bf Diffusion Orientation Histogram (DOH) Descriptor:} The DOH descriptor coarsely quantizes EAP information, i.e. $p(r|\angle\Theta,\Delta t, x)$, into histograms over spatial location $x$ and orientation $\angle\Theta$ in a manner reminiscent of the widely used HOG descriptor. Intuitively, coarse quantization provides immunity to noise inherent to keypoint localization, e.g. small shifts or rotations. 

Descriptors are computed at salient 3D keypoint regions defined by location and scale $\{x,\sigma\}$ within the image. We assume keypoints can be identified generic saliency operators in scale-space. Following the approach of~\citep{Toews2013EfficientFeatures}, we experiment with a quantization structure of 8 spatial and 8 orientation bins for a 64-element descriptor. Given the symmetric nature of diffusion, we also consider a half-sphere structure with 8 approximately equally distributed orientations on the half-sphere. We refer to these sampling structures as full-sphere (FS) and half-sphere (HS).

\begin{figure}
	\centering
	\includegraphics[width=0.9\linewidth]{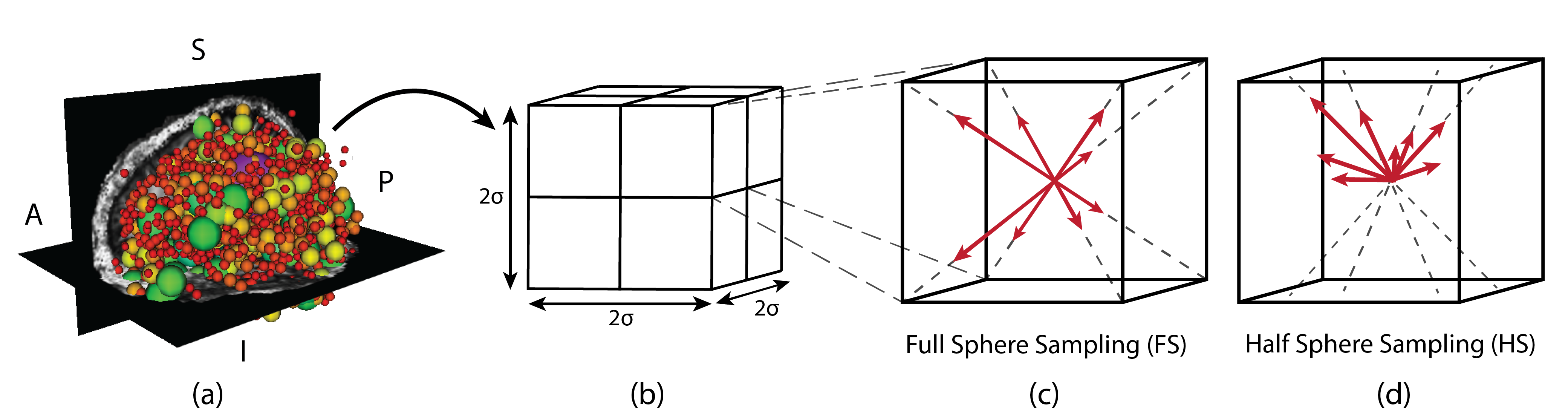}
	\caption[DOH Descriptor]{The DOH descriptor is computed locally at keypoint locations, e.g. (a) keypoints detected in an FA volume. For each keypoint, local diffusion propagators are quantized into histograms of (b) 8 spatial bins and two 8-bin orientation sampling schemes: (c) full sphere (FS) and (d) half sphere (HS)}
	\label{fig:doh_descriptor}
\end{figure}

The primary design question is how to encode gradient information in the form of diffusion coefficients $ADC_{\angle\Theta}$ given the EAP in Equation~\eqref{eq:odf3}, defined by the ratio $\frac{r^2}{\Delta t}$ of squared diffusion magnitude $r$ and time $\Delta t$ values, that may not be known. For this, we consider a soft bin increment value based on the EAP in Equation~\eqref{eq:odf3}, which for an observed diffusion coefficient $ADC_{\angle\Theta}$ is
\begin{equation}
inc( ADC_{\angle\Theta} ) \propto  p(r|\angle\Theta,\Delta t, x).
\label{eq:odf4}
\end{equation}

\section{Experiments}

Experiments assess the utility of the DOH descriptor in feature matching experiments, where nearest neighbor (NN) inter-subject descriptor correspondences are used to automatically infer familial relationships based on a set of DWIs of siblings including twins and non-twin siblings. We use the Human Connectome Project (HCP)~\citep{VanEssen2012HumanPerspective} \index{Human Connectome Project} Q3 Release dMRI data, acquired on a Siemens Skyra $3$T scanner, at isotropic $1.25mm$ resolution. For evaluation here, we consider a subset of 45 unique subject dMRI volumes from 15 families of 3 siblings each, including 2 monozygotic twins (T1, T2) and 1 non-twin sibling (NT), 23-35 years of age.

For each DWI, a fractional anisotropy (FA) is generated using the Dipy software package~\citep{Garyfallidis2014DipyData}. A set of 3D keypoints is then extracted as extrema of the difference-of-Gaussian scale-space~\citep{Toews2013EfficientFeatures} constructed from FA, and is used for all local descriptor evaluations. Keypoint extraction requires on the order of 20 seconds per image, each image results in approximately 2000 keypoints, see Figure~\ref{fig:doh_descriptor}. Using a single set of keypoints ensures that descriptor comparisons are based on precisely the same underlying diffusion information.

To quantitatively assess descriptor performance, we consider the task of automatically recovering familial relationships from the structure of the inter-subject proximity graph, as proposed in~\citep{Toews2016HowFeatures}. 
Let $A=\{a_{i}\}$ and $B=\{b_{j}\}$ be sets of D-dimensional descriptors $a_{i},b_{j} \in R^{D}$, where $|A| \neq |B|$. The distance between two sets of different cardinality $d(A,B)$ is then measured by their Jaccard distance, ranging from 0 to 1:
\begin{equation}
d(A,B) = J(A,B) = 1 - \frac{|A \cap B|}{|A \cup B|},
\label{eq:jaccard}
\end{equation}
where set intersection $|A \cap B|$ is defined by the number of NN descriptor correspondences between sets $A$ and $B$. Note that the distance $d(a_i,b_j)$ between two descriptors $a_{i}$ and $b_{i}$ is defined by the Euclidean distance.
In general, we expect the Jaccard distances from (\ref{eq:jaccard}) to reflect family relationships as siblings share more genetic material (especially twins) than unrelated subjects, such that: 
\begin{equation}
d(T_{1}, T_{2}) < d(T_{1}, NTS) \approx d(T_{2}, NTS) < d(T_{1}, U) \approx d(T_{2}, U) \approx d(NTS, U)
\label{eq:relationships}
\end{equation}
$T_{1}$,$T_{2}$ and $NTS$ being respectively twin and non-twin siblings of the same family, and $U$ an unrelated subject from a different family.
To assess the ability to recover family relationships based on sets of local features, the correct relationship recall rate is calculated for twin-twin $\gamma_{T}$ and twin-non-twin $\gamma_{NT}$ relationships separately, such as:
\begin{equation}
	\gamma_{T} = \frac{|\{ \Gamma_{T} \}|}{|\{ \Omega_{T} \}|}, \gamma_{NT} = \frac{|\{ \Gamma_{NT} \}|}{|\{ \Omega_{NT} \}|},
\end{equation}
$\{\Gamma_{T}\}$ and $\{\Gamma_{NT}\}$ being respectively sets of correct Twin and Non-Twin recovered relationships based on local features, and $\{\Omega_{T}\}$ and $\{\Omega_{NT}\}$ respectively the sets of Twin and Non-Twin relationships in the whole dataset (here $\Omega_{T} = 15$ and $\Omega_{NT} = 30$).

\begin{table}
\caption{Correct family relationship recovery rates for twins $\gamma_{T}$ and non-twins $\gamma_{NT}$ for five descriptor configurations}\label{tab:results}
\begin{center}
\begin{tabular}{c |@{\quad} c@{\quad} c c@{\quad} c@{\qquad} c@{\quad} c@{\quad}}
\hline
\multicolumn{1}{c|@{\quad}}{\rule{0pt}{12pt}Config.}&
\multicolumn{1}{c@{\quad}}{Type}&
\multicolumn{1}{c@{\quad}}{Sampling}&
\multicolumn{1}{c@{\quad}}{Orient. Bins}&
\multicolumn{1}{c@{\qquad}}{D}&
\multicolumn{2}{c@{\quad}}{Relationship Recall}\\
\multicolumn{1}{c|@{\quad}}{}&
\multicolumn{4}{c@{\quad}}{}&
\multicolumn{1}{c@{\quad}}{$\gamma_{T}$}&
\multicolumn{1}{c@{\quad}}{$\gamma_{NT}$}\\
\hline\rule{0pt}{12pt}1 & FA & FS & 8 & 64 & 0.90 & 0.40\\
2 & DWI & FS & 8 & 64 & 1.00 & 0.20\\
3 & FA + DWI & FS + FS & 8 + 8 & 128 & 1.00 & 0.33\\
%4 & DWI & HS & 4 & 32 & 0.97 & 0.27\\
%5 & FA + DWI & FS + HS & 8 + 4 & 96 & 0.93 & 0.33\\
4 & DWI & HS & 8 & 64 & 0.97 & 0.33\\
\textbf{5} & \textbf{FA + DWI} & \textbf{FS + HS} & \textbf{8 + 8} & \textbf{128} & \textbf{1.00} & \textbf{0.40}\\[2pt]
\hline
\end{tabular}
\end{center}
\end{table}

We compare recall rates $\gamma_{T}$ and $\gamma_{NT}$ for five different keypoint descriptors (see Table \ref{tab:results}):  1) standard HOG descriptors computed from scalar FA, 2) DOH descriptors computed from DWI, 3) a concatenated descriptor combining HOG and DOH information 4) a DOH descriptor taking advantage of diffusion symmetry to sample half of the Orientation Distribution Function (ODF) sphere (HS) at twice the sampling resolution and 5) a concatenated descriptor of 1) and 4). We expect more effective encodings to result in higher numbers of correct inter-subject correspondences, and thus inter-subject proximity graphs that more closely resemble the known family structure.

\begin{figure}
	\centering
	\includegraphics[width=0.75\linewidth]{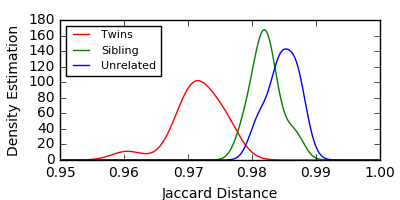}
	\caption[Jaccard Density Estimation]{Kernel density estimates of pairwise Jaccard distances for three groups of subject pairs: Twin, Sibling and Unrelated subject. Experimental results confirm theoretical expectations from (\ref{eq:relationships})}
	\label{fig:jaccard_density_estimation}
\end{figure}

\begin{figure}
	\centering
	\includegraphics[width=0.9\linewidth]{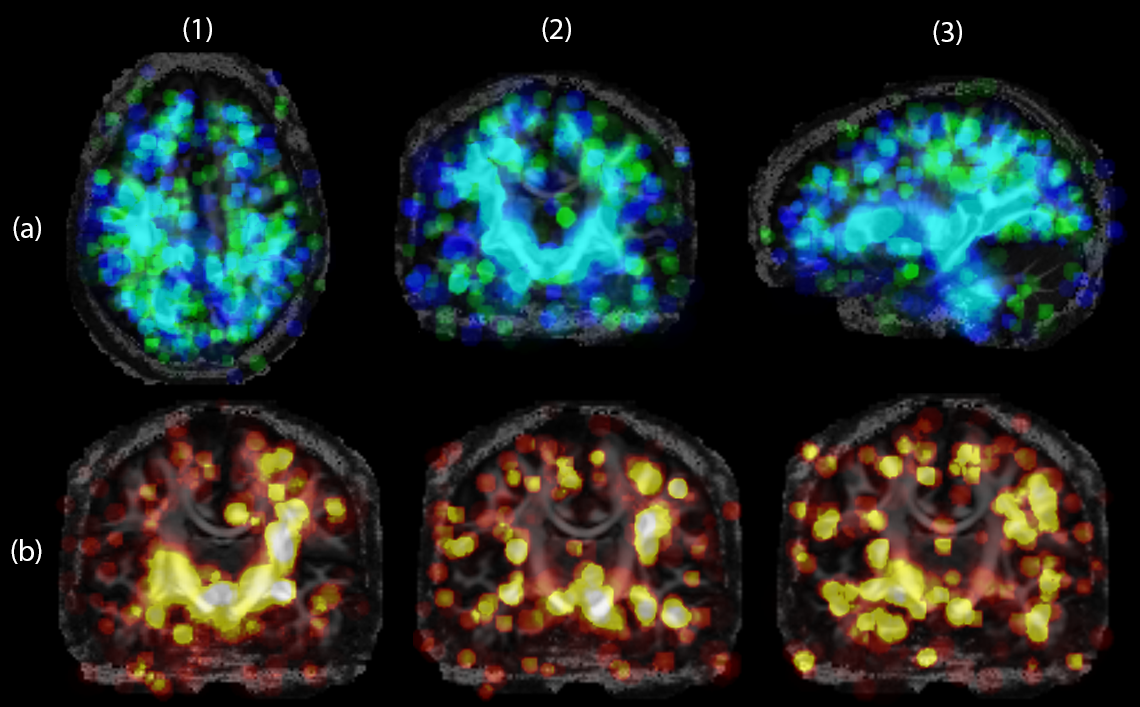}
	\caption[DOH vs FA, Twin vs Non-Twin Descriptor Spatial Location]{Visualizing the spatial distributions correct NN feature correspondences. In row a), for DWI-DOH (green) vs FA-HOG descriptors (blue) in (a1-a3) axial, coronal and sagittal views. In row b), correspondences between twins T1-T2 in (b1) and non-twin siblings pairs in (b2) T1-NT (b3) T2-NT}
	\label{fig:DWI_vs_FA_Twin_vs_NonTwin}   
\end{figure}

Results are shown in Table~\ref{tab:results}, where best results are obtained using concatenated FA-HOG and DWI-DOH descriptors sampled over a half sphere. Figure~\ref{fig:DWI_vs_FA_Twin_vs_NonTwin} illustrates the spatial distributions for NN feature correspondences. In row (a), DWI-DOH and FA-HOG descriptors lead to similar distributions of correspondences throughout the brain, and are thus generally complementary. In row (b), DWI correspondences are generally more concentrated in major white matter bundles for twins (b1) vs non-twins (b2,b3), this may indicate greater heritability. Finally, Figure~\ref{fig:jaccard_density_estimation} show distributions of Jaccard distances for twins vs non-twins vs unrelated subject groups using descriptor configuration 5 (cf Table~\ref{tab:results}), which generally reflect genotypical and phenotypical variations, as described in (\ref{eq:relationships}).

\section{Discussion}

In this paper, we propose the diffusion orientation histogram (DOH) descriptor for DWI analysis, in a manner analogous to intensity gradient descriptors widely used in the computer vision literature. We provide the first preliminary keypoint-based analysis of sibling relationships from DWI data, were results show the DOH encoding significantly augments scale FA feature analysis with complementary information, resulting in more effective identification of between family members based on local patterns of neural connectivity. 

An important aspect is that of keypoint detection - this work adopts keypoints identified in scalar FA images in order to ensure a fair comparison of descriptors. Keypoints derived according to FA saliency are not necessarily optimal for full dMRI data, however, and future work will investigate salient feature detection methods specific to full dMRI data. The fact that DOH descriptors generally outperform FA-HOG descriptors here is thus a further indication of the effectiveness of the encoding. Experiments here are limited to a subset of HCP data with a specific family structure, future work will apply DOH to the entire dataset. We evaluate the DOH representation at sparse keypoints, DOH descriptor may prove a useful representation for dense image analysis.

It should also be noted that by design, the DOH descriptor also have the ability to compare diffusion data acquired under different protocols (e.g. cross-site, cross-scanner), as experimental preliminary results show robustness to variable q-space sampling scheme.

%\bibliographystyle{splncs03}
%\bibliography{references}

\chapter{Laplacian-of-Gaussian Particle Theory}
\chaptermark{Laplacian-of-Gaussian Particle Theory}
\label{app:particle}
%The Gaussian kernel is the solution to the heat equation and the diffusion equation Brownian motion~\citep{einstein1905} $\frac{dI(\vect{x},\sigma)}{dt}=D\nabla^2 I(\vect{x},\sigma)$, where $D$ is a constant diffusion rate and $\nabla^2$ is the Laplacian operator. 

%The scale-space $I(\vect{x},\sigma)$ may thus thought of as simulating the isotropic diffusion of image intensity information $I(\vect{x})$ for a time period $t=\sigma^2$, where the intensity $I(\vect{x},\sigma)$ represents the image content according to a receptive field of spatial extent proportional to scale $\sigma$.

% The via the following differential equation:
% \begin{align}
%     \frac{\partial I(\vect{x},t)}{\partial t}=D\nabla^2 I(\vect{x},t),\\
%     -\dfrac{\hslash^2}{2m} \, \dfrac{\mathrm{d}^2 \psi}{\mathrm{d} x^2}
% \end{align}

Our work is based on Laplacian-of-Gaussian, here we emphasize the connection between fields and to scale-space. A single particle may be defined as a complex-valued wave function $I(\vect{x},t)$ in space time $\vect{x},t$ satisfying the Schrodinger Equation~\citep{Schrodinger1926UndulatoryMolecules}:
\begin{align}
    \frac{i\partial I(\vect{x},t)}{\partial t}=D\nabla^2 I(\vect{x},t)
    \label{eq:schro}
\end{align}
where $\frac{\partial I(\vect{x},t)}{\partial t}$ is the partial derivative of $I(\vect{x},t)$ with respect to time, $\nabla^2 I(\vect{x},t) = \frac{d^2I}{dx^2}+\frac{d^2I}{dy^2}+\frac{d^2I}{dz^2}$ is the Laplacian operator over spatial coordinates $\{x,y,z\}$, $i=\sqrt{-1}$ is the imaginary unit and D is a constant. In the case of a single real 3D image $I(\vect{x})$, the partial derivative $\partial I(\vect{x},t)$ in Equation~\eqref{eq:schro} may not be evaluated. However the assumption of a passive diffusion process may be used to model the evolution of the image function over time $t$, and is expressed via the 3D heat or diffusion equation (Brownian motion~\citep{Einstein1905UberTeilchen}), equivalent to Equation~\eqref{eq:schro} without $i$ and where $D$ is a constant diffusion rate. The solution is the Gaussian function where the variance is proportional to $t$:
\begin{align}
I(\vect{x},t) = \frac{1}{\sqrt{(4\pi D t)^{3}}}\exp{ \left(-\frac{\|\vect{x}-\vect{u}\|^2}{4D t}\right)}
\end{align}

The scale-space $I(\vect{x},\sigma)$ may thus thought of as simulating the isotropic diffusion of image intensity information $I(\vect{x})$ for a time period $t=\sigma^2$, where the intensity $I(\vect{x},\sigma)$ represents the image content according to a receptive field of spatial extent proportional to scale $\sigma$.

The isotropic Gaussian scale-space $I(\vect{x},\sigma)$ is then defined as: 
\begin{align}
I(\vect{x},\sigma)=I(\vect{x}) \ast N(\vect{x},\sigma)    
\end{align}
by convolution of the image $I(\vect{x})$ with Gaussian kernel $N(\vect{x},\sigma) \propto e^{-(x^2+y^2+z^2)/2\sigma^2}$ with variance $\sigma^2$.

The Laplacian-of-Gaussian operator is defined as follows:
\begin{align}
\nabla^2 I(\vect{x},\sigma) &= \frac{\partial^2}{\partial x^2} I(\vect{x},\sigma) + \frac{\partial^2}{\partial y^2} I(\vect{x},\sigma) + \frac{\partial^2}{\partial z^2}I( \vect{x},\sigma) \notag \\ 
&= \frac{x^2+y^2+z^2-3\sigma^2}{\sigma^4}e^{-(x^2+y^2+z^2)/2\sigma^2}
\end{align}
The Laplacian-of-Gaussian operator may be approximated by the difference-of-Gaussian operator:
\begin{align*}
\nabla^2 I(\vect{x},\sigma) \approx I(\vect{x},\sigma)-I(\vect{x},\kappa\sigma)
\end{align*}
where $\kappa$ is a multiplicative difference in scale sampling.

In digital image processing, early work by Marr and Hildredth used the Laplacian-of-Gaussian as an edge detector via zero crossings~\citep{Marr1980TheoryDetection}. Witkin in 1D audio signals~\citep{Witkin1984ScalespaceDescription} and Koenderink in 2D image processing~\citep{Koenderink1984StructureImages} demonstrated that a scale-space $I(\vect{x},\sigma)$ may be constructed via convolution of the image $I(\vect{x},\sigma)=I(\vect{x})*G(\sigma)$ with an isotropic Gaussian kernel $G(\sigma) \propto exp-\|\vect{x}\|^2/2\sigma^2$ of standard deviation $\sigma$. Burt described the Laplacian pyramid, where the Laplacian-of-Gaussian (LoG) function may be evaluated via a difference-of-Gaussian (DoG)~\citep{Burt1987LaplacianCode}. Lindeberg showed the Gaussian kernel was uniquely consistent with scale-space axioms~\citep{Lindeberg1998FeatureSelection} including non creation or enhancement of local image maxima, and thus served as a mechanism for identifying salient image features in a manner invariant to image rotation, translation and scaling. Lowe proposed identifying Difference-of-Gaussian maxima as distinctive image keypoints~\citep{Lowe2004DistinctiveKeypoints}, which identified distinctive keypoints which maximize the Laplacian-of-Gaussian operator $\nabla^2 I(\vect{x},\sigma)$ 
Note also that center-surround filters reminiscent of the Laplacian are found in the receptive field of the mammalian visual cortex~\citep{Hubel1962ReceptiveCortex}.

\chapter{CPD Registration Details}
\chaptermark{CPD Registration Details}
\label{app:solve}
In the case of rigid transform, the $solve$ function is defined for rigid and affine transforms as~\citep{Myronenko2010PointDrift}:
\begin{algorithm}
    \DontPrintSemicolon
    \SetKwFunction{FMain}{solve\_rigid}
    \SetKwProg{Fn}{}{:}{}
    \Fn{\FMain{$\mathcal{F}$, $\mathcal{M}$, $P$}}{
        $\mathit{N}_{\bf{P}}=\bf{1}^{\mathit{T}}\bf{P1},~ \mu_\mathcal{F}=\frac{1}{\mathit{N}_{\bf{P}}}\bm{\mathcal{F}}^{\mathit{T}}\bf{P}^{\mathit{T}}\bf{1},~\mu_\mathcal{M}=\frac{1}{\mathit{N}_{\bf{P}}}\bm{\mathcal{M}}^{\mathit{T}}\bf{P1},$\;
        $\bm{\hat{\mathcal{F}}}=\bm{\mathcal{F}}-\bf{1}\mu_{\mathcal{F}}^{\mathit{T}},~\bm{\hat{\mathcal{M}}}=\bm{\mathcal{M}}-\bf{1}\mu_{\mathcal{M}}^{\mathit{T}},$\;
        $\bf{A}=\bm{\hat{\mathcal{F}}}^{\mathit{T}}\bf{P}^{\mathit{T}}\bm{\hat{\mathcal{M}}}$, compute SVD of $\bf{A}$ such as $\bf{A}=\bf{U}\textit{SS}\bf{V}^{\mathit{T}},$\;
        $\bf{R}=\bf{UCV}^{\mathit{T}}$, where $\bf{C}=\mathrm{d(1,..,1,det(\bf{UV}^{\mathit{T}}))},$\;
        $s=\frac{tr(\bf{A}^{\mathit{T}}\bf{R})}{tr(\bm{\hat{\mathcal{M}}}^{\mathit{T}}d(\bf{P1})\bm{\hat{\mathcal{M}}})},$\;
        $\bf{t}=\mu_{\mathcal{F}}-s\bf{R}\mu_{\mathcal{M}},$\;
        $\sigma^2 = \frac{1}{\mathit{N}_{\bf{P}}D}(tr(\bm{\hat{\mathcal{F}}}^{\mathit{T}}\mathrm{d(\bf{P}^{\mathit{T}}\bf{1})}\bm{\hat{\mathcal{F}}})-\mathnormal{s~tr(\bf{A}^{\mathit{T}}\bf{R}))}.$\;
        \KwRet $\{s,\bf{R},\bf{t}\},\sigma^2$
    }
    \label{alg:solve_rigid}
    \SetAlgoCaptionLayout{centerline}
    \caption{Solve for rigid transform}
\end{algorithm}

\begin{algorithm}
    \DontPrintSemicolon
    \SetKwFunction{FMain}{solve\_affine}
    \SetKwProg{Fn}{}{:}{}
    \Fn{\FMain{$\mathcal{F}$, $\mathcal{M}$, $P$}}{
    $\mathit{N}_{\bf{P}}=\bf{1}^{\mathit{T}}\bf{P1},~ \mu_\mathcal{F}=\frac{1}{\mathit{N}_{\bf{P}}}\bm{\mathcal{F}}^{\mathit{T}}\bf{P}^{\mathit{T}}\bf{1},~\mu_\mathcal{M}=\frac{1}{\mathit{N}_{\bf{P}}}\bm{\mathcal{M}}^{\mathit{T}}\bf{P1},$\;
    $\bm{\hat{\mathcal{F}}}=\bm{\mathcal{F}}-\bf{1}\mu_{\mathcal{F}}^{\mathit{T}},~\bm{\hat{\mathcal{M}}}=\bm{\mathcal{M}}-\bf{1}\mu_{\mathcal{M}}^{\mathit{T}},$\;
    $\bf{B}=(\bm{\hat{\mathcal{F}}}^{\mathit{T}}\bf{P}^{\mathit{T}}\bm{\hat{\mathcal{M}}})(\bm{\mathcal{M}}^{\mathit{T}}\mathrm{d(\bf{P1})}\bm{\hat{\mathcal{M}}})^{-1},$\;
    $\bf{t}=\mu_{\mathcal{F}}-\bf{B}\mu_{\mathcal{M}},$\;
    $\sigma^2 = \frac{1}{\mathit{N}_{\bf{P}}D}(tr(\bm{\hat{\mathcal{F}}}^{\mathit{T}}\mathrm{d(\bf{P}^{\mathit{T}}\bf{1})}\bm{\hat{\mathcal{F}}})-\mathnormal{tr(\bm{\hat{\mathcal{F}}}^{\mathit{T}}\bf{P}^{\mathit{T}}\bm{\hat{\mathcal{M}}}\bf{B}^{\mathit{T}}))}.$\;
    \KwRet $\{\bf{B},\bf{t}\},\sigma^2$
    }
    \label{alg:solve_affine}
    \SetAlgoCaptionLayout{centerline}
    \caption{Solve for affine transform}
\end{algorithm}

%%%%%%%%%%%%%%%%%%%%%%%%%%%%%%%%%%%%%%%%%%%%%%%%%%%
% BIBLIOGRAPHY AND REFERENCES
%%%%%%%%%%%%%%%%%%%%%%%%%%%%%%%%%%%%%%%%%%%%%%%%%%%

% TODO: CLEAN THE REFERENCES (SOME ARE DOUBLED BETWEEN MENDELEY AND TMI/NEUROIMAGE)

%%- Bibliography -%%
\newpage
% Single spacing for the bibliography
\begin{spacing}{1}
    \setlength{\bibsep}{\baselineskip}
	%\nocite{*} % The "nocite" command can be used to print references that haven't been used in the document. The "*" option specifies that every reference should be printed
	\bibliographystyle{bibETS} % ETS bibliography style
	\addcontentsline{toc}{chapter}{BIBLIOGRAPHY} % Addition of the bibliography in the table of contents

	\bibliography{Zotero} % List of bibliography files, biblio.bib is an example

\end{spacing}

%%- Other list of references, "refs" example --%
%%%%%%%%%%%%%%%%%%%%%%%%%%%%%%%%%%%%%%%%%%%%%%%%%%%
% IMPORTANT: HOW TO COMPILE AND PRINT ADDITIONAL REFERENCES (replace "refs" by the chosen name)
%%%%%%%%%%%%%%%%%%%%%%%%%%%%%%%%%%%%%%%%%%%%%%%%%%%
% Follow these three steps:
%   1. Compile the document once, to save the used references in refs.aux
%   2. Compile the references
% 		- On Linux: Use the "bibtex refs" command in the document folder
%		- On MacOSX (MacTex distribution): Use the "/usr/texbin/bibtex refs" command in the document folder
%		- On Windows: Edit the "update_refs.bat" script to put the right suffix ("refs" here), and launch the script
%   3. Recompile the document TWICE
%%%%%%%%%%%%%%%%%%%%%%%%%%%%%%%%%%%%%%%%%%%%%%%%%%%

\newpage
% Same commands than for the bibliography, only with the "refs" suffix
\begin{spacing}{1}
    \setlength{\bibsep}{\baselineskip}
	%\nociterefs{*}
	\bibliographystylerefs{bibETS}
	\addcontentsline{toc}{chapter}{LIST OF REFERENCES}

	\bibliographyrefs{refs}

\end{spacing}

\end{document}